%% file: Infant_BH_revised_clean2.tex
\documentclass[longauth]{aa}  

\usepackage{subcaption}
\usepackage{graphicx}
\usepackage{xcolor}
\usepackage{pdflscape}
\usepackage{multirow}
\usepackage[figuresright]{rotating}
\usepackage{txfonts}
\usepackage{hyperref}
\usepackage{xspace}
\hypersetup{
    colorlinks=true,
    linkcolor=blue,
    citecolor=blue,
    filecolor=magenta,      
    urlcolor=cyan,
    pdftitle={Overleaf Example},
    pdfpagemode=FullScreen,
    }

\makeatletter
\renewcommand*\aa@pageof{, page \thepage{} of \pageref*{LastPage}}
\makeatother

\let\oldAA\AA
\renewcommand{\AA}{\oldAA\xspace}

\begin{document}

\title{JADES. The diverse population of infant Black Holes at 4$<$z$<$11: merging, tiny, poor, but mighty}

 \titlerunning{Infant Black Holes}

\input{authors.tex}


\abstract{
Spectroscopy with the James Webb Space Telescope has opened the possibility to identify moderate luminosity Active Galactic Nuclei (AGN) in the early Universe, at and beyond the epoch of reionization, complementing previous surveys of much more luminous (and much rarer) quasars. We present 12 new AGN at 4$<$z$<$7 in the JADES survey (in addition to the previously identified AGN in GN-z11 at z=10.6) revealed through the detection of a Broad Line Region as seen in the Balmer emission lines. The depth of JADES, together with the use of three different spectral resolutions,  enables us to probe a lower mass regime relative to previous studies. In a few cases we find evidence for two broad components of  H$\alpha$, which suggests that these could be candidate merging black holes (BHs), although a BLR complex geometry cannot be excluded. The inferred BH masses range between $\rm 8\times 10^7 ~M_\odot$  down to $\rm 4\times 10^5 ~M_\odot$, interestingly probing the regime expected for Direct Collapse Black Holes. The inferred AGN bolometric luminosities ($\sim 10^{44}-10^{45}$~erg/s) imply accretion rates that are $< 0.5$ times the Eddington rate in most cases. However, small BHs, with $\rm M_{BH} \sim 10^6~M_\odot$, tend to accrete at Eddington or super-Eddington rates. These BHs at z$\sim$4-11 are over-massive relative to their host galaxies stellar masses when compared to the local $\rm M_{BH}-M_{star}$ relation, even approaching $\rm M_{BH}\sim M_{star}$, as expected from heavy BH seeds and/or super-Eddington accretion scenarios. However, we find that these early BH tend to be more consistent with the local relation between $\rm M_{BH}$ and velocity dispersion, as well as between  $\rm M_{BH}$ and dynamical mass, suggesting that these are more fundamental and universal relations. On the BPT excitation-diagnostic diagram these AGN are located in the region that is that is locally occupied by star-forming galaxies, implying that they would be missed by the standard classification techniques if they did not display broad lines. Their location on the diagram is consistent with what expected for AGN hosted in metal poor galaxies ($\rm Z \sim 0.1-0.2~Z_\odot$).
The fraction of broad line AGN with $\rm L_{AGN}>10^{44}~erg/s$ among galaxies in the redshift range $4<z<6$ is about 10\%, suggesting that the contribution of AGN {\it and} their hosts to the reionization of the Universe is $>$10\%.
}
 \keywords{Galaxies: active - 
 Galaxies: high-redshift - Galaxies: luminosity function - quasars: general}
 
\maketitle

\section{Introduction} \label{sec:intro}

Evidence for supermassive black holes, with masses ranging from a few million to several billion solar masses, has been found in the nuclei of most galaxies in the local Universe. The tight relation with many of the host galaxy properties, and in particular with the central velocity dispersion, has been regarded as indication of co-evolution between black holes and their host galaxies \citep{Kormendy13,Greene20}. Models and cosmological simulations envisage different possible co-evolutionary (as well as not co-evolutionary) scenarios, possibly involving galaxy and BH merging, as well as mutual self-regulation via feedback processes \citep[e.g.][]{Sijacki09,volonteri2010,valiant2016,
inayoshi+2020,
greene+2020, Trinca22,
Fan22, Volonteri23, Bennett23,Sassano2023,Koudmani2022}.
Most of these scenarios are degenerate in explaining the scaling relations observed locally. Yet, at high redshift different models and simulations expect different properties for the population of black holes and their relations with their host galaxies \citep[e.g. ][]{Visbal2018,Valiante2018a,Schneider23,Volonteri21rev,Habouzit22,Trinca22}. Therefore, in order to validate, test and discriminate among different scenarios, it is crucial to explore the population of (accreting) black holes at high redshift, along with their host galaxies.

The search and characterisation of accreting black holes and their host galaxies at high redshift has made tremendous progress during the past 20 years
\citep[e.g.][]{Merloni10, Bongiorno14, Trakhtenbrot17,Mezcua2018,
Lyu2022}. However, at z$>$4, until recently, observational constraints limited the identification and characterisation of sources primarily to the very luminous-quasar regime \citep[see ][ for a review]{Fan22}. Within this context,  the discovery of black holes with masses in excess of several billion solar masses already in place at z$>$6-7 has been unexpected \citep[e.g.][]{Banados18,Wang20}, since models and cosmological simulations found it challenging to reproduce the growth of such massive black holes within the limited amount of time since the Big Bang. Different scenarios have been invoked, such as direct-collapse black holes (DCBH), merging of stars and black holes in nuclear clusters, and super-Eddington accretion from stellar mass black hole seeds, possibly originating from `Population III'  remnants 
\citep{inayoshi+2020, greene+2020,Ferrara2014,
Trinca22, Volonteri23, Mckee08,
Banik19,Sassano2021,Singh23,haidar2022,Ni2022,Weller2023,beckman2023,degraf2020}. Testing and differentiating between these different scenarios requires the detection and characterisation of smaller black holes at high redshift.

Major progress has been made thanks to JWST \citep{Gardner2023,Rigby2023}. Indeed, while several AGN candidates have been identified through JWST imaging and broad band, photometric Spectral Energy Distribution \citep{Furtak2022,Onoue2023,Barro2023,Yang2023,Bogdan2023,Ignas2023J}, JWST spectroscopy has revealed broad line AGN at high redshift with moderate to low luminosities. Specifically, both NIRSpec MOS and IFU observations, as well as slitless NIRCam grism spectroscopy, have revealed a population of AGN at z$>$4
and out to z=10.6 with luminosities ($\rm 10^{44}-10^{45} ~erg~s^{-1}$; \citealt{Kocevski23,ubler+2023,Ubler24,
Harikane23BH,Matthee23,maiolino_bh_2023,Juodzbalis2024,Schotlz2023NLAGN}) lower than those of classical QSOs ($\gtrsim 10^{46}$~erg/s). The estimated black hole masses are between $\rm 10^6$ and $\rm 10^8 ~M_\odot$,  significantly lower than those typically inferred for quasars at similar redshifts. Interestingly, based on their narrow line ratios, these systems would not be classified as AGN in classical diagnostic diagrams, such as the BPT diagram \citep{Baldwin81}, since they are primarily located in the region populated by star forming galaxies in the local Universe. This offset compared to their lower redshift counterpart, in the above diagrams, is primarily interpreted as due to high-redshift AGN being hosted in a low metallicity environment \citep{Kocevski23,ubler+2023}.

Although less luminous, but much more common than quasars, these early AGN are likely playing an important role in the evolution of their host galaxies by exerting feedback processes \citep{koudmani+2022}. An intriguing example is the detection of a prominent AGN-driven outflow in the most distant AGN, GN-z11 \citep{maiolino_bh_2023}, which is observed to eject gas and metals in its Circum-Galactic Medium \citep{Maiolino23heii}, while also heating and ionizing it \citep{Scholtz23}. These phenomena may result in rapid suppression of star formation and lead to the early emergence of quiescent galaxies, or contribute to short-term quenching and to the burstiness of star formation \citep{Carnall23a,Carnall23b,Looser23a,Looser23b,Dome23,Strait2023}.

There have been differing claims about the number of  AGN in early galaxies, with fractions ranging from 1\% to 5\% \citep{Harikane23BH,Matthee23}. Even more unclear is the potential contribution of AGN to the reionisation of the Universe, with some estimates claiming that they could contribute up to 50\% \citep{Giallongo19,Harikane23BH} and other indicating that they are unlikely to contribute significantly \citep{Matthee23}.

In this paper, we present the discovery of a sample of 12 new broad line AGN at z$>$4 in the first DEEP tier and two of the MEDIUM tiers of the JADES survey \citep{Eisenstein23}, by using the NIRSpec MSA spectroscopic observations. These observations are  deeper than previous observations and were performed with multiple dispersers providing different spectral resolutions. Therefore, these datasets enable us to unveil AGN with a diversity of broad line widths and often in a lower luminosity regime than previous surveys, either associated with lower mass black holes, lower accretion rates, or more obscured AGN. We will show that this data uncovers the properties of the early phases of black hole formation and their connection with their host galaxies, some of which nicely confirm expectations from models and simulations, while others are unexpected and prompt further theoretical modelling.

Throughout this work, we use the AB magnitude system and assume a flat $\Lambda$CDM cosmology with $\Omega_m=0.315$ and $H_0=67.4$ km/s/Mpc \citep{Planck20}.  With this cosmology, $1''$ corresponds to a transverse distance of 5.84 proper kpc at $z=6$.

\section{Sample, observations and data processing}\label{sec:data}

\subsection{Observing strategy and target selection}

The data used in this paper have been obtained as part of the JADES survey \citep{Eisenstein23}. This survey combines nearly 800 hours of NIRCam, NIRSpec-MSA and MIRI observations in parallel mode, in the GOODS-S and GOODS-N fields. This is a multi-tiered survey reaching different depths (down to AB$\sim$30.5 in imaging and AB$\sim$29 in spectroscopy) in multiple bands, multiple dispersers, and over different areas (for a total  of 175~arcmin$^2$). An extensive description of the survey is given in \citet{Eisenstein23}. Here we only discuss briefly the three specific spectroscopic tiers that are used in this paper: Deep/HST in GOODS-S, Medium/JWST in GOODS-N and Medium/HST in GOODS-N.

A detailed description of the target selection and of the spectroscopic observations is given in \citet{Eisenstein23} and \cite{bunker_jades_2023}. 
Here we only summarise that in the Deep/HST tier spectroscopic targets were selected giving higher priority to the highest redshift candidates, according to their photometric redshifts, and primarily relying on the Ly$\alpha$ dropout signature, and then gradually lower priorities to galaxies at lower redshifts. 
The targets selected in Deep/HST were primarily obtained from previous HST imaging.  However, 
a number of high priority targets were also selected from NIRCam imaging \citep{Rieke2023} obtained shortly before the NIRSpec observations. We note that the parent HST targets list was leveraging on previously published catalogues, which may have discarded high-z AGN based on their ``stellarity'', i.e. point-like appearance. Moreover, the initial selection from NIRCam images may have discarded some high-z AGN because of their colours being similar to brown dwarfs. Therefore, the Deep/HST tier is likely biased against (type 1) AGN.

The observations in Deep/HST were obtained with three dithered configurations
of the MSA \citep{Jakobsen22,Ferruit22,boker23}, each with 3-shutters nodding. The low-resolution prism, the three medium-resolution gratings (G140M/F070LP, G235M/F170LP, G395/F290PL), and the high-resolution G395H/F290LP grating were used, for a total observing time of up to $\sim$27 hours with the prism and up to $\sim$7 hours with each of the gratings (the specific exposure for each target depends on whether they could be accommodated in the three dithered MSA configurations or not). Overall, spectra for about 250 galaxies were obtained in Deep/HST \citep{bunker_jades_2023}.

In the case of the Medium/JWST GOODS-N tier, the targets were primarily selected from NIRCam images, with a similar set of selection criteria and priorities as for Deep/HST, although targeting brighter magnitudes on average. In contrast to previous observations in JADES, this was the first tier that was not biased against the selection of AGN (and it is the reason why we chose it for this paper). Actually, three targets were specifically selected because of their imaging (very compact) and color properties, similar to those of other AGN identified in other JWST observations \citep{ubler+2023,Harikane23BH, Matthee23}.
Specifically, for identifying these specific candidate AGN, we initially considered well-detected sources (F444W$>$250 nJy) with F090W$<$F200W$<$3$\times$F444W. From this initial selection we inspected the SEDs (to reject unreliable measurements and obvious brown dwarfs). We further inspected the cutout images to reject contaminated sources and sources with proper motions. A considerable fraction of these candidates ($>$30\%) has one or more blue sources/features within 0.1--0.2 arcsec, similar e.g. to GS-3073 \citep{ubler+2023}.
The final sample was prioritised with objects having nearby, blue companions first, because these objects were considered having the lowest risk of being brown-dwarf contaminants.

The observing strategy in the Medium/JWST tier was very similar to that used for Deep/HST, but with shorter overall exposures, resulting in 2.6 hours in each of the five dispersers (prism, the three medium resolution gratings, and G395H). Four different pointings were planned in GOODS-N for this tier, but one of them failed twice because of a telescope guide problem and MSA shorts; hence, the fourth pointing is planned for a repeated observation in 2024. Overall, in the three successful pointings 712
targets were observed.

Finally, we also explored spectra from the Medium/HST GOODS-N tier. This consisted of four pointings (two dithers) with targets selected from HST imaging, with selection criteria similar to Deep/HST, but brighter magnitudes. In this tier we used the prism (1.7 hours on source) and the three medium-resolution gratings (0.8 hours on source for each of them). About 660 sources were observed in this tier.

GN-z11 is part of the JADES sample and was observed both in the Medium/HST and Medium/JWST tiers in GOODS-N. It was specifically targeted because previously identified as a galaxy at z$\sim$11 based on HST and ground-based observations \citep{Oesch16}. The first spectrum was obtained by \cite{Bunker2023GNz11} (within Medium/HST), and additional MSA (Medium/JWST) and IFS observations were obtained by \cite{maiolino_bh_2023}, \cite{Maiolino23heii} and \cite{Scholtz23}, which identified it as a type 1 AGN (specifically an analogue of NLSy1). Although specifically observed because its previously known properties, it is part of the JADES sample and it is included in the analysis of this paper.

\subsection{Data processing}

The processing of the MSA data is also described in \cite{bunker_jades_2023} and other MSA-JADES papers \citep[e.g.][]{Carniani23}. A detailed description of the data processing will be presented in Carniani et al. (in prep.). Here we only recall that we have used the pipeline 
developed by the ESA NIRSpec Science Operations Team and the NIRSpec GTO Team. As we are primarily interested in the detection of the broad component of the Balmer lines emitted from the unresolved Broad Line Region of AGN, we use the spectra extracted from the central 3 pixels of each 2D spectrum (i.e. the central 0.3$''$), which maximises the S/N for compact sources.
We also mention that the pipeline automatically corrects for the wavelength dependent slit losses (by also taking into account the source position in the slit), by assuming a pointlike source. The latter assumption is certainly appropriate for the radiation coming from the AGN, but more broadly also for their host galaxies, as these systems are extremely compact, as it will be discussed in the next sections.

\subsection{Spectral resolution}

We finally mention that the nominal resolution of the NIRSpec dispersers ($\rm R\sim 30-300$ for the prism, $\rm R\sim 700-1300$ for the medium resolution gratings, and $\rm R\sim 1900-3500$ for the high resolution gratings) applies only in the case of uniformly illuminated shutters. This is rarely the case for galaxies at z$>$4, which are generally more compact than the shutters' width ($0.2''$), even when convolved with the telescope's PSF for most wavelengths, and it is certainly true for the Broad Line Region, which is totally unresolved. As a consequence, the effective resolution of MSA observations is generally significantly higher, and primarily driven by the telescope PSF rather than the shutters' width.

Although the resolution slightly depends on the position on each quadrant (due to the slightly varying PSF) and on the position of the target in the shutter, for a point source the effective resolution spans the following ranges \citep{de_Graaff2024}:
$\rm R\sim 60-500$ for the prism, $\rm R\sim 1200-2500$ for G140M, $\rm R\sim 1100-2300$ for G235M, $\rm R\sim 1300-2000$ for G395M,  and
$\rm R\sim 3600-5400$ for G395H.

\begin{figure*}[h]%
\centering
\includegraphics[width=0.65\columnwidth]{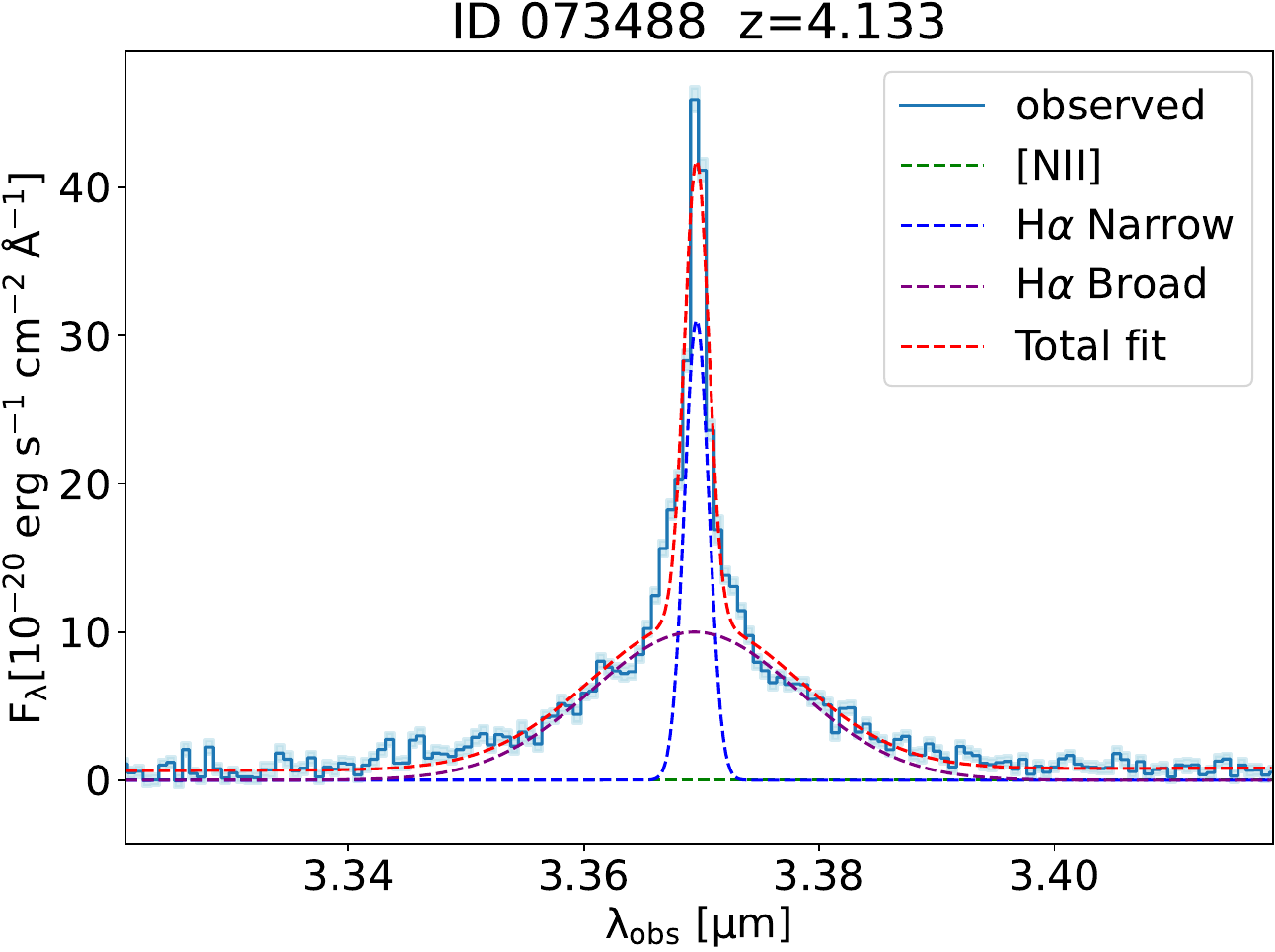}
\includegraphics[width=0.65\columnwidth]{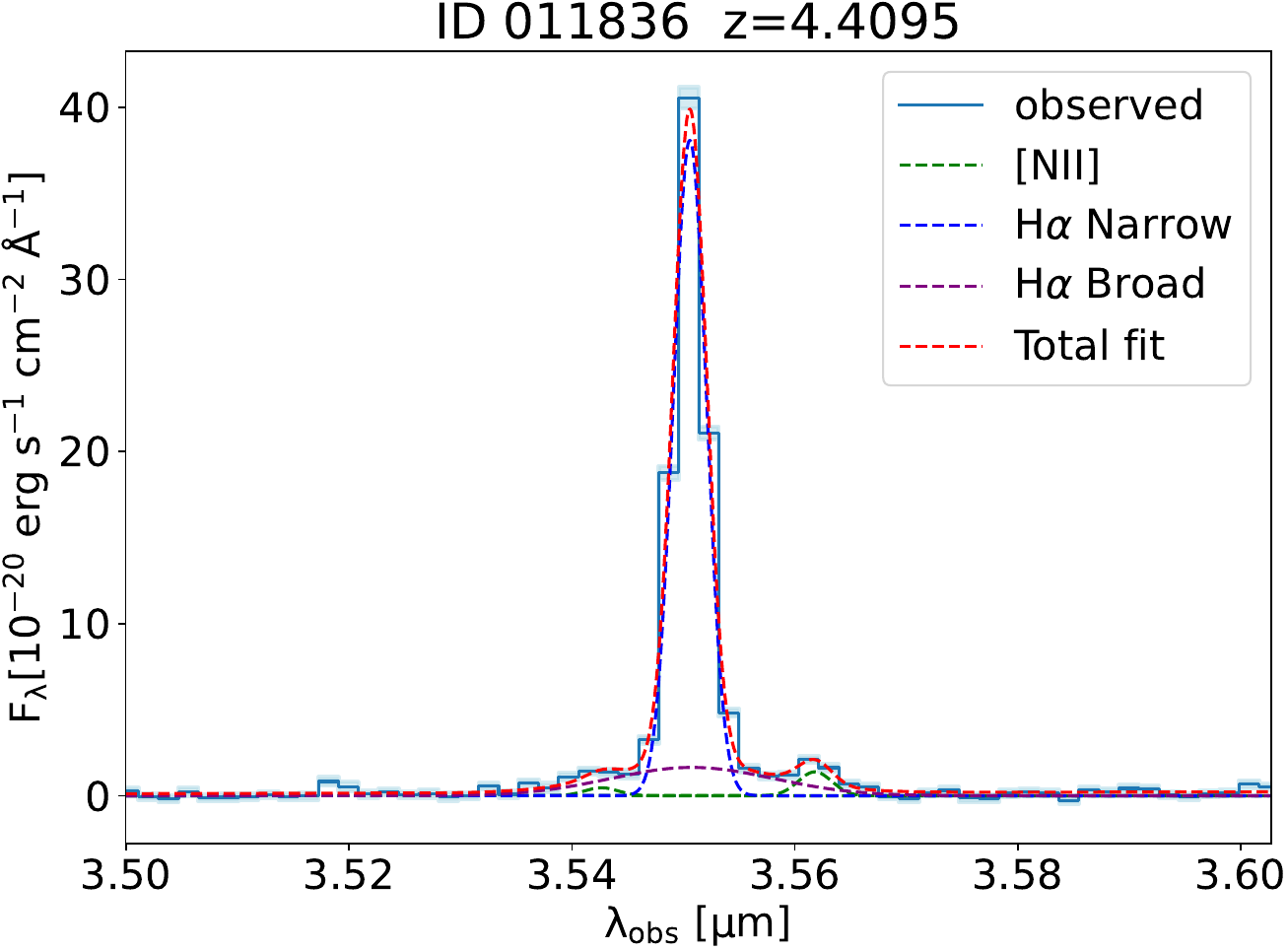}
\includegraphics[width=0.65\columnwidth]{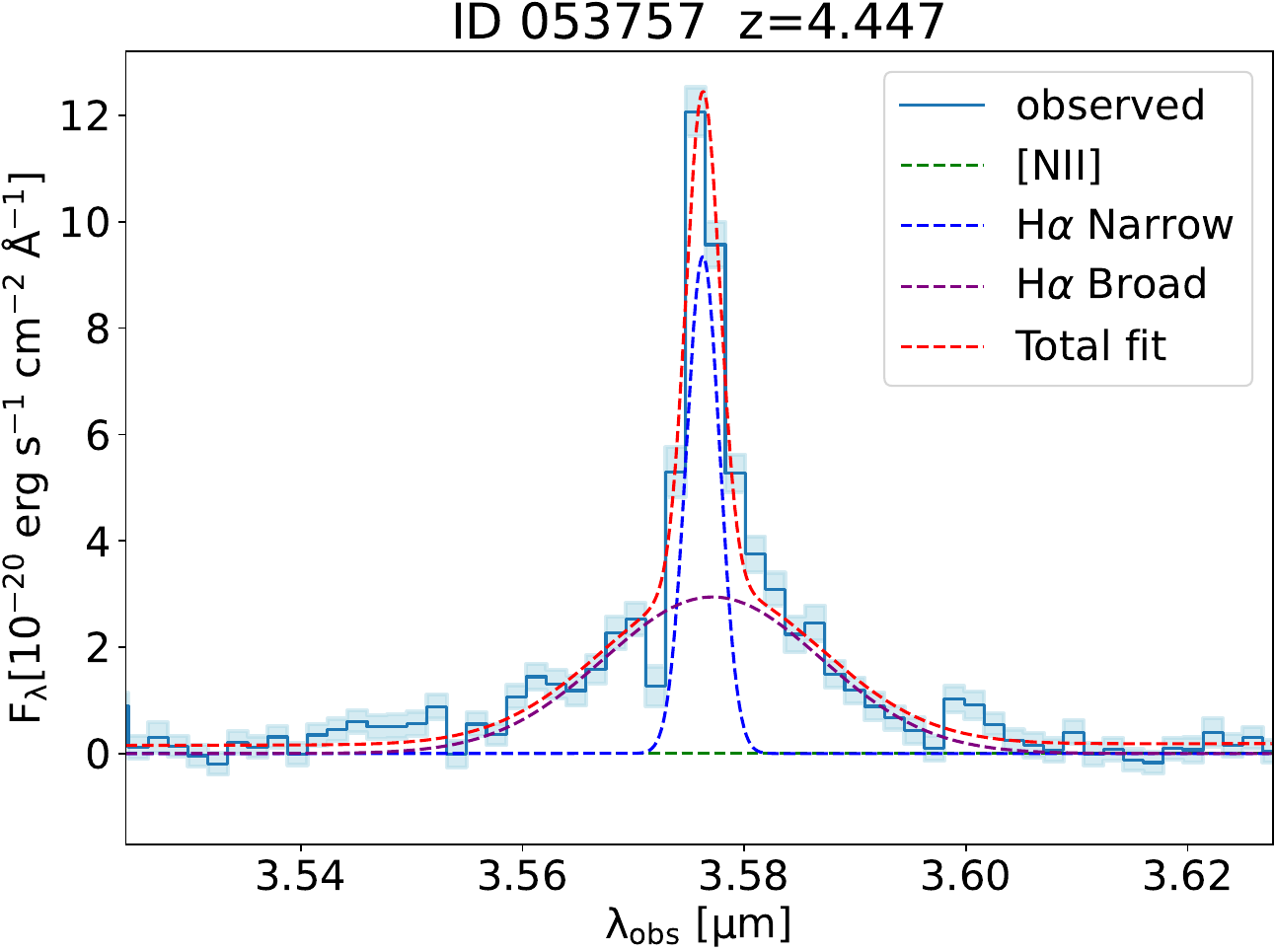}
\includegraphics[width=0.65\columnwidth]{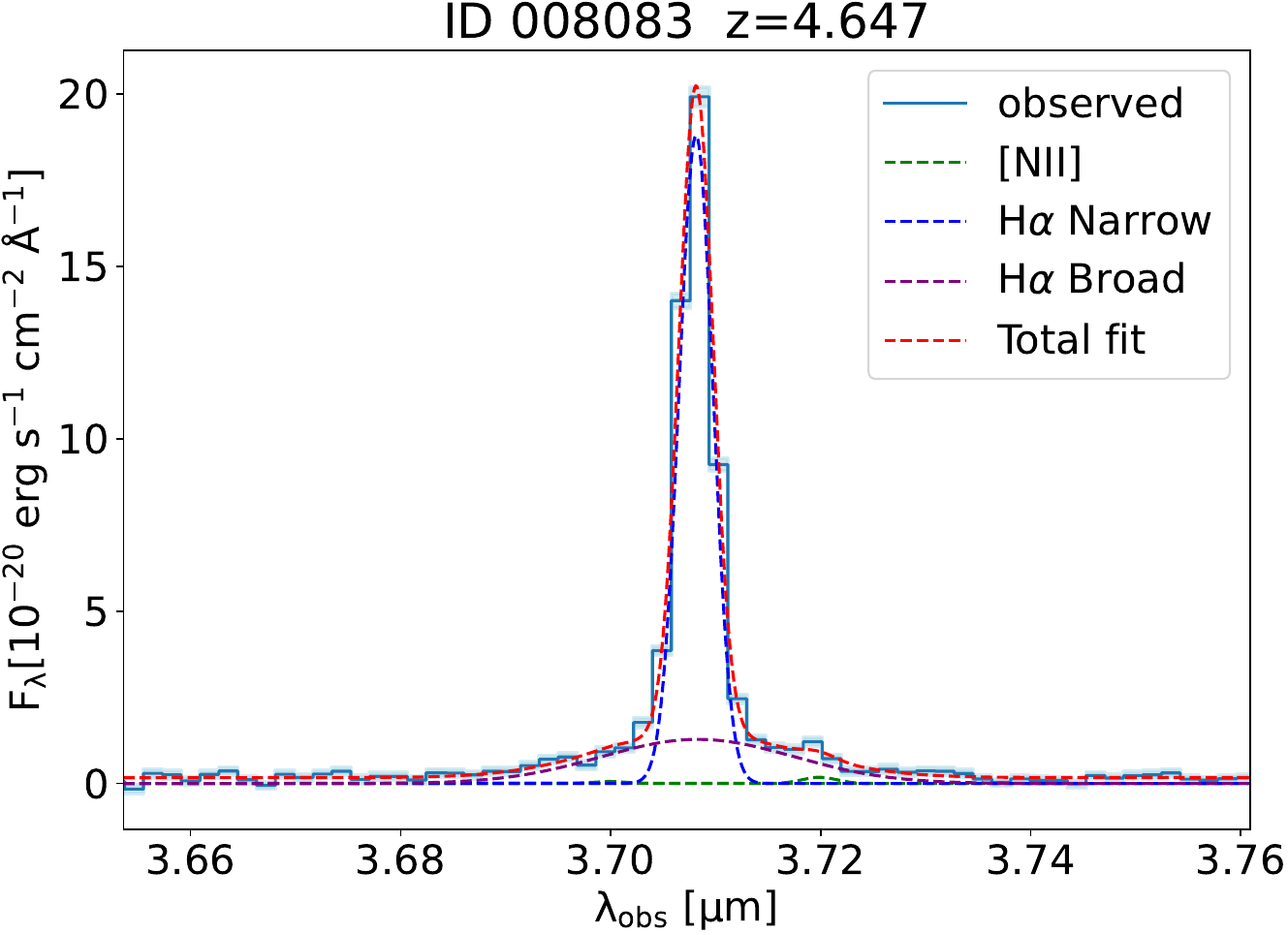}
\includegraphics[width=0.65\columnwidth]{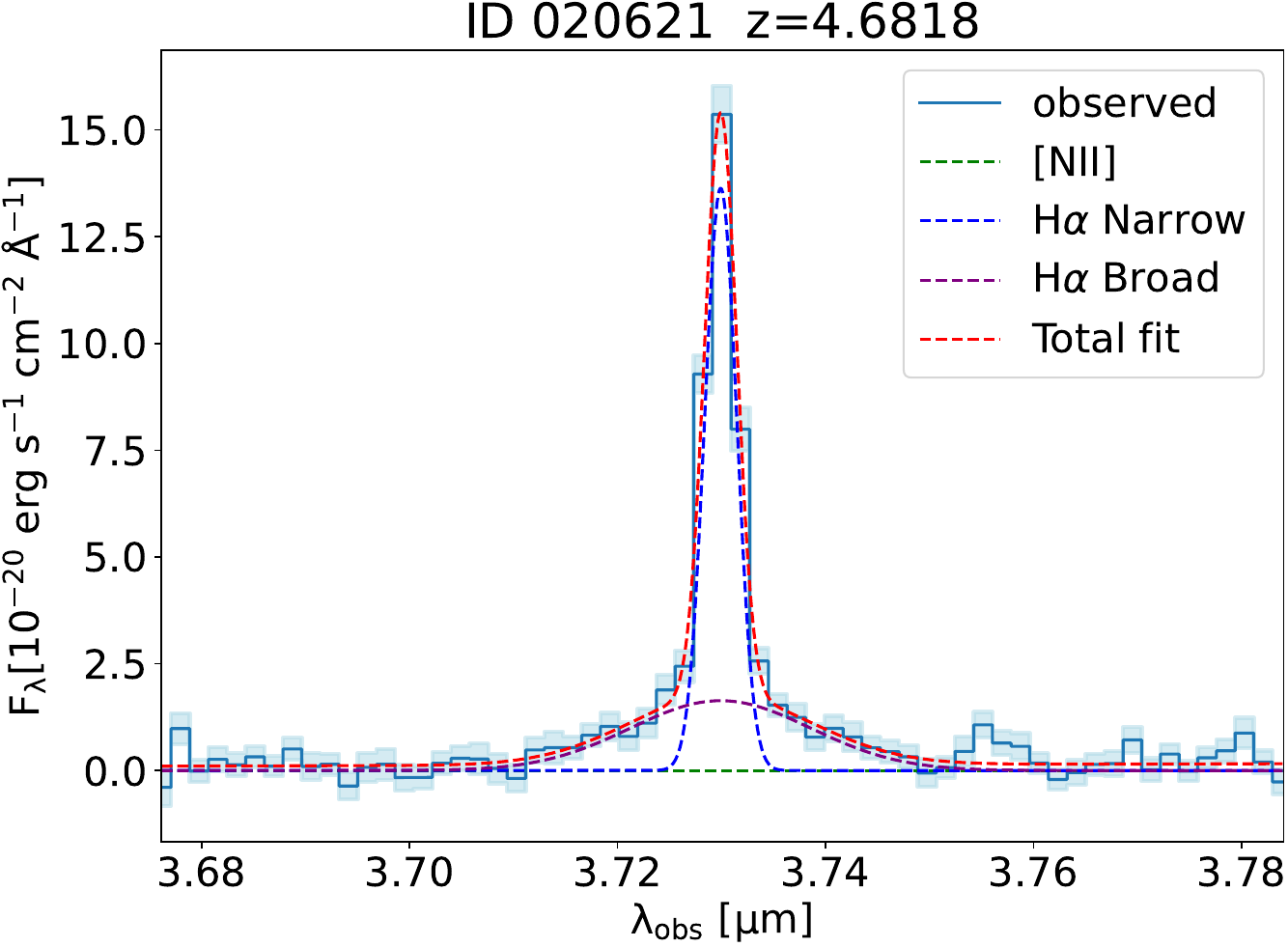}
\includegraphics[width=0.65\columnwidth]{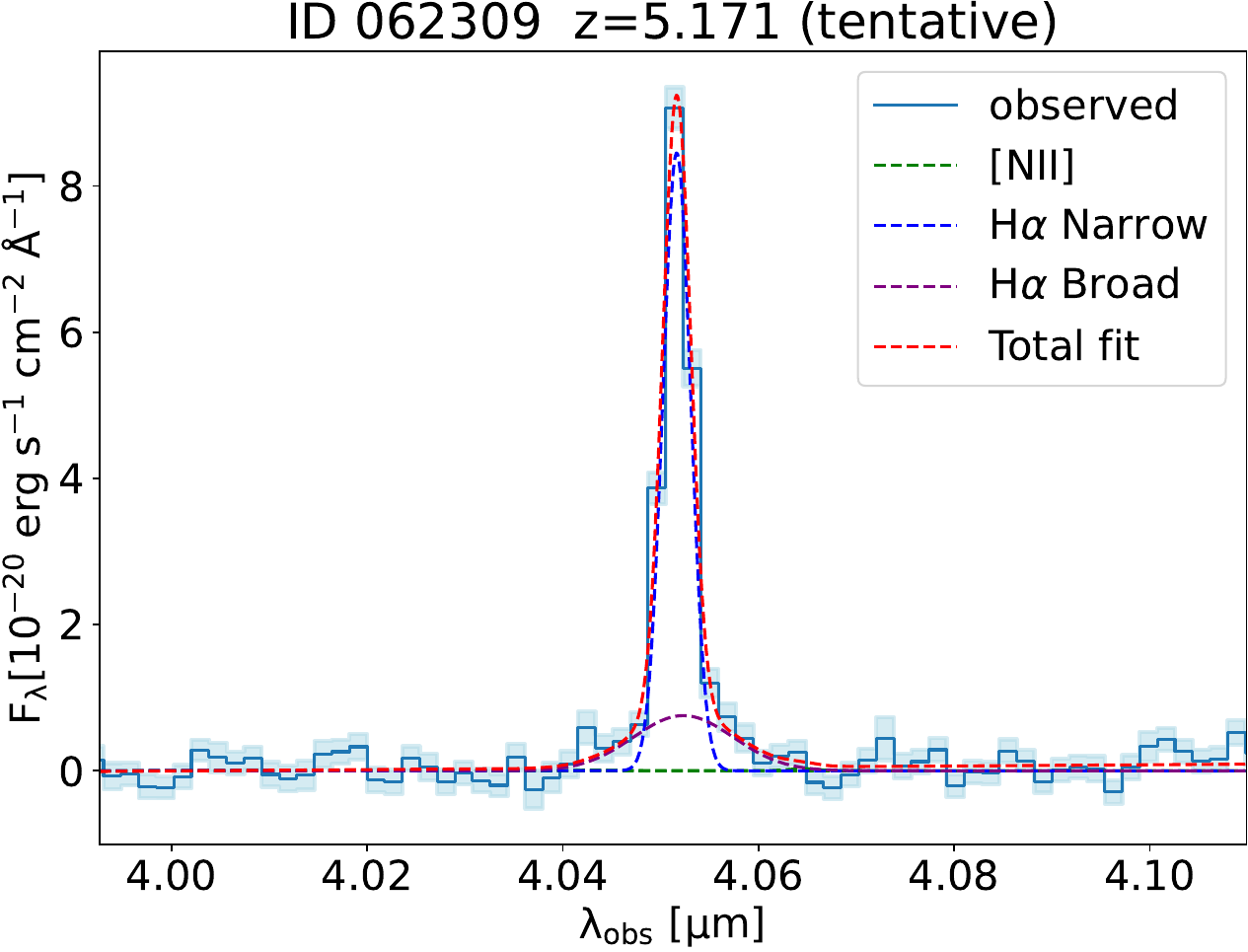}
\includegraphics[width=0.65\columnwidth]{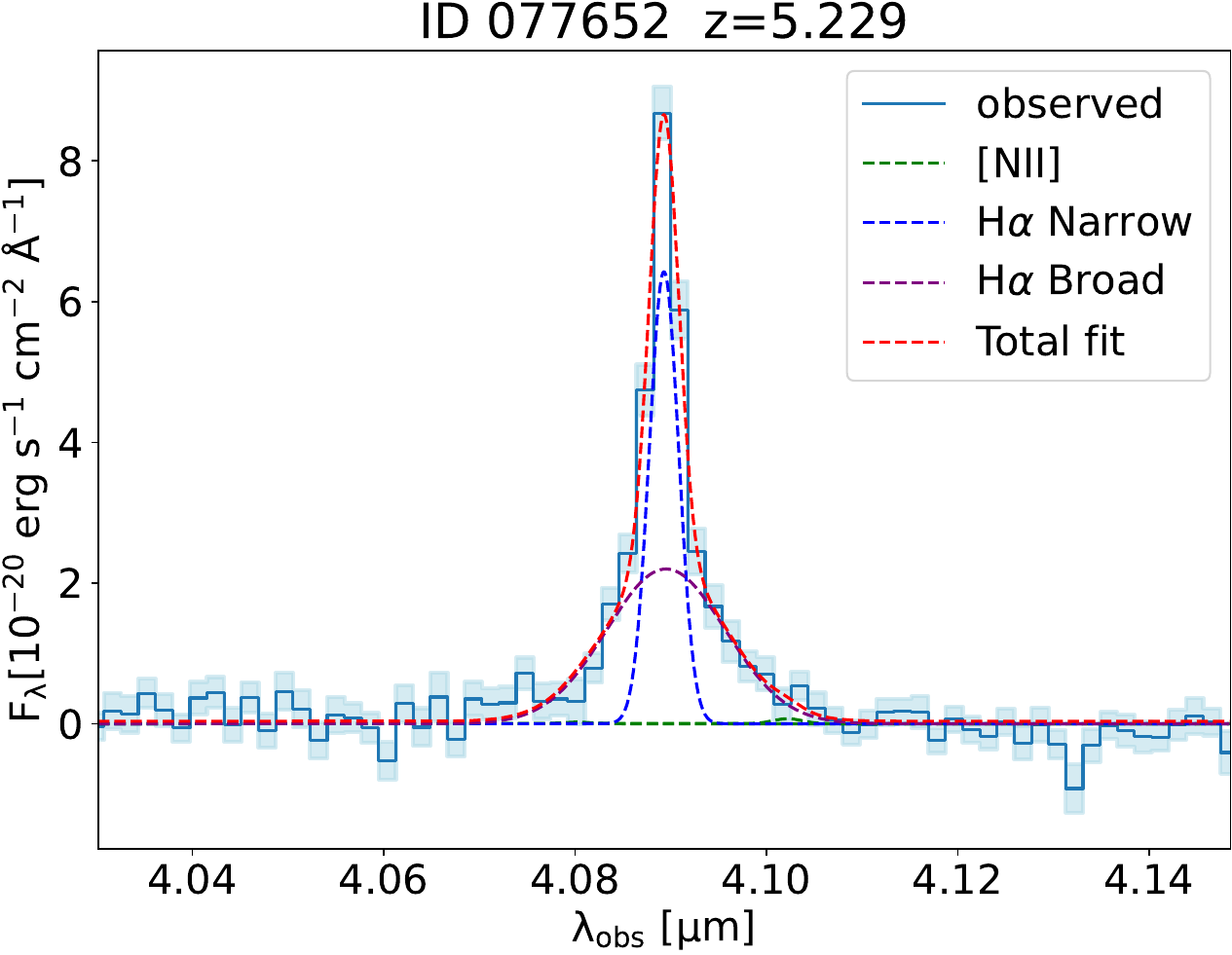}
\includegraphics[width=0.65\columnwidth]{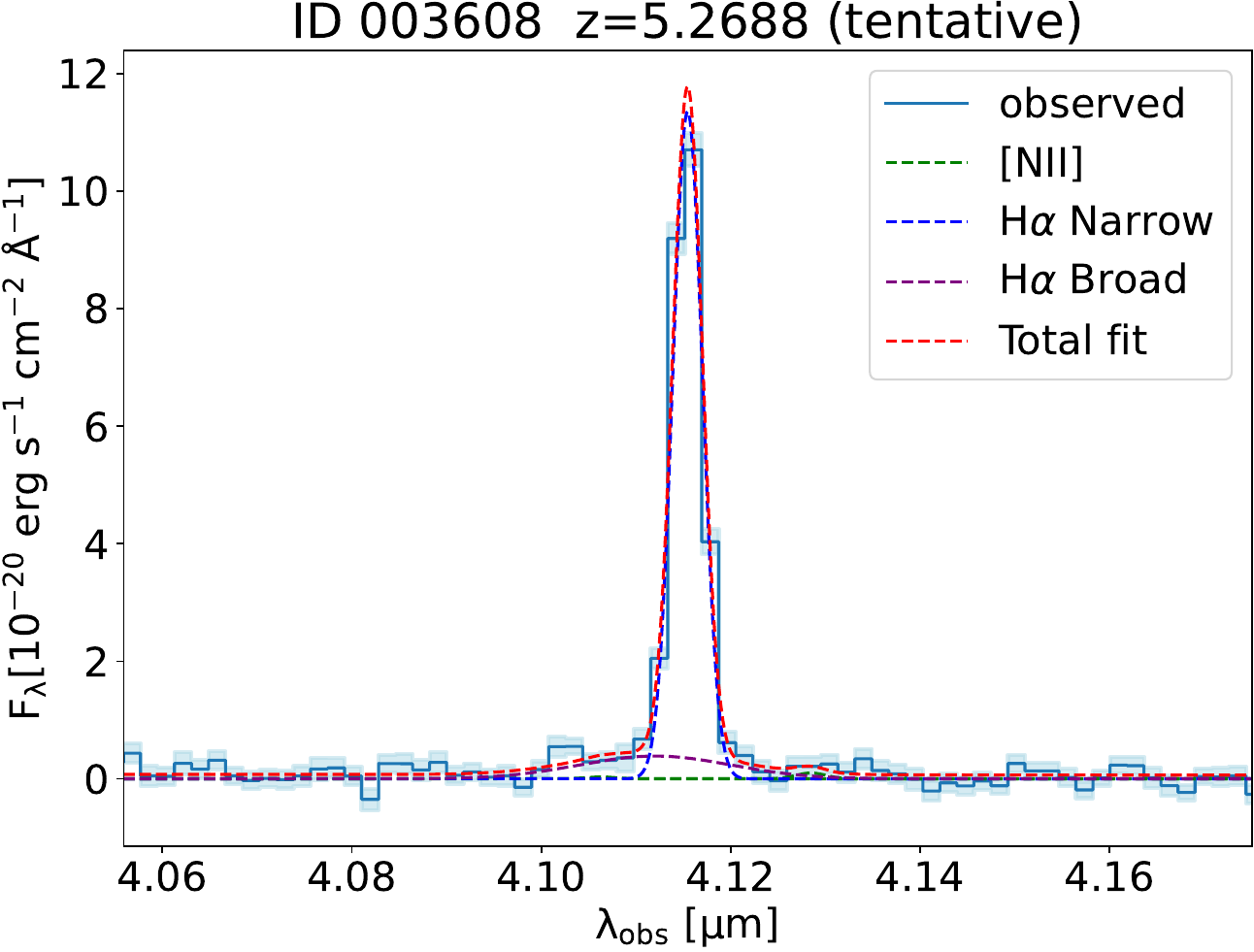}
\includegraphics[width=0.65\columnwidth]{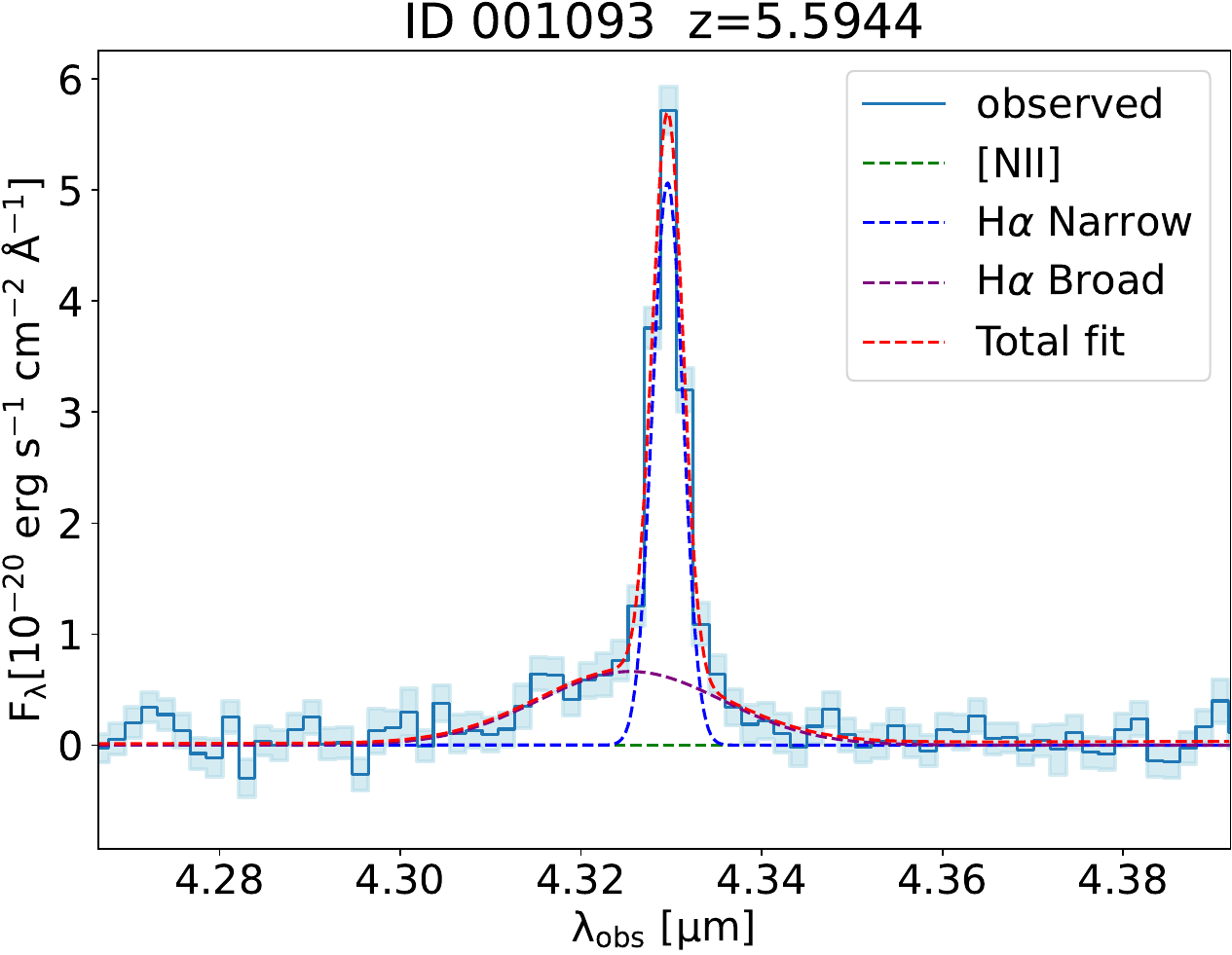}
\includegraphics[width=0.65\columnwidth]{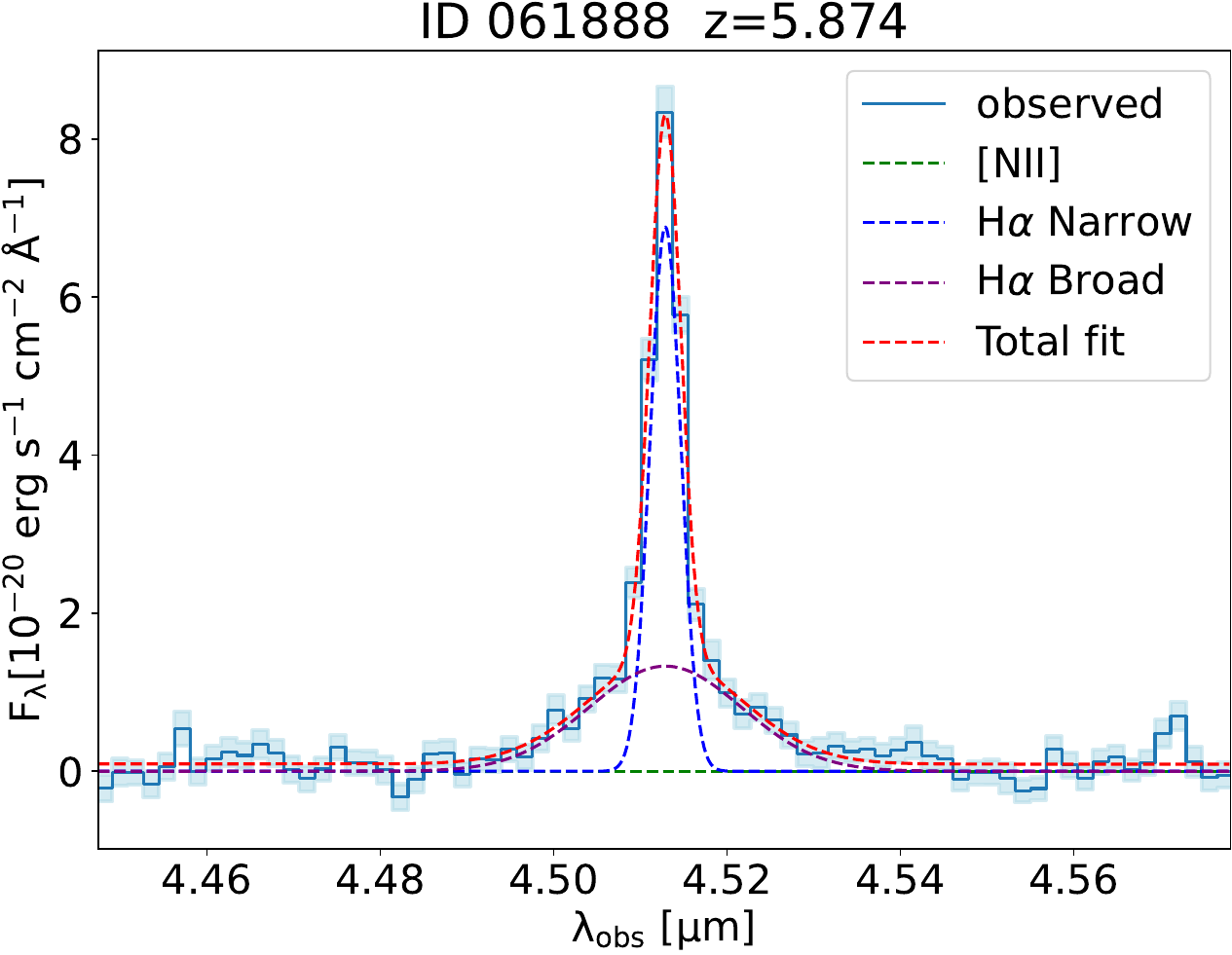}
\includegraphics[width=0.65\columnwidth]{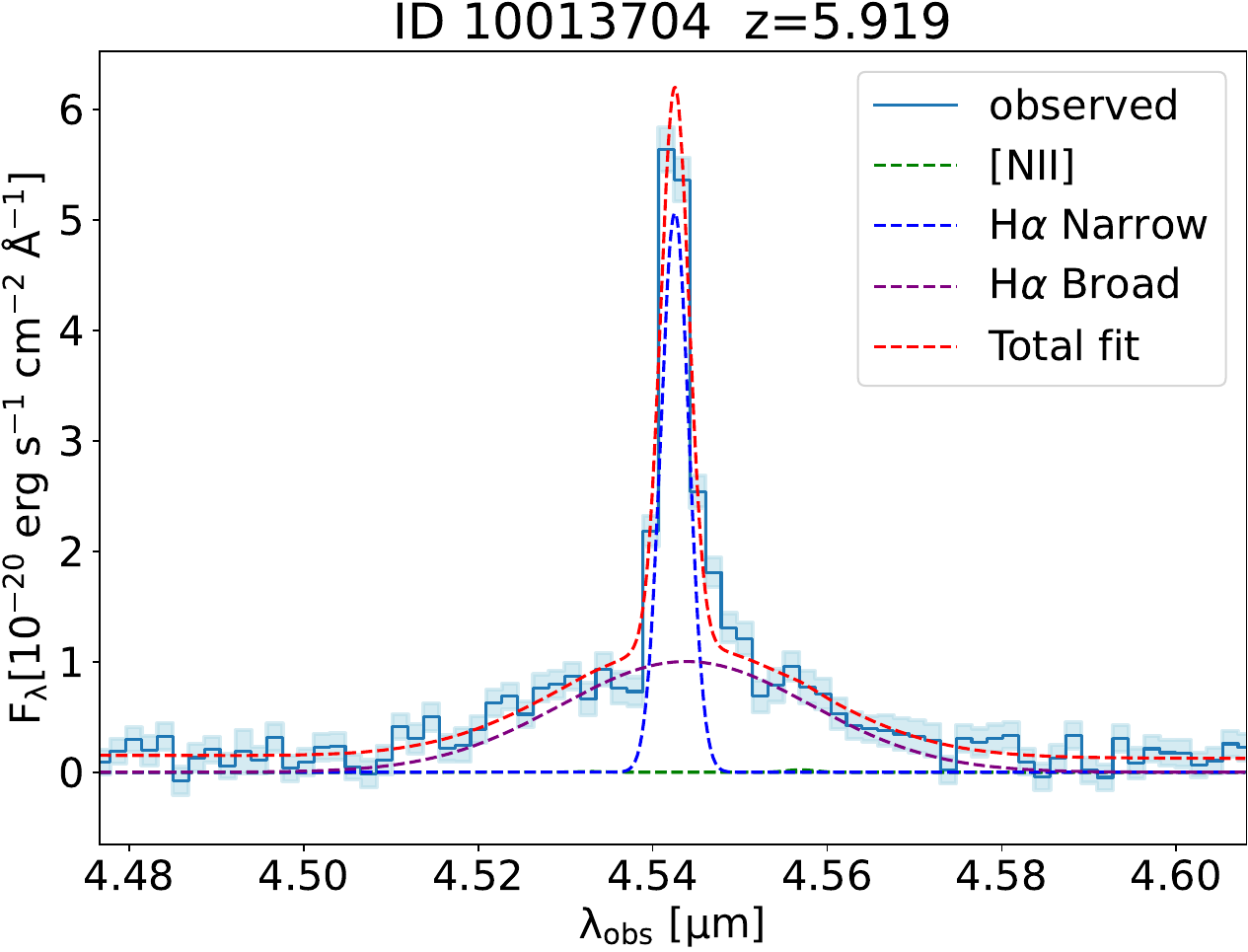}
\includegraphics[width=0.65\columnwidth]{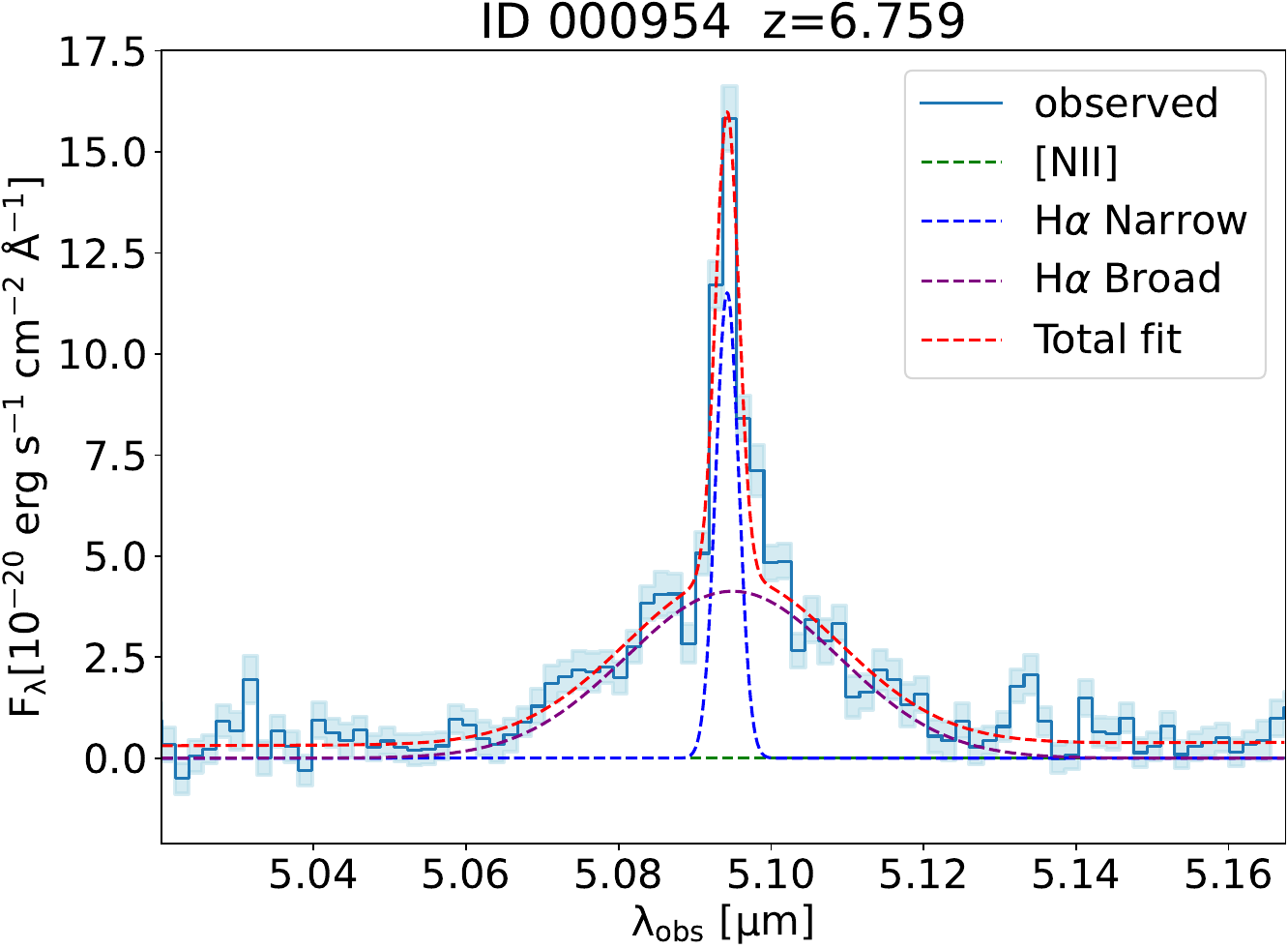}
\caption{Medium resolution spectra of the 12 new JADES galaxies with evidence for a broad component of H$\alpha$ ascribed to the BLR of an AGN. We show a zoom around H$\alpha$. The blue solid line shows the spectrum along with the errors (light blue shaded area). The red dashed line shows the total multi-component fit; the blue and purple dashed lines show the narrow and broad components of H$\alpha$, respectively, while the green dashed lines show the  components fitting the [NII]
doublet (often undetected).
}\label{fig:spectra}
\end{figure*}

\begin{figure*}[h]%
\centering
\includegraphics[width=1.5\columnwidth]{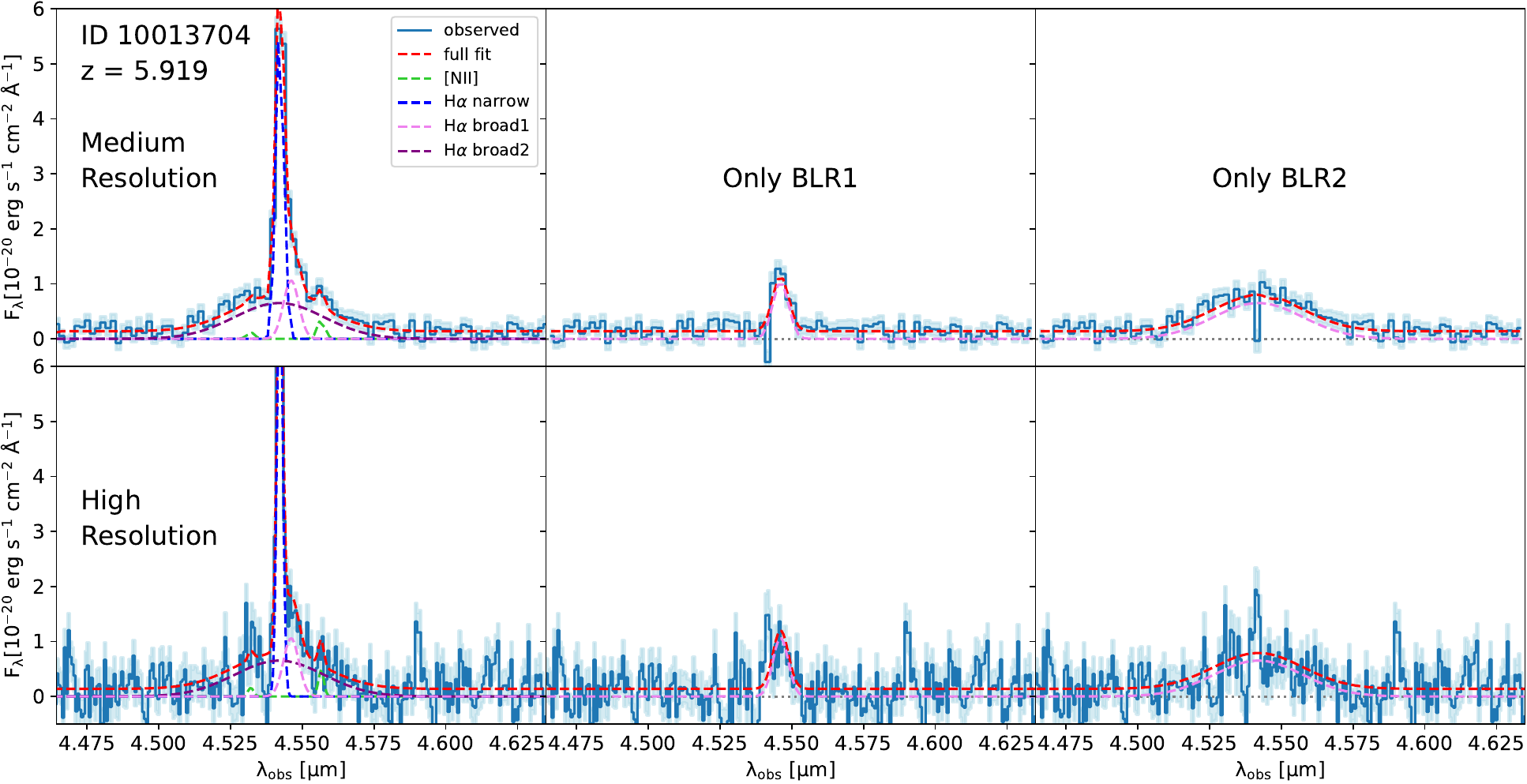}
\caption{
Spectra around H$\alpha$ of ID 10013704, the highest redshift AGN showing indication for a dual BLR.
The top and bottom panels show the medium and high resolution spectra, respectively. The line coding is the same as in Fig.~\ref{fig:spectra}, but in this case the violet dashed line shows the intermediate broad component that is needed to  properly reproduce the observed profile. The central and rightmost panels show the spectrum from which the narrow components, as well as one of the two broad components have been removed, to better highlight the significance of the other broad component.
}\label{fig:clara_gemma}
\end{figure*}

\section{Identification of broad line AGN}\label{sec:identification}

In this section we describe the identification of broad line AGN and their basic properties. As mentioned, we focus  on galaxies at z$>$4, as this is the new regime probed specifically by JWST.

\subsection{Criteria for the detection of a BLR}\label{sec:blr_crit}

The presence of a broad line region (BLR) is assessed via the detection of a broad component of either H$\alpha$ or (at z$>$7) H$\beta$  line emission, without a broad counterpart in the forbidden transitions (in particular the bright [OIII]5007).

We fit the Balmer lines with two Gaussian component, narrow and broad. The [NII] doublet around H$\alpha$ is forced to have the same width and velocity as the narrow component of H$\alpha$ and, similarly, the [OIII] doublet near H$\beta$ is forced to have the same width and velocity as of the narrow H$\beta$. 
For each of the [NII] and [OIII] doublets the intensity ratios are  fixed by to the corresponding Einstein coefficients.

While for each line we simultaneously fit the spectra in the available dispersers, we do not simultaneously fit the H$\alpha$+[NII] and H$\beta$+[OIII] groups because of two main reasons: small wavelength calibration uncertainties (primarily associated with uncertainty in the location of the target in the slit, both because of residual astrometric uncertainties and because of MSA target acquisition uncertainties of $\sim 0.05''$, Jakobsen et al. in prep., Carniani et al. in prep.), may result in slight artificial wavelength shifts and dispersion solutions (and also slightly different resolutions) for the two groups; secondly, due to the different PSF at the two wavelengths, the spectrum may sample slightly different regions of the host galaxy. While we do not fit the H$\alpha$+[NII] and H$\beta$+[OIII] groups simultaneously, we checked that the two separate fits are fully consistent within uncertainties.

In order to claim the detection of a BLR,
we require that the second, broader component of the Balmer lines should be at least a factor of two broader than the narrow component and have a significance of at least 5 $\sigma$. Furthermore, we require that the Bayesian Information Criterion (BIC) parameter \citep{Liddle2007}, defined (in case of Gaussian noise) as:
$$
 BIC = \chi ^2+k~\ln{n}
$$
(where $\rm k$ is the number of free parameters and $\rm n$ is the number of data points), for the fit with the broad component is at least a factor of 6 smaller than the value for fit with only the narrow component, i.e.
$$
BIC_{only-Narrow}
-BIC_{Broad+Narrow}
=\Delta BIC_{NB} 
>6
$$
We conservatively mark two cases with
$6<\Delta BIC_{NB}<10$ as `tentative' (IDs 3608 and 62309) although their broad components are detected at $>5\sigma$; their removal from our analysis would not change the conclusions.

Finally, we note that while the broad lines of low and intermediate luminosity type 1 AGN are often well fitted with a Gaussian profile \citep[e.g.][]{Marziani2019}, for very luminous quasars it has been suggested that a double powerlaw profile may reproduce better the profile of the BLR  permitted lines \citep{Nagao06}. Our targets are certainly not in the category of luminous quasars and indeed the broad H$\alpha$ is not fit better with a double powerlaw profile. This aspect is not of particular interest in the context of the detection of the broad components, but it is relevant for the interpretation of the complex H$\alpha$ profiles, which will be discussed in the next section.

\subsection{Ruling out outflows}\label{sec:outflows}

For many objects, the case for
 a BLR is pretty much obvious, with a prominent and nearly symmetric broad component of H$\alpha$ as in classical type 1 AGN. However, in cases where the broad wings of the Balmer lines are fainter or asymmetric,  these could in principle be associated also with galactic outflows, and indeed the JADES survey has revealed a number of galactic outflows by inspecting the emission line profiles \citep{Carniani23}. However, high-velocity ionized gas in the host galaxy, especially if associated with outflows, should be seen even more prominently in the profile of metal lines, especially the strong [OIII]5007 transition. 
Indeed, the higher excitation of the gas in outflows, along with the fact that galactic outflows are naturally more metal enriched than the host galaxy, typically make the [OIII] line  stronger than both H$\alpha$ and H$\beta$, even in outflows at high redshift \citep{Holden23,Carniani15,Marshall2023}.
Therefore, a requirement for the identification of BLR is that the broad component should not be detected in the [OIII] line. 

We finally note that it is very unlikely that, despite being brighter, an [OIII] outflow component is not detected, while seen in H$\alpha$, because of dust extinction. Indeed, being out of the galactic plane, and also quite extended, is generally less obscured by dust in the galactic disc. Dust extinction is even less prominent at z$>$4, where most galaxies show little or no dust reddening \citep{fiore+2023,Sandles23a}. Specifically, the galaxies with identified broad component of H$\alpha$ in our sample have H$\alpha$/H$\beta$ Balmer decrements generally consisted with the case B value of 2.8, or only slightly higher by a factor of less than 1.5; this indicates, that even in the extreme case of a putative outflow obscured at the same level as the galaxy ISM, the associated broad wings of [OIII] (typically a factor of at least a few stronger than H$\alpha$) would still be detectable.

\subsection{Broad line AGN identified in JADES}\label{sec:sample_blr}

In the three JADES sub-tiers analysed by us we identify 12 new Broad Line AGN at z$>$4 \citep[in addition to GN-z11 reported by ][ which required a dedicated analysis]{maiolino_bh_2023}. These are all identified via the detection of a broad component of H$\alpha$. Although we identified some possible cases of H$\beta$ with broad component at z$>$7, these did not pass our criterion for the identification of a BLR.

Fig.~\ref{fig:spectra} shows the spectral region around H$\alpha$ for the 12 newly identified AGN.
For sake of simplicity in most cases we show only the G395M spectra, which are those showing the broad H$\alpha$  visually seen more clearly, while generally the higher resolution grating spectra (if available for H$\alpha$) have a consistent profile, although noisier.
The only exception is ID073488 for which we directly show the G395H spectrum as it has very high S/N (but we will also show the corresponding G395M spectrum in the next section). In Fig.~\ref{fig:spectra} all spectra are fitted with a narrow and a single broad H$\alpha$ component, as detailed below; however, for three objects we will discuss a more complex H$\alpha$ profile fitting in the next section.

In Fig.~\ref{fig:spectra} the observed spectra are shown with a light blue line (and errors  with light blue shading), while the dashed lines show the various components. Specifically, the narrow component of H$\alpha$ is shown with a dark blue dashed line, the [NII] doublets (which we recall are  forced to have the same width as the narrow component of H$\alpha$ and to have doublet ratios fixed by the Einstein coefficients) are shown with green dashed lines (often undetected), and the broad component of the H$\alpha$ lines is shown with a purple dashed line. The total modelled profile, including all the above components and a power-law continuum, is shown with a red dashed line.

The flux, widths and shifts of the main lines of interest resulting from the fit are listed in Tab.~\ref{tab:measures}, and in Tab.\ref{tab:measures_dual} for the three cases with more complex H$\alpha$ profiles. The $\rm \Delta BIC_{NB}$ values, as described above, are also reported in Tab.~\ref{tab:measures}
(in Tab.\ref{tab:measures_dual} the meaning of $\rm \Delta BIC$ is different, as discussed in the next section).

For the mentioned two tentative detections, IDs 3608 and 62309, the significance of the broad component of H$\alpha$ is 8.1$\sigma$ and 5.4$\sigma$, respectively, and the $\Delta BIC_{NB}$ are 6.2 and 8.0, respectively. However, for all other AGN in the sample the detection of the broad component of H$\alpha$ is between 11$\sigma$ and 67$\sigma$, while the $\Delta BIC_{NB}$ are between 50 and 330.

In appendix \ref{app:oiii} we also show the spectra around the H$\beta$+[OIII] region, primarily to illustrate the absence of a broad counterpart of [OIII], hence excluding an outflow origin. A broad component of H$\beta$ is only detected in the case of ID 954 at z=6.759, which is the most luminous AGN in our sample.

Grating spectra of [OIII] are not available for IDs 8083 and
53757. However, in these two cases the identification of a BLR is unambiguous given that the H$\alpha$ broad component profile is nearly symmetric (in contrast to outflow profiles, which are generally blueshifted) and the flux of the broad component is similar or higher than the flux of the narrow component, which would imply that the ionized gas in the outflow is more massive than the whole ionized ISM in the host galaxy.

\begin{figure*}[h]%
\centering
\includegraphics[width=1.5\columnwidth]{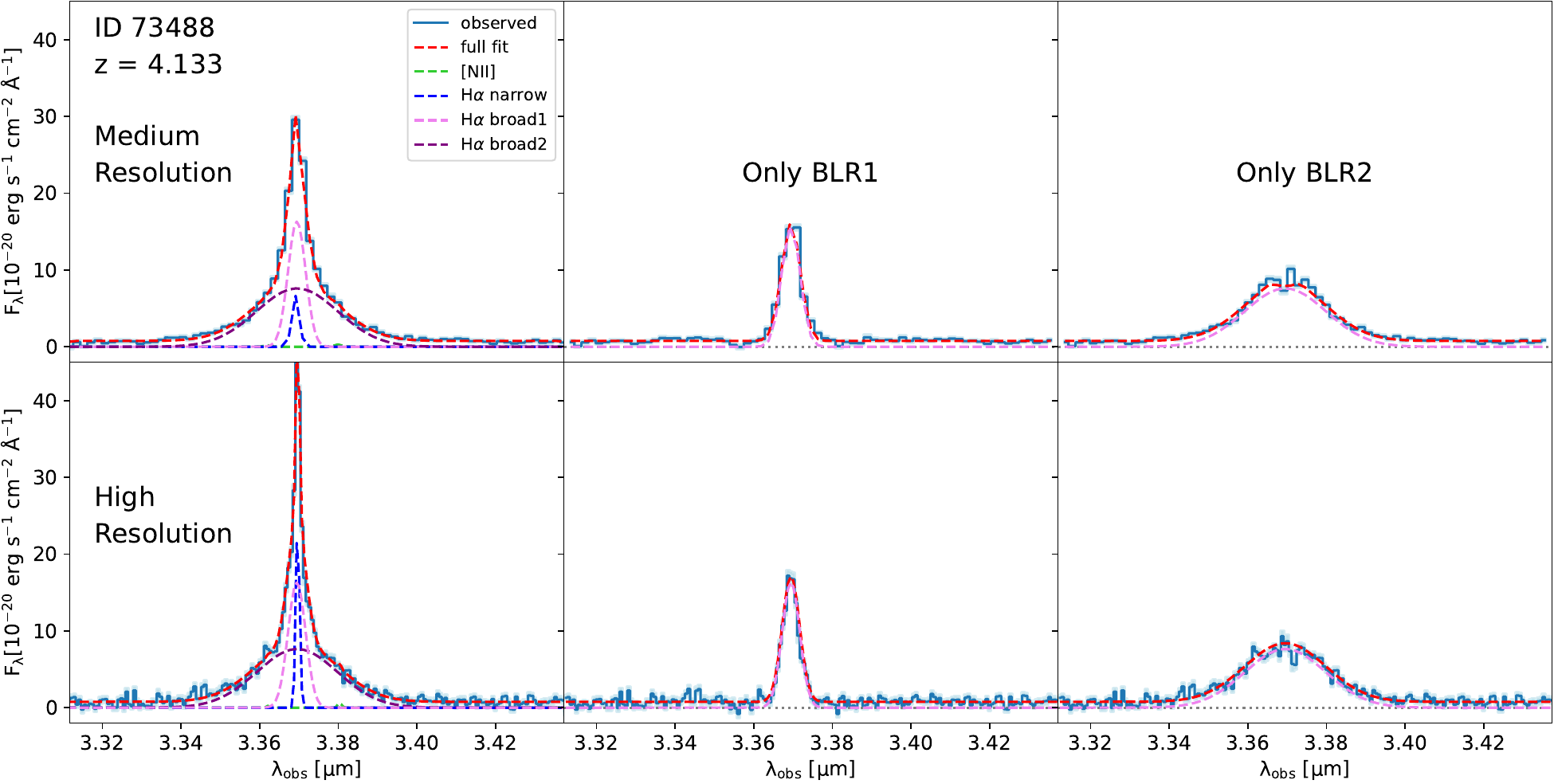}
\caption{
As Fig.\ref{fig:clara_gemma} but for the additional candidate dual BLR AGN ID 73488.
}\label{fig:lara_emma}
\end{figure*}

\subsection{NIRCam images of broad line AGN}\label{sec:images}

\begin{figure*}[h]%
\centering
\includegraphics[width=1.9\columnwidth]{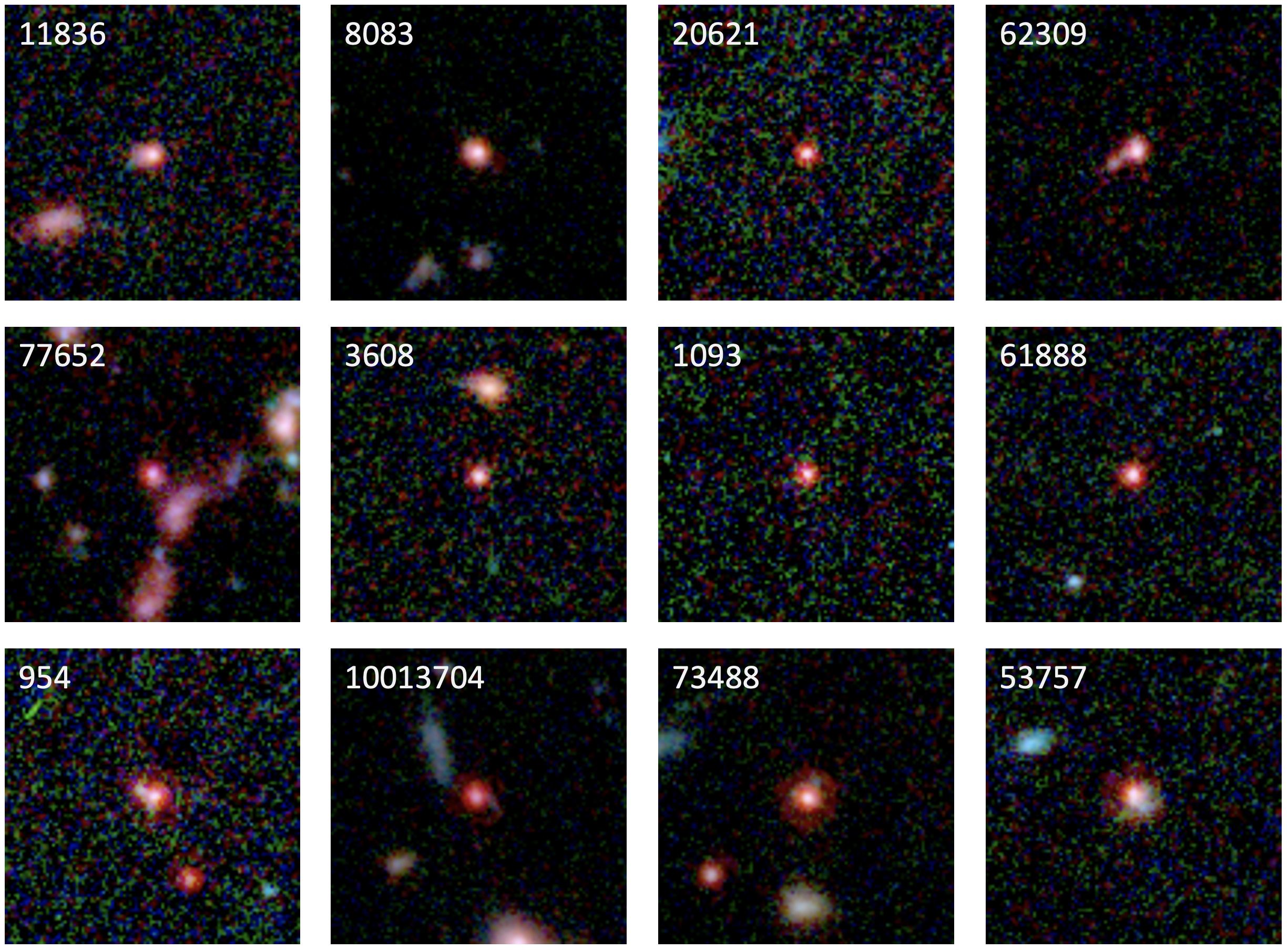}
\caption{
NIRCam images of the selected broad-line AGN. The following false-color coding was adopted: blue -
F115W; green-F200W; red-F444W. Each thumbnail is $3''\times 3''$ in size.
}
\label{fig:images}
\end{figure*}

Fig.\ref{fig:images} shows the NIRCam images \citep{Rieke2023} of the selected broad-line AGN (each thumbnail is $3''\times 3''$ in size). In most cases these targets are very compact, often dominated by a central point source, as expected for type 1 AGN. Some of them have red colors and point-like appearance similar to those identified by \cite{Matthee23} and found to be hosting reddened AGN. We will see in Sect.\ref{sec:mbh_mstar} that indeed the AGN in these targets tend to be reddened, although not by a large amount. Along with the central point source tracing the AGN, more extended structures are often also seen.

We have quantified the presence of an underlying host galaxy by using the ForcePho software to perform such a point-source (aka the central AGN) and host-galaxy decomposition in our sample. ForcePho
(Johnson et al., in prep.) fits multiple PSF-convolved profiles simultaneously to all spectral bands by sampling the joint posterior distribution via Markov Chain Monte Carlo (MCMC).
The fitting of the components was done in the individual NIRCam tiles and observations, even before their combination and mosaicing, which allows a much more accurate control of the PSF and avoids issues with correlated noise in the mosaics. This software and methodology has already beeen successfully employed in similar cases using the same set of NIRCam images \citep[e.g. ][]{tacchella_jades_2023,Baker23bulge,Robertson2023}.

In those cases for which a host galaxy could be detected by ForcePho, the resulting radii and S\'ersic indices are reported in Table \ref{tab:inferred_properties}.
Clearly in most cases the host galaxies of these AGN  are extremely compact, typically with effective radii of a few 100~pc. The S\'ersic indices are typically disc-like (n$\sim$1), with the exception of ID 8083 which has an extremely high S\'ersic index, and which may indicate the presence of an early, compact spheroid.

These parameters will be useful both for assessing the dynamical mass and for interpreting the velocity dispersion in Sec.\ref{sec:mbh_sigma_dyn}.

\begin{figure*}[h]%
\centering
\includegraphics[width=1.5\columnwidth]{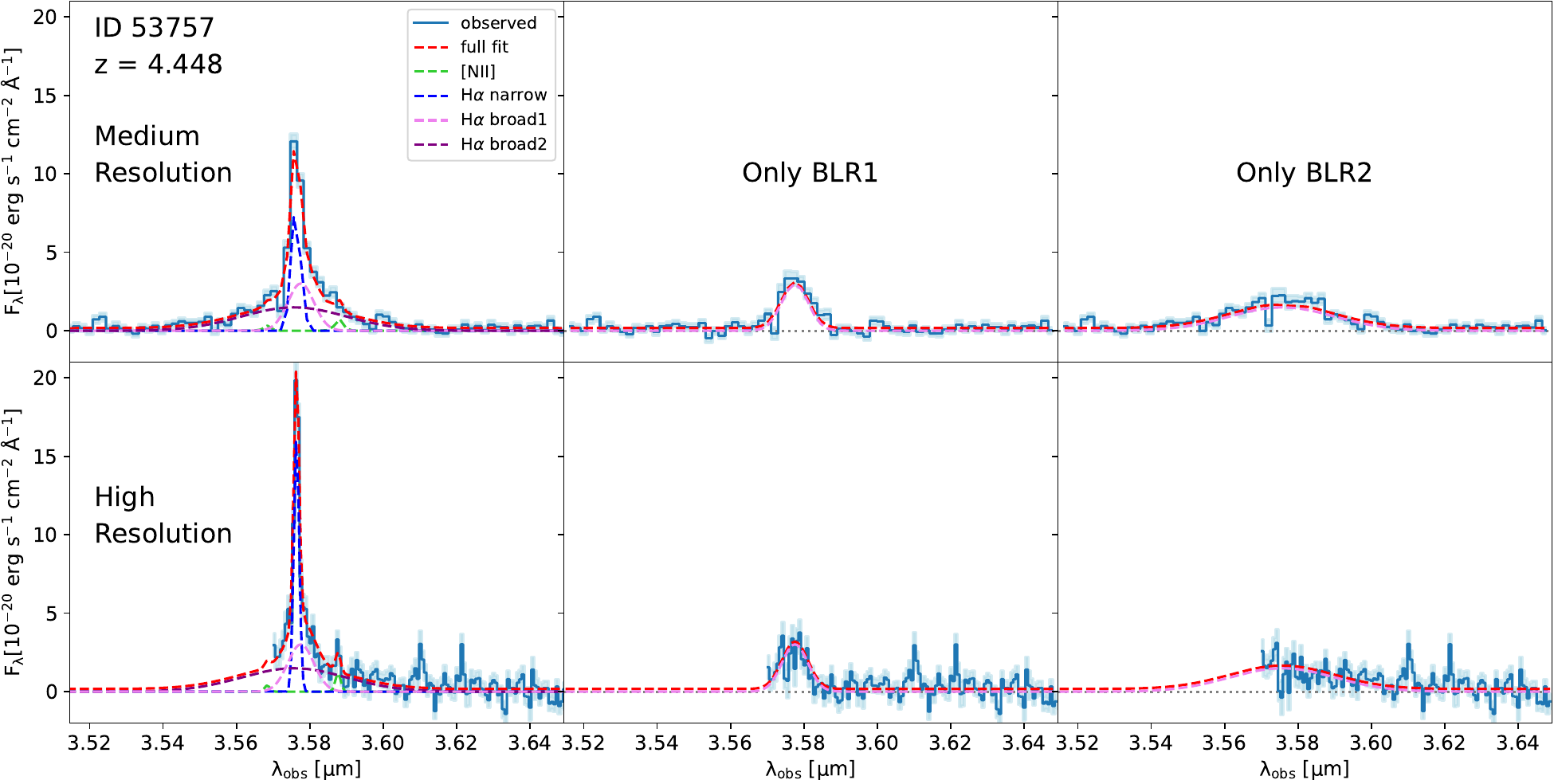}
\caption{
As Fig.\ref{fig:clara_gemma} but for the additional candidate dual BLR AGN ID 53757.
}\label{fig:laura_julia}
\end{figure*}

\section{Candidate merging black holes}\label{sec:merging}

\subsection{Identification of candidate dual AGN}

Among the galaxies for which we have identified a broad component of the Balmer lines, three cases stand out for the peculiar profile of H$\alpha$, which is not fitted properly with a single broad component.
The first (and most distant) one is ID 10013704 at z=5.919, in Deep/HST GOODS-S, whose medium and high resolution spectra are shown in Fig.~\ref{fig:clara_gemma}. In addition to the narrow component, having the same width as [OIII] (140~km/s), the medium resolution grating shows a clear, broad component of H$\alpha$ with a width of 2400 km/s (Fig.~\ref{fig:clara_gemma}, top-left panel). Not surprisingly, such a broad component is nearly lost in the noise in the high resolution spectrum (Fig.~\ref{fig:clara_gemma}, bottom-left panel). However, the medium resolution spectrum also shows a prominent redshifted wing of the narrow component. Such a wing is seen also in the high resolution spectrum, where, rather than a wing, it is more clearly resolved as a bump, slightly redshifted by $\sim$250~km/s and a width of 415~km/s. The central and right panels of the same figure show the medium and high resolution spectra after subtracting the narrow and one of the two broad components, to better illustrate the significance of the remaining broad component.

Tab.\ref{tab:measures_dual} gives the best fit parameters for the two components of H$\alpha$ that are broader than the narrow component, showing that the very broad component is detected at $>10\sigma$ and the second broad component is detected at $>6\sigma$. In this case the $\Delta BIC$ given in the table is not the difference of the BIC with only a narrow component and after adding a broad component (which is very large), it is instead the difference of the BIC with only one broad component and with two broad components:
$$
\Delta BIC_{1B2B}=
BIC_{1BLR}-BIC_{2BLRs}$$
In the case of ID 10013704, $ \Delta BIC_{1B2B}=77$, clearly indicating the strong need of a second broad component.

Neither the very broad component nor the second broad and slightly redshifted bump are seen in the bright [OIII]5007 profile (a detailed comparison of the [OIII] and H$\alpha$ profiles is discussed in Appendix \ref{app:clara_gemma_oiii} and shown in Fig.\ref{fig:clara_gemma_oiii} of that Appendix), implying that it is very unlikely that either of them is associated with an outflow, as galactic outflows, as discussed in Sect.\ref{sec:outflows}, are generally more metal enriched than the host galaxy and, especially if AGN-driven, more prominent in [OIII]. Moreover, asymmetries in the wings associated with outflows are typically blueshifted, as a consequence of dust extinction in the host galaxy disc, which absorbs preferentially the receding (redshifted) component of the outflow. It is also very unlikely that the bump is tracing star formation in a merging companion. The lack of broad [OIII] emission would imply extremely low metallicity \citep[$< 0.01~Z_\odot$, ][]{Curti23, Laseter23,Vanzella23}, but the width of the hump (415~km/s) would indicate that the putative merging galaxy, despite being extremely metal poor, is much more massive than the primary galaxy, which would completely contrast with any formulation of any mass-metallicity relation \citep{Maiolino2019,Curti2023}.

We suggest that the second broad, redshifted bump in ID 10013704 is tracing the BLR of a second, fainter AGN, probably associated with a secondary BH in the process of merging with the BH in the primary galaxy.

We note that morphologically ID 10013704 shows the presence of a possible weak tidal tail (Fig~\ref{fig:images}), indicating that very likely it went through a recent merger, and that the black hole of the secondary galaxy is now approaching the nucleus of the primary galaxy, while accreting gas and being detectable as a secondary AGN\footnote{We note that the tidal tail cannot be responsible for the intermediate broad component of H$\alpha$, as it is far too broad even for the tidal tails of merging massive galaxies, hence even more unlikely for the host of ID 100133704, which, as we will see, has a mass of $<10^9~M_\odot$. Moreover, being the flux of the intermediate component comparable with the flux of the narrow component, it would imply that the amount of ionized gas in the tail is comparable with the  ionized mass in the whole galaxy.}

A similar case is seen in the spectrum of ID 73488 at z=4.133, in the Medium/JWST tier in GOODS-N, as illustrated in Fig.~\ref{fig:lara_emma}. Also in this case there are a clear narrow (68~km/s) and a very broad ($\sim$ 2400~km/s) component of H$\alpha$. However, the line profiles, especially in the high resolution spectrum, cannot be explained with only two components and reveal the presence of an intermediate component with a width 460~km/s (the results of the double component fitting are shown in Fig.~\ref{fig:lara_emma} and reported in Tab.~\ref{tab:measures_dual}). In this case, the velocity shift of the intermediate component is much smaller, but the signal-to-noise is much higher than in the case of ID 10013704. Unfortunately, in this case, we do not have the high resolution grating covering [OIII], yet this transition is observed in a spectral region where the G235M grating has fairly high resolution (resolution FWHM$\sim$150 km/s, or $\sigma \sim 63$, for compact sources), and the presence of the intermediate component in [OIII] should be clearly seen if present (at S/N>20 even in the unlikely case of F(H$\alpha$)=F([OIII])), but is undetected. Therefore, following the same arguments as for ID 10013704, we suggest that this is
an additional case of two BLRs associated with a dual AGN.

Also in the case of ID 73488 both broad components are detected at very high significance ($>35\sigma$), as reported in Tab.\ref{tab:measures_dual}. Introducing the second broad component of H$\alpha$ the BIC  improves by more than 500 ($\Delta BIC_{1B2B}=547$, Tab.\ref{tab:measures_dual}).

We note that, being the additional broad component less shifted in velocity relative to ID 10013704, this could be a case in which the profile of H$\alpha$ is better reproduced with a double powerlaw. However, in Appendix \ref{app:lara_emma_doublepowlaw} we illustrate that this is not the case. While in that Appendix we rule out the double powerlaw scenario more quantitatively, here we simply notice that the need for two BLR separate components is visually clear from the H$\alpha$ profile in the high resolution spectrum (Fig,\ref{fig:lara_emma} bottom-left), which shows a clear inflection of the profile at about $\pm 400~km/s$ from the line centre.

Finally, in Fig.\ref{fig:laura_julia} we show an additional case of candidate dual BLR in the galaxy ID 53757. Unfortunately, in this case we do not have any grating spectrum of [OIII]. However, the intensity of the broad component of H$\alpha$ (more than two times the flux of the narrow component) makes it very unlikely that it is associated with outflow, as it would imply more gas mass in the outflow than in the host galaxy. Moreover, the nearly symmetric profile of H$\alpha$ (if anything slightly redshifted, as discussed below), argues against the outflow interpretation (which generally requires a blueshift of the high velocity gas). Also in this case, as for ID 73488, the H$\alpha$ broad profile, especially the inflection seen in the high resolution spectrum strongly suggest the presence of a second BLR. Additionally, the second broad H$\alpha$ component is slightly redshifted (by $\sim 100~km/s$) relative to the narrow component, as in the case of ID 10013704.  The result of the fit with two BLRs is shown in Fig.\ref{fig:laura_julia} and the resulting parameters reported in Tab.\ref{tab:measures_dual}: the second broad component, with width $\sim 700~km/s$, is detected at $>5\sigma$, while the broader component, with FWHM$\sim$2800~km/s, is detected at $10\sigma$. Introducing the second broad component results into a $\Delta BIC_{1B2B}=39$.

In terms of morphologies, similarly to ID 10013704, both ID 73488 and ID 53757 show the presence of weak asymmetric features, which are likely remnants of recent mergers, hence fitting in the scenario in which these systems may be hosting dual BHs, in the process of merging.

We note that these three cases are not ``double-peaked'' broad lines that have sometimes been invoked in the past as possible signatures of dual AGN, and whose interpretation is ambiguous since BLRs in disk-like configurations can potentially display a double-peaked profile \citep{Eracleous97,Eracleous03,Krolik2019}. Instead, what we detect in these three objects are broad components of the H$\alpha$ line with very different profiles, on top of each other. 

Other recent works on the profile of the H$\beta$ of quasars have been fitted with multiple components \citep{Yang_2023_quasar}, however these are systems that
(beside being in a totally different luminosity regime than our targets) have a H$\beta$ profile that is heavily blended with very broadened [OIII] lines and with strong FeII multiplet emission. Therefore the profile of the H$\beta$ from the BLR cannot be really disentangled properly from the other emission features, including H$\beta$ outflowing components. \cite{Bosman2023} fit the H$\alpha$ of a z$\sim$7 quasar with a double Gaussian, however: 1) it may well be that this is a double BLR too (JWST is indeed revealing that high-z quasars are nearly ubiquitously in merging systems, so the possibility of dual BH is very high); 2) they have not attempted a double-powerlaw profile fit, which is generally more appropriate for quasars; 3) they do not have access to [OIII], hence they cannot assess whether one of the two broad components is due to outflowing gas. In our case, the absence of any broad component in the [OIII] profile rules out that any of the two BLR components is actually associated with an outflow.

Summarizing,
while it might be possible to envisage peculiar geometries of a single BLR which can mimic these profiles, we are not aware of models predicting a narrower profile on top of a broader profile around a single black hole. Even if possible from future models, we think that the interpretation of a dual AGN is a plausible one.

It is interesting that in galaxies ID 1093 (z=5.59) and ID 3680 (z=5.27) the broad component is significantly blueshifted. These could potentially be additional cases in which an accreting black hole is merging, while the black hole of the primary galaxy is not actively accreting. Alternatively, in these cases the two putative black holes already merged and the resulting black hole received a recoil velocity kick, which is expected to happen after BH coalescence \citep{Blecha2011,Blecha2016,Civano2010,Chiaberge2018,Morishita2022}.

In all three cases presented in this paper the putative dual AGN must be located within the MSA shutter (0.2$''$) hence they must be separated by less than $\sim 1$~kpc. There is no clear evidence for a double nucleus in the NIRCam images of these galaxies, indicating that the separation is likely less than about $500$~pc (which is the NIRCam projected resolution at 2$\mu$m at z=5). However, the nucleus with the smaller black hole has lower luminosity (as discussed in the next section) and it is possible that it is outshone by the larger black hole in the NIRCam images, even if at separation larger than 500 pc. However, we are clearly probing a regime of dual AGN much closer and much less luminous than dual quasars found at lower redshifts (z$\sim 0.5-3$) via imaging and spatially resolved spectroscopy \citep[e.g.][]{Mannucci22,Mannucci2023,Ciurlo2023,Scialpi2023}.
Yet, NIRSpec-IFU spectroscopy 
is starting to reveal dual AGN {\it spatially resolved} on scales of a few 100~pc at higher redshift \citep{Ubler24}, providing further support that dual accreting black holes on relatively small scales may not be uncommon.

\subsection{Fraction of merging black holes}\label{sec:merg_frac}

We have found  three candidate merging BH out of the 11 AGN in the redshift range 4$<$z$<$6, which is where we have the best statistics. This is excluding the two AGN with shifted BLR, which may also be merging BHs (in which the primary black hole is inactive), but which may also be recoiled BHs after merging.

As discussed in the next section, the smaller black holes have masses of about $10^6~M_\odot$, while the companion larger black holes have masses of a few/several times $10^7~M_\odot$.

Multiple simulations and semi-analytical models have predicted merging black holes and dual AGN at various cosmic epochs
\citep{Dimatteo2022a,Dimatteo2022,Chen23_dualbh_astrid,Volonteri20_merging,Volonteri22_dualAGN,Barausse20,Barai2018,valentini2021,Vito2022,Dimascia2021,Mannerkoski2022}.
It is however difficult to compare our findings of dual, close-pair AGN with predictions from simulations, as most of them provide predictions of dual AGN and/or merging black holes at lower redshifts (z$<$4) and/or at larger separations ($>$1~kpc) and/or more massive black holes ($M_{BH}> 10^7~M_\odot$) \citep{Chen23_dualbh_astrid,Volonteri20_merging,Volonteri22_dualAGN,Barausse20,
Dimatteo2022a}. Alternatively, other works explore a wider range of BH masses and redshifts, but only provide the expected merger rates \citep{Sesana2004,Sesana2007}, for the goal of establishing the detectability of their gravitational wave signals, which are difficult to compare with our results. 

However, it is useful to compare our findings with simulations in other regimes published so far.
At  redshifts z$\sim$2--3, the simulations mentioned above predict a fraction of dual AGN of only about 1-2\%, while this fraction increases to 10-20\% if considering BH pairs in which only one of the two BH is accreting (hence identified as AGN). This may appear in  tension with our fraction of $\sim$25\% dual AGN (i.e. in which both BH are accreting). However, assessing in simulations whether a BH is accreting or not, at the time of observation, is regulated by assumption at the sub-grid physics level, and therefore there is probably scope to increase the fraction of expected active dual BH in simulations to levels similar to those inferred by us. 

Moreover, while  the smallest black holes in our putative merging systems have masses of only $\rm 10^6~M_\odot$ or a few times $\rm 10^5~M_\odot$, the simulations do not provide predictions on the merging fraction for  black holes smaller than $10^7~M_\odot$;
so the fraction of dual AGN predicted in simulations would probably increase significantly if including lower mass BHs.

Finally, the fraction of dual AGN and black hole mergers is seen to increase at high redshift in some of these simulations.

It is important that future simulations provide predictions specifically in the luminosity, M$_{BH}$, and redshift ranges explored by us, in order to compare exactly the same regimes.

It is also important that future spectroscopic surveys, possibly including  IFS observations, assess the fraction of merging BHs, with higher statistics and expanding the luminosity and BH mass ranges, also to guide the comparison with ongoing and future gravitational wave experiments  \citep[e.g.][]{Amaro2012,Amaro2023,Nanograv2023}

\subsection{Comparison with previous surveys}\label{sec:comparison}


In their sample of 10 Broad Line AGN, from NIRSpec medium resolution spectroscopy, \cite{Harikane23BH} do not find evidence for peculiar H$\alpha$ profiles that could be ascribed to dual BLRs. However, they lack information from high resolution spectroscopy, which we have seen to be crucial to identify and explore the presence of dual BLRs. Specifically, in all cases (possibly with the exception of ID 10013704) we would have not identified the presence of a peculiar BLR profile, if it was not for the profile seen in the high resolution grating.

\cite{Matthee23} used NIRCam slitless spectroscopy, at resolution R$\sim$1600, to identify 20 broad line AGN. Their H$\alpha$ broad lines  do show some peculiar profiles that could be in principle be ascribed to dual BLRs.
We note that the spectral profile measured through slitless spectroscopy is convolved with the spatial distribution of the line emission. If the spatial profile of a line emitting source is not point-like, then it can be difficult to discriminate an intrinsically complex spectroscopic profile from a profile induced by spatially resolved distribution of the nebular emitting gas.
However, there are two cases in their sample with double peaked broad H$\alpha$ profiles, which are intriguing. \cite{Matthee23} interpret these as due to H$\alpha$ absorption. However, H$\alpha$ is not a resonant line and the n=2 is not a metastable level. As a consequence, it is  difficult to see H$\alpha$ in absorption, as it requires extremely high densities, temperatures of $\sim$10$^4$~K and high column densities. Indeed, although a dozen cases of H$\alpha$ absorption have been seen in some type 1 AGN and quasars \cite[e.g.][]{Williams2017,Shi2016,Zhang2015, Schulze2018}, these are extremely rare at low/intermediate redshifts,  less than about 0.1\% of the AGN population. It might be unlikely that 10\% of the type 1 AGN found by \cite{Matthee23} are characterized by such a rare phenomenon. It is possible that the peculiar broad H$\alpha$ profiles might be instead associated with dual AGN. Yet, before making any claim on the H$\alpha$ profile of these targets, slit or IFS spectroscopy would be needed, so to avoid any potential profile artifacts resulting from the slitless spectroscopy, as discussed above. Within this context we note that the detailed spectral analysis of some AGN discovered by JWST is actually revealing Balmer absorption lines that are clearly not associated with a double BLR \citep[e.g.][]{Juodzbalis2024b}, but their occurrence has yet to be assessed. This clearly prompts for more extensive and more detailed spectroscopic observations (especially at higher resolution) of the type 1 AGN that are being discovered by JWST.

\section{Stellar masses and dust extinction}\label{sec:beagle_fitting}

We have used the low-resolution prism spectra from NIRSpec, along with the multi-band photometry from NIRCam, to model the UV to optical rest-frame spectral energy distribution with \textsc{beagle} \citep{Chevallard2016}. We have added a power-law continuum component to \textsc{beagle} to account for the contribution of an AGN.  We fit the spectra with a delayed-exponential star formation history, with a burst of constant star formation lasting 10Myrs prior to observation.  We model the dust attenuation of the star-forming component with the two-component dust law of \cite{CF00}.  The simple model of the emission from the accretion disc consists of a single power-law component and is parameterised by the slope and the fractional contribution to the luminosity at a rest-frame wavelength at 1500\AA.  Since \textsc{beagle} does not include  broad-line region models, we mask all emission lines with significant emission in the prism spectra, limiting the fits to the shape of the continuum. Since the accretion disc emission is likely to be reddened, we included extinction of the power-law component with an SMC dust-law \cite{Pei1992}, which is often found to be appropriate for high-z AGN \citep{Richards2003,Reichard2003A}.  To avoid full degeneracies between reddenning and the power-law slope, we tried fitting with two fixed power-law slopes, allowing the attenuation to vary freely. Specifically, we use both the power-law of a standard Shakura-Sunyaev accretion disc, i.e. $\beta=-7/3\approx-2.33$
(with $\rm F_{\lambda} \propto \lambda ^{\beta}$) \citep{shakura+sunyaev1973}, as well as the power-law inferred from the Sloan composite quasars template obtained by \cite{Vandenberk2001}, i.e. $\beta=-1.56$. The latter is  redder than the Shakura-Sunyaev slope as probably it already incorporates some dust reddening (in addition to the convolution with the accretion disc turnover at high energies in the individual quasar spectra). Of course, quasars and AGN with redder slopes are observed, but generally ascribed to various degrees of reddening \citep{Richards2003}. Therefore our choice of two powerlaws and dust reddening, should cover most of the observed cases, while miniminzing degeneracies.
Both powerlaws generally give acceptable fits. In the following we use the average stellar masses and attenuation inferred from the two cases, and the errorbars will reflect the difference between the two cases (in addition to the errors associated with the individual fitting).

Note that in the case of candidate merging BHs the AGN component was modelled with a single (reddened) powerlaw, as attempting to include two separate powerlaws would result in strong degeneracies.

 Tab. \ref{tab:inferred_properties}  lists the stellar masses and dust attenuation inferred from \textsc{beagle} as discussed above. In Appendix \ref{app:beagle_prism} we show the \textsc{beagle} fits to the prism spectra.
 The inferred stellar masses span from  $\sim 10^8~M_\odot$ up to a few times $10^{10}~M_\odot$.
 We will discuss that, when compared with the dynamical masses, in the case of ID 954 the stellar mass is probably overestimated (likely because of this is the most luminous AGN in our sample and therefore it is more difficult to disentangle host galaxy and AGN components), while for ID 11836 the stellar mass might be underestimated (but see discussion in Sect.~\ref{sec:mbh_mdyn}).
 
 The inferred dust extinctions are similar to those inferred by \cite{Harikane23BH} for the AGN in CEERS and GLASS, but significantly lower than those inferred by \cite{Matthee23}. The latter result may indicate that, as a consequence of the JADES selection function, we may be missing type 1 AGN that are very reddened. This has implications for the AGN census, as discussed later on.

 In Tab. \ref{tab:inferred_properties} 
 we also include GN-z11, whose properties were inferred in \cite{maiolino_bh_2023} and \cite{tacchella_jades_2023}. As mentioned,
 GN-z11 will be included in the rest of the analysis, as part of the JADES sample in GOODS-N (Medium-HST and Medium-JWST tiers).

\begin{figure}[h]%
\centering
\includegraphics[width=1.\columnwidth]{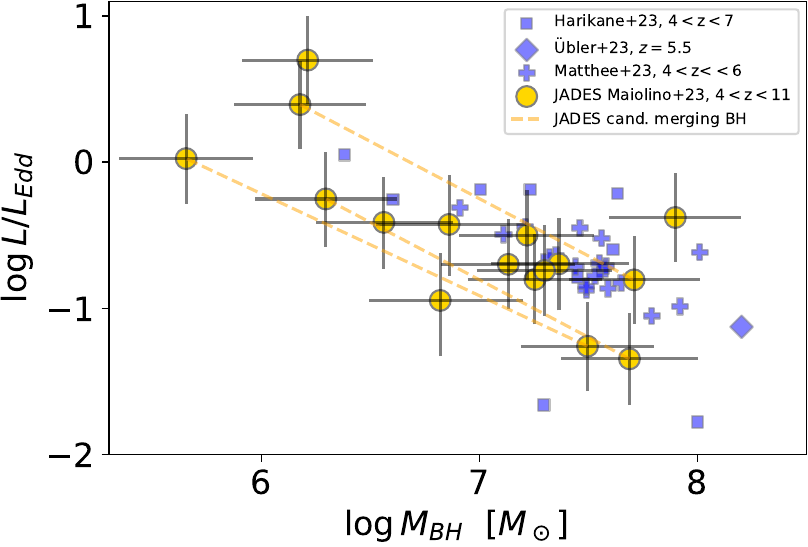}
\caption{
Distribution of black hole masses and Eddington ratios (L/L$_{\rm Edd}$) for the broad line AGN in JADES (large golden circles). The orange dashed lines connect candidate dual AGN. We also show the results from other JWST surveys with blue symbols (see legend, only detections at $>3\sigma$ are shown). Note that the apparent anticorrelation is probably spurious, as the BH mass is at the denominator of the Y-axis quantity. The plot has simply the purpose of visually illustrating the distribution of the two quantities.
}\label{fig:mbh_ledd}
\end{figure}

\section{Black Hole masses and accretion rates}\label{sec:mbh_ledd}

We can estimate the black hole masses in these systems by using
the local virial relations, which link black hole masses with the width of the broad lines and the continuum or line luminosity. We specifically use the relation provided by \cite{Reines13} and \citep{Reines15}, which provides the black hole mass in terms of width and luminosity of the broad component of H$\alpha$:

\begin{multline}\label{eq:mbh}
\log\left(\frac{M_{\rm BH}}{M_\odot}\right)=6.60 \\ + 0.47 \log\left(\frac{L_{\rm H\alpha}}{10^{42}~{\rm erg/s}}\right) + 2.06 \log\left(\frac{\rm FWHM_{\rm H\alpha}}{10^{3}~{\rm km/s}}\right)
\end{multline}

 The advantage of using this relation is that we can consistently compare with the local scaling relations provided by \cite{Reines15}. When comparing with the results from other surveys, for consistency we will re-calculate the black hole masses by using this relation.
The H$\alpha$ luminosity used in Eq.~\ref{eq:mbh} was corrected for dust extinction, as inferred in the previous section.

Obviously, the local virial relations, such as Eq.~\ref{eq:mbh}, are derived locally and there is no guarantee that they apply also to high redshift AGN and quasars. These relations are associated with the small scale ($<$pc) physics and dynamics around the BH, and there is no reason to think that this would change with redshift. The only potential concern is the drop in metallicity and the associated reduction  of the dust content. Indeed, it is thought that the radius of the BLR is primarily set by the dust sublimation radius, which gives the squared root dependence on the luminosity in the virial relations. However, in the dusty ``torus'' the medium remains optically thick to the UV radiation even if the dust-to-gas ratio is very low, because of the extremely large column densities  \citep[typically  $>10^{23}~cm^{-2}$, ][]{Risaliti1999}, while the sublimation radius 
remains set by the dust micro-properties, which do not seem to change drastically in these low mass systems at z$\sim$4--7 \citep{Witstok23UVhump,Witstok23ALMA}. Even if there was some change in the dust properties \citep[which may be possible in much more massive and more luminous systems ][]{Maiolino2004,Gallerani2010}, the weak (squared root) dependence of the BH mass on the AGN luminosity in the virial relations would not affect strongly the estimates of the BH masses at high-z.

The inferred black hole masses are reported in Tab.~\ref{tab:inferred_properties} and span from $\rm 8\times 10^7~M_\odot$ down to $\rm 4 \times 10^5~M_\odot$. The uncertainties on the black hole masses include the propagation of the errors of the quantities involved in calculating them; however, we also take into account the scatter of the virial relations, which contribute to about 0.3 dex (in quadrature) to the black hole mass uncertainty.

The black hole associated with BLR1 of ID 10013704, at z=5.9, has the lowest black hole mass, $\rm 4 \times 10^5~M_\odot$.
Interestingly, this is in the range expected for DCBH ($\rm 10^4-10^6~M_\odot$). Obviously, this does not mean that such a small BH is a DCBH, but it is nonetheless encouraging  that with JWST we are starting to probe this regime. More statistics in this range from future data, and in comparison with expectations from models, may provide constraints on the black hole seeding scenarios.

We can potentially also infer the AGN bolometric luminosity from the monochromatic luminosity of the AGN continuum at a given UV or optical rest frame wavelength and then using bolometric corrections \citep[e.g.][]{Netzer19,Duras20,Saccheo23}. However, this requires a proper deblending of the (reddened) AGN component from the light of the host galaxy, which subject to significant uncertainty. Moreover, in the case of dual AGN, it is not possible to disentangle the contribution to the continuum from the two AGN. An alternative is to use the scaling relations between the luminosity of the broad component of H$\alpha$ and the AGN bolometric luminosity, which also allow the disentangling of the two companions in the case of dual AGN. Specifically, we have used the scaling relation between broad H$\alpha$ (extinction corrected) and AGN bolometric luminosity provided by \cite{Stern12}, as given by their Equation 6.
The resulting AGN luminosities are reported in Tab.~\ref{tab:inferred_properties}.

We can in principle also compare the inferred bolometric luminosity with the Eddington luminosity of the black holes, although one should be aware that, since the H$\alpha$ luminosity has been used to calculate both quantities, there are unavoidably spurious correlations, although the black hole mass (used to infer the Eddington luminosity) primarily depends (quadratically) on the width of the line and its dependence on the H$\alpha$ luminosity is only with the power of 0.5. Aware of these caveats, Tab.~\ref{tab:inferred_properties} reports the inferred
$\rm L/L_{Edd}$ ratios, and these are also visually reported in Fig.~\ref{fig:mbh_ledd}, together with the associated black hole masses. We warn that the apparent anti-correlation between $\rm L/L_{Edd}$ ratio and black hole mass in Fig.~\ref{fig:mbh_ledd} is primarily resulting from the fact that the black hole mass is included at the denominator of the y-axis quantity. Hence, Fig.~\ref{fig:mbh_ledd} should only be used for a quick visualization of the distribution of Eddington ratios and black hole masses. 
Results from previous JWST spectroscopic surveys \citep{ubler+2023,Harikane23BH,Matthee23} are also shown with blue symbols, as indicated in the legend (we only show results with a significance larger than 3$\sigma$ and for which the relevant information is available).

Despite the uncertainties, it is clear that most black holes in our sample accrete at  sub-Eddington rates, mostly $\rm L/L_{Edd}<0.5$. However, the smallest black holes, with masses around or below $\rm 10^6~M_\odot$, tend to accrete at Eddington or super-Eddington rates. One of these is GN-z11, already discussed in \cite{maiolino_bh_2023}, and the other  are two of the candidate small merging black holes. 

As already discussed in \cite{maiolino_bh_2023}, these small BH accreting at high-z are very similar to local/low-z Narrow Line Seyfert 1's, which are indeed found to have low-mass BH, hosted in low-mass galaxies, accreting at super-Eddington rates \citep{Greene2004,Mathur2012}

Our finding of such small and vigorous BHs at high redshift, may suggest that the early phases of black hole accretion happen through Eddington or super-Eddington phases, as suggested by various models \citep{Trinca22,Bennett23}. It is also interesting to note that simulations of dual/merging black holes predict that the smaller black hole might often accrete at a higher rate than the more massive one \citep{Chen23_dualbh_astrid}.

However, it is important also to be aware of possible observational biases. For instance, the detection of small black holes with accretion rates significantly below Eddington would likely make them undetectable in our spectroscopic observations. Therefore, although the finding of low-mass BH at high-z with Eddington/super-Eddington accretion rates is exciting, the lack of their sub-Eddington counterparts might be an observational effect.

Finally, we note that the lack of BHs with masses below a few times $10^5~M_\odot$ is also an observational effect, as such small BHs would have a broad component of H$\alpha$ (only a few 100 km/s) which would be difficult to disentangle from the narrow component of the host galaxy and from even low-velocity outflows.

\begin{figure*}[h]%
\centering
\includegraphics[width=2\columnwidth]{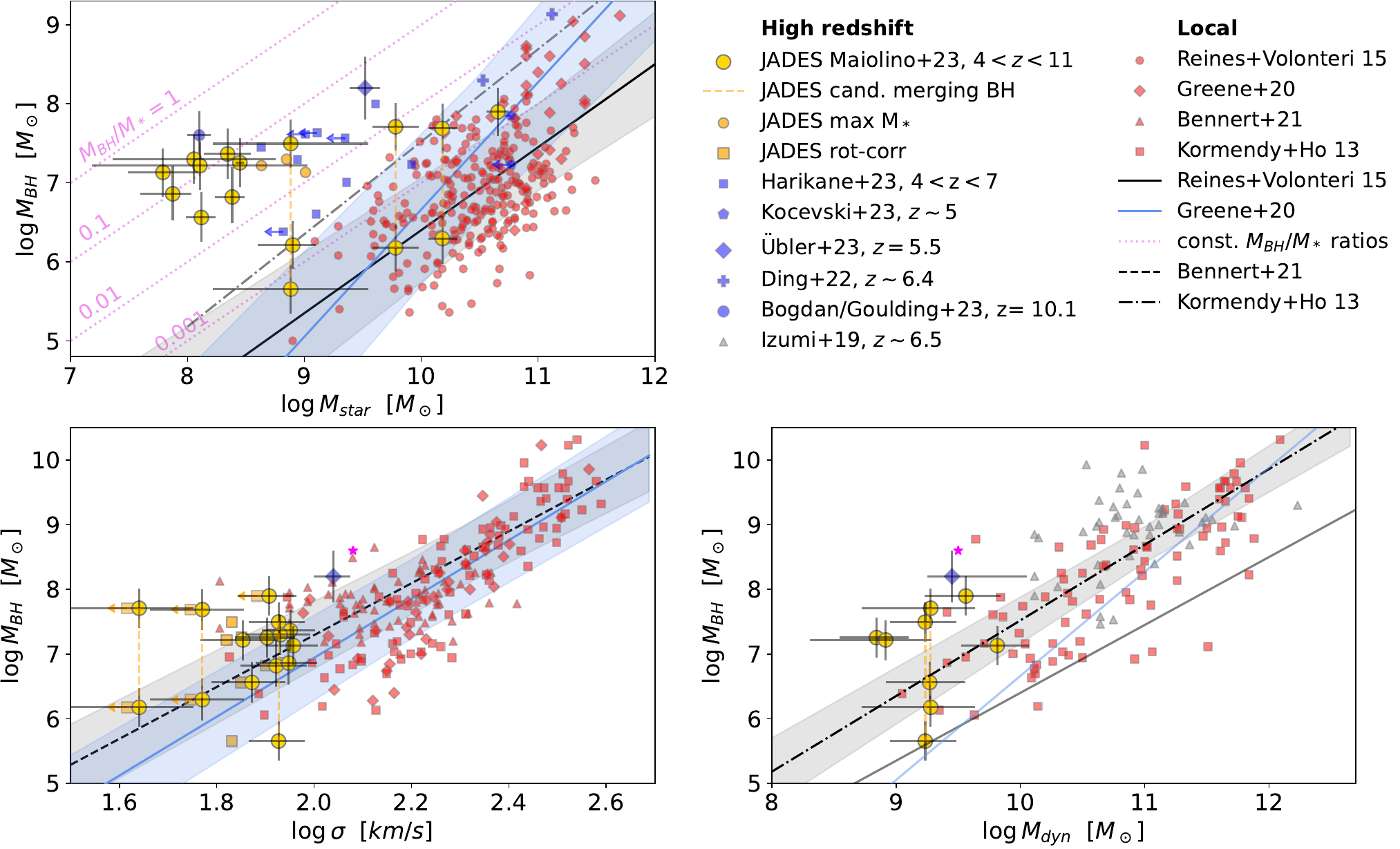}
\caption{
BH mass as a function of host galaxy properties, specifically: stellar mass (top-left),  velocity dispersion (bottom-left), dynamical mass (bottom-right). The JADES results presented in this work are shown with large golden circles. Orange dashed vertical lines connect candidate merging BHs. Blue symbols indicate measurements from other JWST surveys at high-z, as indicated in the legend (only detections at $>3\sigma$ are shown). Gray triangles are measurements of high-z QSOs using ALMA data.  Small red symbols show the distribution of local galaxies as indicated in the legend; straight lines show the local relation fits (with shaded regions providing the scatter and slope uncertainty), using the samples that are best matching our high-z systems, in each panel (see text for details). 
In the top-left panel small orange circles show the  maximum stellar masses estimated for a few JADES AGN (see text) and the dotted violet lines show constant values of the $\rm M_{BH}/M_{star}$ ratio. In the bottom-left panel small squares indicate the effect of correcting for rotation velocity broadening within the slit.
}\label{fig:mbh_host}
\end{figure*}

\section{Overmassive black holes relative to the host galaxy stellar mass}\label{sec:mbh_mstar}

Fig.\ref{fig:mbh_host}, top-left panel, shows the black hole mass as a function of stellar mass of the host galaxy. The small red circles show the  distribution of local active galaxies obtained (consistently with the same calibration) by \cite{Reines15}; their best-fitting relation is shown by the black solid line, while the gray shaded region shows the dispersion and uncertainty of the fit.
This is probably the best local relation for comparing the $\rm M_{BH}-M_{star}$ with our results, as \cite{Reines15} use the same method as ours for estimating the BH masses in AGN, and also because their host galaxies are mostly star forming, late type systems as in our sample (given that the bulk of the host galaxies of the AGN in our sample have Sersic index of $\sim 1$, as shown in Tab.\ref{tab:inferred_properties}).
The red diamonds show additional measurements provided by \cite{Greene20}; the best-fit relation provided by them is shown with the blue line and blue shaded region (giving the dispersion and uncertainty of the relation); we caution that \cite{Greene20} uses different BH estimations relative to our prescription, so the comparison with our results has this additional uncertainty factor. For completeness, we also show the local relation by \cite{Kormendy13} with dash-dotten line, which is however for a sample made primarily of early type galaxies, hence probably not directly comparable with our sample in terms of stellar mass \citep{Sturm2024} given that our AGN are primarily hosted in late type galaxies, but may be more adequate for comparing with dynamical masses, as discussed in the next section. The blue squares indicate high redshift broad-line AGN recently discovered with JWST by various AGN surveys, where we report only those detections that are at least more significant than 3$\sigma$ and which have a stellar mass reported \citep{Kocevski23,Harikane23BH,ubler+2023,Ding2022,Bogdan2023,Goulding2023}. The new JADES broad lines AGN, including GN-z11, are shown with large golden circles. The vertical, orange dashed lines connect the candidate merging black holes. 

Most black holes at these early epochs are significantly over-massive relative to the stellar mass in their host galaxies, when compared with the local scaling relation. This was already found by previous surveys (blue squares), but for the newly identified AGN this phenomenon becomes even more prominent. In some extreme cases we even find black hole masses approaching the stellar masses of their host galaxies \citep[see also ][]{Bogdan2023,Goulding2023}.

The strongest deviations occur at low stellar masses ($M_{star}<10^9~M_\odot$), although we caution that at such low masses the local scaling relation is actually poorly explored and we mostly rely on the extrapolation from higher masses.

Such a strong offset may be partly due to the stellar masses being significantly underestimated. However, even if we use masses obtained 
without accounting for the AGN in the continuum modelling (i.e. assuming that all continuum emission is due to stars), the JADES AGN are still located well above the local relation; this is illustrated by the orange circles in the top-left of Fig.~\ref{fig:mbh_host}, which show the maximum stellar mass inferred in this way (for sake of clarity we only show those few cases for which the maximum stellar mass exceeds the mass inferred with the standard stellar+powerlaw method by more than 0.1 dex).  

BHs being overmassive relative to the local $M_{BH}-M_{star}$ relation may be partly due to selection effects.
Specifically, given a scatter around the $M_{BH}-M_{galaxy}$ relation, the sensitivity limit of quasar/AGN surveys may favour the detection of more massive BHs (which, for a given  average $L/L_{Edd}$, are more luminous) \citep{Willott2015,Lauer2007}. Although this effect was thought to be less important at low AGN luminosities \citep{Izumi19}, especially below the quasar regime, it may still play a role, as also recently highlighted by \cite{Volonteri23}. 
The BH selection bias on this relation was explored more recently and more in detail by \cite{Li2024bias} and \cite{Juodzbalis2024} specifically for the JWST-selected broad line AGN. \cite{Li2024bias} claim that the offset can entirely explained with a $M_{BH}-M_{star}$ relation consistent with the local one, but a much larger dispersion (about 1 dex) and selection effects on the BH mass. However, \cite{Juodzbalis2024} illustrate that the finding of overmassive black holes that are dormant (i.e. with very low accretion rate) does not fit in the scenario outlined by \cite{Li2024bias}. Additionally, in the next section we will show that the JWST-discovered broad line AGN are fairly consistent with the local $M_{BH}-\sigma$ relation (where $\sigma$ is the velocity dispersion) - if the offset on the $M_{BH}-M_{star}$ were dominated by selection effects then the same offset should be seen on the $M_{BH}-\sigma$ relation. There are also other issues related to the space density of AGN that makes unlikely the selection bias being the dominant effect in the $M_{BH}-M_{star}$ offset; however, this is beyond the scope of this paper and will be discussed in a separate, dedicated work (Maiolino et al. in prep.).

Finally, it is interesting to note that black holes above the local scaling relation at early epochs, and even approaching $\rm M_{BH}\sim M_{star}$, are expected by various theoretical models, especially scenarios envisaging black holes accreting at super-Eddington rates or with very high efficiency, and are also expected to be associated with heavy seeds (aka DCBHs) \citep{Trinca22,Schneider23,Koudmani2022}.

\section{The $M_{BH}-$\texorpdfstring{$\sigma$}{sigma} and $M_{BH}-M_{dyn}$ relations: fundamental and universal?}\label{sec:mbh_sigma_dyn}

\subsection{The BH--velocity dispersion relation}\label{sec:mbh_sigma}

In the local Universe, studies of the black hole scaling relations with the properties of the host galaxies, have unambiguously found that the tightest relation is with the central (stellar) velocity dispersion
\citep{Kormendy13,Terrazas17,Piotrowska22}, suggesting that this is the most fundamental (causal) relation, while other relations may simply be an indirect byproduct.

As discussed, most of the AGN in our sample have also high resolution spectra, covering both [OIII]5007 and H$\alpha$, or at least one of these two bright emission lines. For compact sources the spectral resolution in terms of velocity dispersion is about 25~km/s (FWHM$\sim$60~km/s) in the spectral range of interest (de~Graaff et al. in prep). We can therefore  accurately measure the velocity dispersion of the gas in the host galaxies of these AGN (taking into account the instrumental resolution, although this is generally a small correction). This is not necessarily the same as the stellar velocity dispersion used in the local scaling relations; however, various studies have shown that the ionized gas velocity dispersion can be used as a good proxy, although it may require some small correction \citep{Bezanson18b}. Following a similar approach as in \cite{ubler+2023} for another JWST identified AGN at z=5.55, we implement an upward correction of 0.12-18~dex \citep{Bezanson18b} to the gas velocity dispersion in order to get a close estimate of the stellar velocity dispersion.

Fig.\ref{fig:mbh_host}-bottom-left shows the resulting relation between black hole mass and velocity dispersion. As in the top-left panel, the red small circles indicate local galaxies from \cite{Kormendy13}, \cite{Bennert21} and \cite{Woo2015}, while the black solid line shows the best fit relation from \cite{Bennert21}, with the shaded region showing the uncertainty and dispersion.
Red diamonds show the new local measurements reported by \cite{Greene20}, while the blue line and shaded region show the local relation and dispersion  obtained by them.
The previous JWST measurement on a single AGN from \cite{ubler+2023} is shown with a blue diamond, while our new JADES AGN are shown with large golden circles. Also in this case, the vertical, orange dashed lines show the candidate merging BHs. 

The AGN at z$\sim$4-6 are mostly located on the same relation as local galaxies, most of  them fully consistent with the local scatter, at least when compared with the local relation obtained by \cite{Bennert21}. There is a small offset relative to the relation obtained by \cite{Greene20}, although still within the scatter, and anyhow the offset is far smaller than for the $\rm M_{BH}-M_{star}$ relation.
The only exception is the more massive component of the putative merging system ID 73488 (the leftmost JADES point in the figure), which is significantly above the relation, but  whose smaller companion is, interestingly, on the relation. However, even \cite{Bennert21} report some local cases with a similar large deviation (see their lowest velocity disperion point in the figure).

The fact that the $\rm M_{BH}-\sigma$ relation at z$>$4 is essentially consistent with the local relation,  confirms that this relation is not only fundamental, but also universal, as it holds across most cosmic epochs, at least out to z$\sim$6. The implications of this finding for the co-evolution of galaxies and black holes are profound. It is beyond the scope of this paper to explore the physics behind this relation and why it is so stable across the Universe. Here we only comment that, combined with the previous finding of the dissolving relation between $\rm M_{BH}$ and $\rm M_{star}$ at high-z, these results indicate that the black holes at the center of galaxies have essentially no knowledge of, nor connection with, the star formation history of their host galaxies, while they are connected with the mass assembly history.

The tight relation between black hole mass and velocity dispersion is often interpreted as a causal link between the formation of black holes and the merging history of the central part of the galaxy, responsible for the formation of the spheroidal component \citep{Kormendy13}. The finding that such a relation was already in place around the epoch of re-ionization, suggests that early merging and the early spheroidal formation \citep{Baker23bulge} was linked to the early evolution of black holes. Our finding of candidate merging BHs, presented in Sect.~\ref{sec:merging} consistently supports this scenario, although quantifying the scenario requires a close comparison with simulations.

We conclude by noting that our measurements of the line-of-sight gas velocity dispersion include contributions from both the random motions from the gas as well as, potentially, projected rotational velocities of the gas within the slit. This velocity measurement is therefore a good approximation of the second moment of the velocity distribution of the system, which enters the virial theorem \citep[e.g. ][] {Cappellari2006}. As a consequence, the measured velocity dispersion is a good proxy of the gravitational potential and dynamical mass (as we will discuss in the next section). However,
measurements of the velocity dispersions in local galaxies (and used in the local $\rm M_{BH}-\sigma$ relation) span a wide range of projects apertures, from 100 pc scale to kpc scale (hence overlapping with the scales probed by us), and may or may not include the contribution from unresolved rotation.
Therefore, it is not clear whether our measurements should be corrected for the effect of unresolved rotation in order to be compared with the local relation.
However, we also explore the extreme case in which we should only consider the intrinsic velocity dispersion. \cite{Carniani23} estimates the effect of line broadening due to potentially unresolved rotation curves, when information on the mass of the galaxy, radius and inclination is available. Based on the parameters of the host galaxy that we could derive from imaging and from prism spectroscopy (Sects.~\ref{sec:images} and \ref{sec:beagle_fitting}), we estimate the line broadening from putative rotation and subtract it in quadrature to the observed gas velocity dispersion. The resulting values  are shown with offset orange large squares in the bottom-left panel of Fig.~\ref{fig:mbh_host}. In most cases the effect is minimal, although there are a few cases for which we could only infer an upper limit on the intrinsic velocity dispersion (marked with orange upper limits). We also note that the previous result from \cite{ubler+2023} on an AGN at z=5.55 (blue square) was obtained via spatially resolved NIRSpec IFS and in that case it is directly verified that the small amount rotation has a negligible contribution to the velocity dispersion (and it is accounted for).  Summarising, the result of the $M_{BH}-\sigma$ relation at z$\sim$4--7 being mostly consistent with the local one remains valid even if considering only the intrinsic velocity dispersion, although there might be a few outliers.

\subsection{The black hole-dynamical mass relation}\label{sec:mbh_mdyn}

Given that the velocity dispersion is also linked to the dynamical mass through the virial theorem, it is interesting to explore also the relation between black hole mass and dynamical mass.

To infer the latter we need the information on the radius of the galaxy and on its mass profile (probed by the S\'ersic index), which were inferred through the morphological analysis in Sect.~\ref{sec:images}. 
We then estimate the dynamical mass by following the same approach as in \cite{ubler+2023} through the equation
\begin{equation}\label{eq:mdyn}
    M_{\rm dyn} = K(n)K(q)\frac{\sigma^2 R_e}{G},
\end{equation}

where $K(n)=8.87-0.831n+0.0241n^2$ with Sérsic index $n$, following \cite{Cappellari06}, $K(q)=[0.87+0.38{\rm e}^{-3.71(1-q)}]^2$, with axis ratio $q$ following \cite{vdWel22}, and $R_e$ is the effective radius.

We note that, although the broadening associated with unresolved rotation is likely not contributing significantly to the observed velocity dispersion, the advantage of the dynamical mass is that Eq.~\ref{eq:mdyn} {\it does require $\sigma$ to include both intrinsic and unresolved rotation components} \citep{cappellari2013a}, therefore it does not suffer the potential issues discussed in the previous section.
However, it is also true that, given the uncertainties on the several quantities involved in the determination of the dynamical mass (and in particular the radius), this quantity is more uncertain than the simple velocity dispersion measurement. 

The resulting dynamical masses are reported in Table \ref{tab:inferred_properties}, limited to those cases for which we have enough information to derive this quantity. In most cases the dynamical mass is larger than the stellar mass. In  some cases the two are consistent with each other within the uncertainties. However, in some cases the dynamical mass is significantly larger than the stellar mass, even by more than one order of magnitude. This is not totally surprising, as it is known that high redshift galaxies (already at z$\sim$1--4) can have gas fraction even higher than 90\%, especially in the low mass regime explored here \citep{Tacconi2020,Santini2014,Scoville2017,Liu2019,Dessauges2020,Zhang2021}. The specific case of ID 11836 is remarkable, as the inferred dynamical mass is about two orders of magnitude higher than the stellar mass. It might be that in this case the stellar mass is significantly underestimated owing the difficulty to account for the AGN contribution, and indeed not including the power-law component gives a stellar mass that is $\sim$15 times higher; however, it is also true that previous studies at z$<$4 have found some galaxies with inferred gas masses that are nearly two order of magnitudes higher than the stellar masses \citep{Scoville2017,Zhang2021,Liu2019}.

\begin{figure*}[h]%
\centering
\includegraphics[width=1.5\columnwidth]{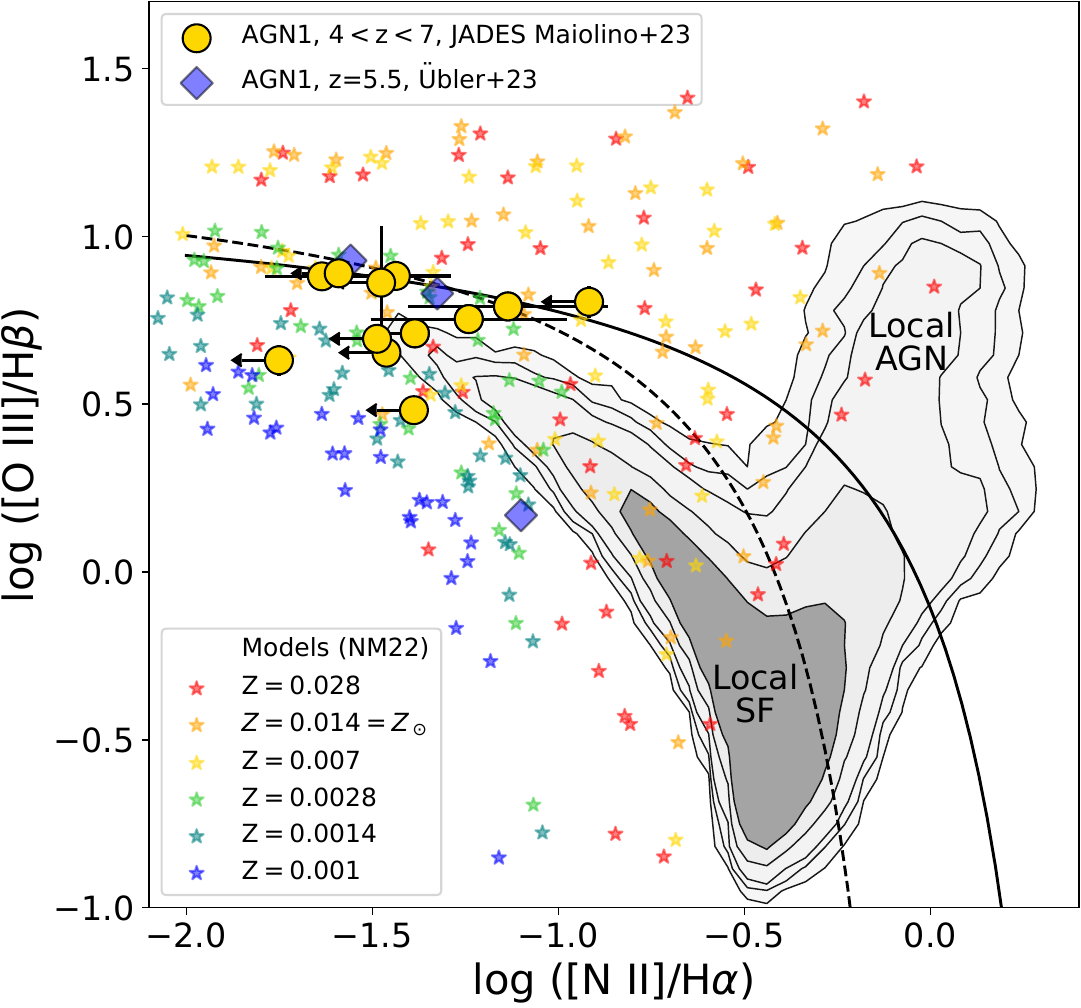}
\caption{
[NII]-BPT diagram. Contours show the distribution of  galaxies in the local Universe from SDSS (DR7), showing the AGN sequence (right) and the star forming sequence (left). The solid and dashed lines are the corresponding dividing lines from \cite{Kewley01} and \cite{Kauffmann03}, respectively. The JADES broad line AGN presented in this paper are shown with golden circles, which blue diamonds show the values obtained by \cite{ubler+2023} for an AGN at z=5.5. Stars show predictions by AGN photoionization models from \cite{nakajima_diagnostics_2022} color-coded by metallicity.
}\label{fig:bpt}
\end{figure*}

ID 954 is also another extreme case in which the inferred dynamical mass is about one order of magnitude lower than the stellar mass. This is the most luminous AGN in our sample, and it is likely that the spectral decomposition struggled to disentangle the stellar component. 

The bottom-right panel of Fig.\ref{fig:mbh_host} shows the black hole mass as a function of dynamical mass, where the values obtained for the JADES broad line AGN are shown with golden large circles. In the local Universe there is shortage of proper equivalent measurements of the scaling relation in terms of dynamical masses. This was attempted by \cite{Beifiori12}, however most of the BH in their sample have upper limits on their masses and the sample has a limited mass range. We follow \cite{ubler+2023} by taking, as a good approximation, the $\rm M_{BH}-M_{bulge}$ relation obtained by \cite{Kormendy13}, 
where their measurement of $\rm M_{bulge}$ in ellipticals is close to a dynamical mass measurement. The values for local galaxies obtained by them are shown with small red squares and their best-fit relation is shown with a dash-dotted black line and the uncertainties and scatter with the shaded region. For completeness we also show the relations by \cite{Reines15} (black solid line) and \cite{Greene20} (blue solid line), although, as discussed, these are likely less appropriate for comparing with the dynamical mass. We also plot the estimates of  $\rm M_{BH}$ and $\rm M_{dyn}$ for a sample of quasars at z$\sim$6.5 (grey triangles), for which the dynamical masses were inferred from the velocity dispersion of far-IR/submm transitions (typically [CII]158$\mu$m) with ALMA \citep{Izumi19}. We also report the previous measurement obtained with the NIRSpec IFU observation of an AGN at z=5.5 by \cite{ubler+2023}. 

The $\rm M_{BH}-M_{dyn}$ relation at high-z is, unfortunately, less populated than the $\rm M_{BH}-\sigma$ relation (due to the lack of information to derive the dynamical mass in a number of galaxies). The larger scatter of the $\rm M_{BH}-M_{dyn}$ relative to the $\rm M_{BH}-\sigma$ relation is probably partly a consequence of the additional uncertainties in deriving $\rm M_{dyn}$. However, the interesting result is that the broad line AGN found in JADES are not strongly offset from the local relation, in contrast with the case of the $\rm M_{BH}-M_{star}$ relation. Broad line AGN at z$>$4 are generally scattered around the local relation. There is a tendency for more AGN being offset above the local relation, but not by more than two times the local dispersion, and consistent with the deviations seen in other local galaxies. Part of the better agreement may be a consequence of using here the \cite{Kormendy13} relation (which, as discussed above, is more adequate in this case), instead of the \cite{Reines15} relation. However, even considering only the \cite{Kormendy13} local relation, it remains true that the offset is much larger on the  $\rm M_{BH}-M_{star}$ than on the $\rm M_{BH}-M_{dyn}$ relation.

More accurate measurements are required, possibly with IFS spectroscopy. However, based on the current results, at face value we can say that the $\rm M_{BH}-M_{dyn}$ is generally consistent with the same relation in local galaxies.

Overall, these findings indicate that, at these early epochs, the BH mass follows well the mass assembly of the host galaxy, in a similar way as local galaxies. However, at these early epochs most of the mass is still in gas, which has not yet been converted to stars, and therefore explaining the large offset in the $\rm M_{BH}-M_{star}$ relation.

\section{High-z AGN are elusive in standard diagnostic diagrams}\label{sec:bpt}

For AGN at z$<$7 the JADES spectra cover both [OIII] and H$\beta$ as well as [NII] and H$\alpha$. Therefore, from the narrow components of these lines it is possible to locate these AGN on the so-called BPT diagnostic diagrams \citep{Baldwin81,Veilleux1987}. Generally in these high-z AGN the [NII] doublet is very faint and in many cases undetected, so we can often only set an upper limit on the [NII]/H$\alpha$ ratio. The resulting distribution of the JADES AGN on the [NII]-BPT diagram is shown with golden circles in Fig.\ref{fig:bpt}, in which the local distribution of galaxies from the SDSS survey is shown with shaded contours (where the lowest contour includes 99\% of the local galaxies). The solid and dashed lines indicate the demarcation between AGN and star forming galaxies provided by \cite{Kewley01} and by \cite{Kauffmann03}, respectively, for local galaxies.

As already found by other studies based on JWST spectroscopic data \citep{Kocevski23,ubler+2023,Harikane23BH}, AGN at z$>$4 are completely offset from the AGN locus in the local Universe, and are mostly overlapping with the region that is locally occupied by star forming galaxies. Clearly, these diagnostic diagrams cannot discriminate AGN from star forming galaxies in these early systems.

\section{Very low metallicity host galaxies}\label{sec:Z}

One possible explanation for the offset in the BPT diagram is that the narrow emission lines are not primarily associated with the AGN NLR, but actually dominated by star formation in the host galaxy. 
Recently, \cite{Maiolino24X} have suggested that the covering factor of the BLR clouds in these JWST-discovered AGN could be very high, hence leaving few ionizing photons escaping to produce a NLR on larger scales.
However, we know that in at least a few cases the narrow lines are certainly dominated by the AGN. For instance, in the broad-line AGN GS-3073, explored in detail and with very high signal to noise by \cite{ubler+2023}, the narrow lines also have the clear detection of strong HeII4686 (with high EW typical of AGN), and other high ionization lines, such as [ArIV] and coronal lines, which are typically tracing the NLR of AGN. Despite this, GS-3073 is clearly completely offset from the AGN local locus on the BPT diagram and is instead located on the local SF sequence (blue diamonds in Fig.~\ref{fig:bpt}).

Another likely possibility, also pointed out by previous works \citep{Kocevski23,ubler+2023,Harikane23BH}, is that the NLR of high-z AGN is characterized by low metallicities. This is indicated by the starred symbols in Fig.\ref{fig:bpt}, which are the result of the photoionization models for the NLR of AGN obtained by \cite{nakajima_diagnostics_2022} and color-coded by metallicity. Clearly, as the metallicity of the NLR decreases from super-solar to sub-solar, the expected location on the BPT diagram shifts from the local AGN locus to overlap the local star forming locus, and beyond. Most of the AGN on the local star forming sequence can be explained in terms of a NLR characterized by a metallicity of about 0.2~Z$_\odot$, although there is large dispersion in the models.

It is interesting that two of the AGN in JADES are even below the local star formation sequence. In these cases, the metallicity of the NLR is probably particularly low,  less than 0.1~$\rm Z_\odot$. These targets are interesting candidates for followup with IFS observation to investigate their very low metallicity environment.

\section{Fraction of AGN at high redshift}\label{ssec:frac_agn}

Previous JWST studies assessing the fraction of type 1 AGN have reached different conclusions. \cite{Matthee23} used NIRCam slitless spectroscopy to assess the fraction of broad line AGN in the EIGER and FRESCO surveys, and infer that AGN in their sample is less than 1\% of star forming galaxies at z$\sim$5. It is important to consider that, due to the limited sensitivity of the slitless surveys used by them, they probe H$\alpha$ broad line luminosities higher than $\rm 2\times 10^{42}~erg/s$ and resulting in AGN bolometric luminosities   $\rm L_{bol} > 5\times 10^{44}~erg/s$.
\cite{Harikane23BH} uses slit spectroscopy from the CEERS, ERO and GLASS surveys to search for AGN, and estimate that 5\% of galaxies at 4$<$z$<$7 host a broad line AGN. Their slit spectra allow them to probe H$\alpha$ broad lines with luminosities down to $\rm 10^{41}~erg/s$ and infer AGN bolometric luminosities down to $\rm L_{bol} > 5\times 10^{43}~erg/s$. Therefore, the higher fraction of broad line AGN inferred by \cite{Harikane23BH} is likely a consequence of the lower luminosity range probed by them.

In the case of JADES
 is not simple to assess the fraction of broad line AGN based on the current JADES NIRSpec spectra, as the selection function of the NIRSpec targets is quite complex, and varying in different tiers. In Deep/HST only a fraction of the targets was selected based on NIRCam data. Moreover, as discussed in Section \ref{sec:data},  Deep/HST, Medium/HST and some of the other early JADES observations have likely been biased against AGN. 
These issues may be responsible for the fact that only two broad line AGN have been found in Deep/HST and two in Medium/HST GOODS-N.

The Medium/JWST GOODS-N tier is the first JADES tier mostly based on NIRCam-selected targets and in which high-z AGN may have not been discarded because of their peculiar colors, and hence not biased against AGN. The fraction of broad line AGN found in this tier is probably more representative of the population of AGN at high redshift. However, in the case of Medium/JWST a few objects (specifically 73488, 77652 and 61888) were targeted just because of their compact morphology and peculiar colours \citep[resembling the type 1 AGN, such as the one identified by ][]{ubler+2023} suggesting their AGN nature (Sect.~\ref{sec:data}), and were indeed confirmed as such in our analysis; therefore, in the case of this sub-sample the selection has obviously been biased in favour of AGN.

Aware of all these caveats, we attempt an assessment of the fraction of AGN in the Medium/JWST GOODS-N tier in the redshift range 4$<$z$<$6, in which the statistics of targeted galaxies with confirmed spectroscopic redshift is high enough. The fraction of broad-line AGN (with $\rm L_{AGN}> 10^{44}~erg/s$, i.e. the luminosity range to which we are sensitive) in this subsample is 11\%. If we exclude the galaxies that were specifically selected because of their AGN-like properties in imaging, then the fraction drops to 7.5\%. However, removing the AGN targeted because of their imaging and photometric properties is actually biasing the sample against AGN (as in the early JADES tiers), as their properties are actually in the range of the JADES selection function. Therefore, actual fraction of galaxies at 4$<$z$<$6 hosting broad line AGN with $\rm L_{AGN}>5\times 10^{43}~erg/s$ must be between the two estimate given above, i.e. between 7.5\% and 11\%. This is higher than what inferred by \cite{Harikane23BH}, who probed a similar sensitivity range. However, the statistics is still modest in both studies (nine targets in this redshift range in both studies). Moreover, the selection functions in the two samples are different and the selection criteria in GLASS and CEERS surveys may have penalised the selection of type 1 AGN.

As discussed in Sect.~\ref{sec:beagle_fitting}, the AGN in our sample have significantly lower dust extinction than the broad line AGN found by \cite{Matthee23} via slitless spectroscopy, and  this may be a consequence of the JADES targets selection function (which has various priority classes that are selected also based on the UV luminosity) that has likely penalised reddened AGN. So the actual fraction of broad line AGN, including the significantly reddened population in our luminosity range, is certainly higher. It is difficult to assess accurately this fraction, as it would require a dedicated, deep MOS spectroscopic survey purely selected based on the red photometry (e.g. F444W).

Finally,  our estimated fraction of AGN refers only to the class of type 1, while the fraction of obscured, type 2 AGN is certainly higher. In the local Universe  type 2 AGN  are about a factor of 4 more numerous than type 1 AGN \citep[e.g. ][]{Maiolino95}. At high redshift the fraction of type 2 AGN has been more difficult to assess, due to sensitivity issues \citep[e.g. ][]{Merloni2014,Maiolino2007,Netzer2016}. These aspects are  explored more extensively, by using the JADES data, in a separate paper \citep{Schotlz2023NLAGN}.

It is important to compare our results with the expectations of some models and simulations. In particular, using the semi-analytical model CAT, \cite{Trinca23seekBH} predict that, at the magnitude limit probed by the Medium/JADES tier, we should detect a few dozen type 1 AGN in the redshift range 5$<$z$<$6. This is not too far from the seven that we have detected in the same redshift range, taking into account that we have only used about 1/3 of the Medium/JWST tier (the GOODS-N component) and that the Medium/HST tier may be biased against AGN, as discussed above. Moreover, we may have still missed a significant fraction of AGN, because too faint relative to their host galaxy, an issue that has been extensively discussed in \cite{Volonteri23} and \cite{Schneider23}. It is interesting to note that \cite{Trinca23seekBH} expect that most of these AGN should host BHs in the mass range $\rm 10^6-10^8~M_\odot$, as found by us. According to their model most of these BH are formed out of `heavy' seeds, i.e. from Direct Collapse Black Holes.

\begin{figure*}[h]%
\centering
\includegraphics[width=1.5\columnwidth]{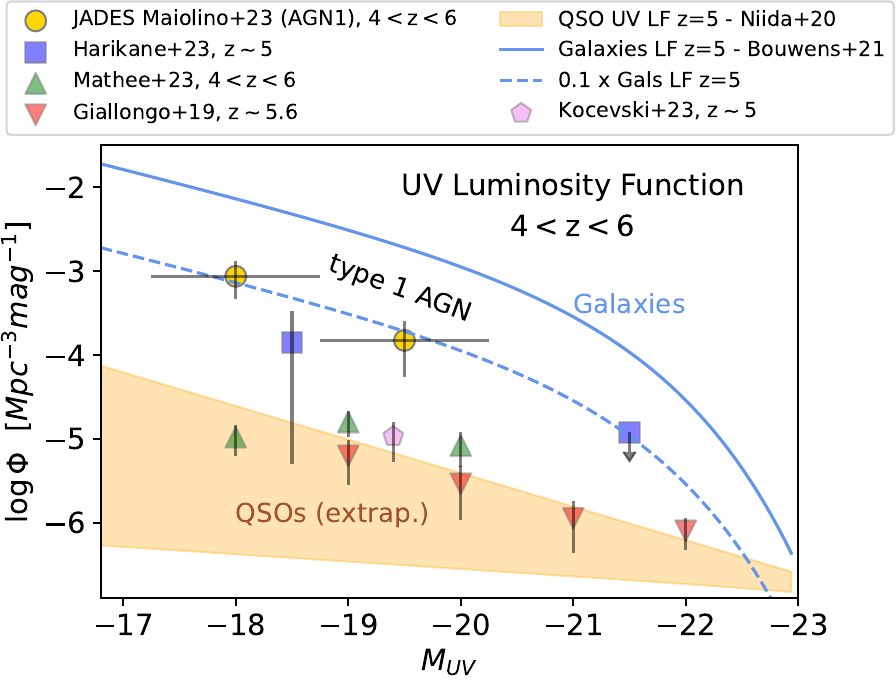}
\caption{
UV luminosity function of galaxies at z=5 from \cite{Bouwens21} (blue solid line) and inferred contribution of galaxies hosting broad line (type 1) AGN with $\rm L_{bol}> 10^{44}~erg/s$, inferred by us from the JADES survey (golden circles). Results from other surveys are also reported, as indicated in the legend. The dashed blue line shows the galaxy luminosity function scaled downward by a factor of 10 and fitting the JADES points. The orange-shaded region shows the range of possible extrapolated luminosity functions for QSOs from \cite{Niida2020}.
}\label{fig:lf}
\end{figure*}

\section{Contribution of AGN to the UV luminosity function}\label{ssec:lf}

We can explore the contribution of the galaxies that host broad line AGN to the UV luminosity function. Assuming that within the 4$<$z$<$6 redshift bin, selection effects within each UV luminosity bin are secondary (i.e. star forming galaxies and AGN in a given UV luminosity bin do not have significantly different probability of being selected for spectroscopy), then the contribution of galaxies hosting broad line AGN in each bin of the UV luminosity function can be inferred from the fraction of AGN in that bin in our spectroscopically targeted sample. As a reference, we take the functional form provided by  \cite{Bouwens21} to describe the UV luminosity function of galaxies at z$\sim$5. 
As discussed above, we have the ambiguity of excluding or including the three AGN specifically targeted because of their AGN appearance in the NIRCam images. We take the mid-point of the two extreme cases and take the deviation from each of the two cases as a contribution of the errorbars, in quadrature to the Poissonian noise and cosmic variance \citep[estimated following ][ where the bias factor is function of redshift and mass]{Somerville2004}.

The UV luminosity function of broad line AGN with $\rm L_{AGN}>5\times 10^{43}~erg/s$ and 4$<$z$<$6 inferred by us 
 from the JADES survey is shown with golden circles in Fig.\ref{fig:lf} and reported in Tab.~\ref{tab:lf} (in the last column of the same table we also provide the minimum density of broad line AGN assuming the most conservative and extreme case that the AGN identified in the current spectroscopic sample are the only AGN in the entire volume surveyed, i.e. that there are no other AGN missed among the vast majority of galaxies that have not been targeted spectroscopically).).
 To obtain a reasonable statistics in UV luminosity bins, we only adopt two bins, one centered at $M_{UV}=-18$ and a second one centered at $M_{UV}=-19.5$. In the lower luminosity bin broad line AGN and their hosts contribute for 12\% to the luminosity function of galaxies, while in the higher luminosity bin the contribution is 8\%. Certainly these contributions are expected to be higher if one could probe lower luminosity AGN.

 We do not attempt to fit a functional form of the luminosity function to these two data points, as the statistics is still too low and errorbars still too large, which would result in different functional forms being largely unconstrained. However, 
 we can reproduce  the values for broad line AGN by simply scaling down 
 the \cite{Bouwens21} UV luminosity function of galaxies by a factor of 10, as illustrated by the dashed blue line.
 
 In Fig.~\ref{fig:lf} we also compare our finding with results from other JWST surveys: \cite{Harikane23BH} blue squares; \cite{Matthee23} green triangles; \cite{Kocevski23} violet pentagon. The red triangles show the luminosity function inferred by  \cite{Giallongo19} based on X-ray surveys.
 We also show the range of extrapolations of the quasar luminosity function at z$\sim$5 as inferred by \cite{Niida2020}.

Our results are, within uncertainties, consistent with the finding of \cite{Harikane23BH}, not surprisingly given that they probe a similar range of AGN luminosities. The luminosity function inferred by \cite{Matthee23} is much lower, which is likely a consequence of the higher luminosities probed by them.
The low value inferred by \cite{Kocevski23}, which uses the same CEERS spectra as \cite{Harikane23BH}, is probably a consequence of the low statistics in that early study. It is interesting that our estimated AGN density is higher than inferred by \cite{Giallongo19} by using some of the deepest X-ray data, indicating that JWST is currently being much more effective in finding AGN than current X-ray surveys and indicating that future X-ray missions are necessary to find and characterise this population of high-z AGN and, most importantly, find the obscured counterparts.

We finally note that the range of possible extrapolations of the QSO luminosity function at z$\sim$5 (orange shaded region) is clearly below the AGN luminosity function estimated by us, indicating that we are probing a quite different population relative to luminous quasar, and not simply their low-luminosity tail.

Given that the yield of ionizing photons ($\xi _{ion}$) by AGN is larger than for star forming galaxies, these findings imply that the contribution to reionization of galaxies hosting broad line AGN (with $\rm L_{AGN}>10^{44}~erg/s$) can potentially be significant.

We note that this result does not necessarily imply that  AGN contribute substantially to the reionization of the Universe, as the UV luminosity that we measure is the sum of the contribution from the AGN and host galaxy. In order to obtain the specific contribution from black hole accretion we should disentangle in each of the selected galaxies the contribution of the (dust-reddened) accretion disc from the light emitted by the young stellar population. The decomposition attempted in Sect.\ref{sec:beagle_fitting} is appropriate to infer the properties of the stellar population, however it is inappropriate to use it to extrapolate the AGN contribution to the extreme UV. A more detailed modelling (also involving the nebular lines) is needed for this goal.
However, if most of the UV emission turns out to be dominated by the AGN, then (given their large escape fraction) these could potentially contribute to a large fraction of the photon budget required for the reionization of the Universe \citep{Madau2024}.


\section{Summary and conclusions}\label{sec:summary}

We have used three tiers of the JADES NIRSpec survey,  specifically Deep/HST (in GOODS-S), Medium/HST and Medium/JWST (in GOODS-N), to search for broad line AGN at z$>$4. The combination of depth and the use of dispersers providing three different resolutions, has enabled us to find this class of AGN more efficiently and has allowed us to explore different regimes relative to previous studies.

In addition to the previously discovered GN-z11 at z=10.6, we have identified twelve new broad line AGN at z$>$4. In these cases the H$\alpha$ line emission shows a broad component (in addition to a narrow component tracing the ISM in the host galaxy) that does not have a counterpart in [OIII]5007, and hence cannot be ascribed to outflowing gas and is most likely tracing the Broad Line Region of an AGN.
Our analysis of the 13 broad line AGN reveals the following findings:

\begin{itemize}

\item In three cases the H$\alpha$ profile requires an additional intermediate width component (FWHM$\sim$400--700~km/s). We interpret this additional component as tracing a secondary accreting black hole with smaller mass, in the same galaxy, which will likely merge with the larger black hole. The finding in the other two galaxies of broad lines that are significantly shifted relative to the narrow component, may indicate that these are also black holes in the process of merging, but in which the more massive black hole is not accreting at the time of  observation; alternatively, these could be black holes recoiled from a recent merger. However, in these cases we cannot exclude that in these cases the complex broad line profile is due to a complex geometry of the BLR. Followup observations with IFU spectroscopy may help to further assess the merging nature of these objects \citep[as it has already successfully happened for other samples, ][]{Ubler2024}.

\item By using local virial relations, we have  inferred black hole masses that are in the range between $\rm 4\times 10^5~M_\odot$ and $\rm 8\times 10^7~M_\odot$. Interestingly, the lowest-mass black hole is in the regime of Direct Collapse Black Holes (DCBH), which is one of the favoured scenarios for the heavy seeds of supermassive black holes. This does not imply that this is a DCBH, as it may have formed from other kind of seeds and gained its mass through various evolutionary paths; however, this result shows that we are now capable of probing this regime potentially populated by DCBH. More detections, and hence more statistics, of black holes in this mass range will help to test different seeding scenarios, especially in merging systems.

\item Although estimating the intrinsic bolometric luminosity is difficult, most black holes in our sample seem to be accreting at sub-Eddington rates, mostly with $\rm L/L_{Edd}<0.5$. However, small black holes, with masses $\rm M_{BH}\sim 10^6~M_{\odot}$, tend to accrete at Eddington or super-Eddington rates. This might be a consequence of selection effects, i.e. small black holes become detectable only when they are accreting at very high rate. However, this finding also provides support to scenarios that envisage phases of super-Eddington accretion in the early phases of black hole formation.

\item We have found that black holes at 4$<$z$<$11 are over-massive relative to their host galaxies, when compared to the local $\rm M_{BH}-M_{star}$ relation. We find cases that are even approaching $\rm M_{BH}\sim M_{star}$.  While selection effects might be partially responsible for this finding, the result may indicate that at early epochs  black holes may form and grow faster than the stellar population in their host galaxies. We also note that high $\rm M_{BH}/M_{star}$ ratios at high redshift  is an expectation of models that envisage super-Eddington accretion at early epochs and/or heavy seeds (aka DCBHs).

\item The high resolution spectra allowed us to estimate the velocity dispersion in the host galaxy. We find that the $\rm M_{BH}-\sigma$ relation of AGN at z$>$4 is generally consistent with the local relation, with only a few exceptions. In the local Universe the
$\rm M_{BH}-\sigma$ relation (in which $\rm M_{BH}$ scales as $\sim \sigma ^4$) is the tightest of all BH scaling relations with the host galaxy properties, and therefore considered the most fundamental (while other relations are possibly an indirect byproduct). Our finding that high-z black holes follow the same  relation,  confirms that the $\rm M_{BH}-\sigma$ is more fundamental than other scaling relations, and that is also universal, i.e. holds at least out to z$\sim$6. 

\item We have also attempted to estimate the dynamical masses, based on measurements of the radius of the host galaxies.
The JADES broad-line AGN are broadly consistent  with the local $\rm M_{BH}-M_{dyn}$ relation (within its 2$\sigma$ scatter), although in this case the uncertainties are larger than for the $\rm M_{BH}-\sigma$ relation.

\item
The large scatter and strong deviation of the $\rm M_{BH}-M_{star}$ relation relative to local galaxies, together with the fact that the $\rm M_{BH}-\sigma$ and 
$\rm M_{BH}-M_{dyn}$ relations at z$>$4 are instead consistent with the local relations, suggest
that black hole formation is little connected to the formation of stars in the host galaxy, while it is tightly connected to the mass assembly history of the central spheroidal component.

\item We have found that the location of the narrow components of these high-z AGN on the [NII]-BPT diagram is completely offset from the local locus of AGN, while overlapping with the local star forming sequence. This confirms that some of the standard nebular optical diagnostics for identifying AGN are ineffective  at high redshift.

\item We have shown that the offset of high-z AGN on the BPT diagram is consistent with the fact that these systems, and their Narrow Line Region, are metal poor (typically $\rm Z\sim 0.2~Z_\odot$). We find two targets which are located even below the local star forming  sequence, and these are likely AGN whose NLR have very low metallicities, below 0.1~$\rm Z_\odot$.

\item We estimate that the fraction of broad line AGN with $\rm L_{bol}>10^{44}$ in galaxies at $4<z<6$ is about 10\%.

\item The luminosity function  of galaxies hosting broad line AGN with $\rm L_{bol}>10^{44}$ at 4$<$z$<$6 is consistent with the luminosity function of galaxies, in the same redshift range, scaled down by a factor of ten.

\item The contribution of galaxies hosting broad line AGN to the re-ionisation  of the Universe is larger than 10\%. Establishing the specific contribution of AGN (without their host galaxies) requires disentangling the AGN and stellar contributions in these galaxies more precisely, as well as larger statistics.

\end{itemize}


\begin{sidewaystable}
    \centering
    \caption{Measured quantities for the broad lines in our JADES sample that require {\it only one} broad component, including the results from our spectral fitting. }
    \begin{tabular}{llcccccc}
    \hline
ID$^a$  & JADES Name & z  & $\rm F(H\alpha)_{broad}$ & $\rm FWHM_{broad}$$^b$ & $\rm F([NII])/F(H\alpha)$ &
 $\rm F([OIII])/F(H\beta)$ & $\rm \Delta BIC_{NB}$$^c$\\
 & JADES- &  & $\rm 10^{-19} erg~s^{-1}~cm^{-2}$ &
 $\rm km~s^{-1}$ & (narrow) & (narrow) & \\
 \hline    
8083 & JADES-GS+53.13284-27.80186 & 4.6482 & $\rm 37.6^{+2.1}_{-2.0}$ & $\rm 1648^{+127}_{-130}$ & $\rm -1.64^{+0.15}_{-0.25}$ & $\rm 0.88^{+0.01}_{-0.01}$ & $\rm 281.7$ \\
1093 & JADES-GN+189.17974+62.22463 & 5.5951 & $\rm 21.3^{+1.9}_{-1.7}$ & $\rm 1662^{+203}_{-165}$ & $\rm <-1.75$ & $\rm 0.63^{+0.05}_{-0.03}$ & $\rm 67.7$ \\
3608$^d$ & JADES-GN+189.11794+62.23552 & 5.26894 & $\rm 9.77^{+2.6}_{-1.8}$ & $\rm 1373^{+361}_{-198}$ & $\rm <-1.46$ & $\rm 0.65^{+0.02}_{-0.02}$ & $\rm 6.2$$^d$ \\
11836 & JADES-GN+189.22059+62.26368 & 4.40935 & $\rm 37.83^{+2.3}_{-2.7}$ & $\rm 1451^{+98}_{-105}$ & $\rm -1.44^{+0.15}_{-0.21}$ & $\rm 0.88^{+0.01}_{-0.01}$ & $\rm 281.7$ \\
20621 & JADES-GN+189.12252+62.29285 & 4.68123 & $\rm 44.58^{+3.0}_{-2.9}$ & $\rm 1638^{+148}_{-150}$ & $\rm <-1.49$ & $\rm 0.69^{+0.04}_{-0.04}$ & $\rm 52.6$ \\
77652 & JADES-GN+189.29323+62.199 & 5.22943 & $\rm 42.93^{+2.5}_{-3.8}$ & $\rm 1070^{+219}_{-180}$ & $\rm <-0.92$ & $\rm 0.81^{+0.05}_{-0.04}$ & $\rm 51.6$ \\
61888 & JADES-GN+189.16802+62.21701 & 5.87461 & $\rm 36.81^{+2.3}_{-1.9}$ & $\rm 1375^{+97}_{-127}$ & $\rm <-1.39$ & $\rm 0.48^{+0.02}_{-0.02}$ & $\rm 107.2$ \\
62309$^d$ & JADES-GN+189.24898+62.21835 & 5.17241 & $\rm 12.15^{+1.7}_{-1.5}$ & $\rm 890^{+107}_{-80}$ & $\rm <-1.48$ & $\rm 0.86^{+0.17}_{-0.04}$ & $\rm 8.0$$^d$ \\
954 & JADES-GN+189.15197+62.25964 & 6.76026 & $\rm 181.2^{+5.0}_{-6.7}$ & $\rm 1931^{+96}_{-96}$ & $\rm <-1.59$ & $\rm 0.89^{+0.03}_{-0.04}$ & $\rm 330.7$ \\
\hline
    \end{tabular}
    \label{tab:measures}
    \\
Notes: $^a$NIRSpec ID. $^b$ Full Width Half Maximum of H$\alpha$ (or H$\beta$), corrected for instrumental broadening (which is however negligible for the broad component). $^c$ Difference between Bayesian Information Criterion (BIC) for the fitting without a broad component and with a broad component. $^d$These cases are marked as ``tentative'' in terms of the BLR detection, as $\Delta BIC<10$, the broad line component is still significant at $>$5$\sigma$.
\end{sidewaystable}

\begin{sidewaystable}
    \caption{Measured quantities for the three cases requiring two broad components of H$\alpha$, candidate to be merging black holes.}
    \begin{tabular}{llclccccc}
    \hline
ID$^a$  & JADES Name & z  & Comp.$^b$ & $\rm F(H\alpha)_{broad}$ & $\rm FWHM_{broad}$$^c$ & $\rm F([NII])/F(H\alpha)$ &
 $\rm F([OIII])/F(H\beta)$ & $\rm \Delta BIC_{1B2B}$$^d$\\
 & JADES- &  &  & $\rm 10^{-19} erg~s^{-1}~cm^{-2}$ &
 $\rm km~s^{-1}$ & (narrow) & (narrow) & \\
 \hline    
\multirow{2}{*}{10013704} &
\multirow{2}{*}{JADES-GS+53.12654-27.81809} &
\multirow{2}{*}{5.9193} & BLR1 & $\rm 7.05^{+1.4}_{-1.1}$ & $\rm 414^{+32}_{-28}$ & \multirow{2}{*}{$\rm -1.24^{+0.26}_{-0.78}$} & \multirow{2}{*}{$\rm 0.75^{+0.01}_{-0.01}$} & \multirow{2}{*}{$\rm 77.7$} \\
 & &  & BLR2 & $\rm 25.43^{+1.9}_{-2.0}$ & $\rm 2416^{+179}_{-156}$ &   &  &  \\
\multirow{2}{*}{73488} &
\multirow{2}{*}{JADES-GN+189.1974+62.17723} & 
\multirow{2}{*}{4.1332} & BLR1 &
$\rm 91.6^{+2.5}_{-2.5}$ & $\rm 464^{+12}_{-15}$ &
\multirow{2}{*}{$\rm <-1.39$} & 
\multirow{2}{*}{$\rm 0.71^{+0.01}_{-0.01}$} &
\multirow{2}{*}{$\rm 547.7$} \\
 &  &  & BLR2 &
 $\rm 197.2^{+3.1}_{-2.9}$ & $\rm 2160^{+45}_{-46}$ & 
 &  & \\
\multirow{2}{*}{53757} & 
\multirow{2}{*}{JADES-GN+189.26978+62.19421} & 
\multirow{2}{*}{4.4480} & BLR1 &$\rm 26.95^{+4.3}_{-5.2}$ & $\rm 699^{+100}_{-96}$ & 
\multirow{2}{*}{$\rm -1.14^{+0.27}_{-0.85}$} & 
\multirow{2}{*}{$\rm 0.79^{+0.03}_{-0.03}$} & 
\multirow{2}{*}{$\rm 39.2$} \\
 &  &  & BLR2 &$\rm 53.6^{+5.1}_{-5.2}$ & $\rm 2834^{+317}_{-282}$  &  & & \\
\hline
    \end{tabular}
    \label{tab:measures_dual}
    \\
Notes: $^a$NIRSpec ID.
$^b$ Name of each of the two BLR components.
$^c$ Full Width Half Maximum of H$\alpha$, corrected for instrumental broadening (which is however negligible for the broad component). $^c$ Difference between Bayesian Information Criterion (BIC) for the fitting with a single broad component and with two broad components.
\end{sidewaystable}

\begin{table*}
    \centering
    \caption{Parameters inferred for the broad line AGN presented in this JADES sub-sample, and including GN-z11 from \cite{maiolino_bh_2023}.}
    \begin{tabular}{llcccccccccc}
    \hline
ID & Comp. & $\rm lg(M_{BH})$ & $\rm lg(L_{bol})$ & $\rm L/L_{Edd}$ & $\rm lg(M_{star})$ & $\rm lg(\sigma)^a$ & $\rm R_*$ & Sersic index & $\rm lg(M_{dyn})$ & $\rm M_{UV}$ & $\rm A_V^b$\\
 & & $\rm [M_\odot]$ & $\rm [erg~s^{-1}]$ & &
 $\rm [M_\odot]$ & $\rm [km~s^{-1}]$ &
 [kpc] & & $\rm [M_\odot]$ & [mag] & [mag] \\
 \hline
\multirow{2}{*}{10013704} &BLR1& $\rm 5.65^{+0.31}_{-0.31}$ & $\rm 43.8$ & $\rm 1.06$ & \multirow{2}{*}{$\rm 8.88^{+0.66}_{-0.66}$} & \multirow{2}{*}{$\rm 1.93^{+0.05}_{-0.06}$} & \multirow{2}{*}{$\rm 0.15$} &
\multirow{2}{*}{$\rm 0.8$} & \multirow{2}{*}{$\rm 9.23^{+0.1}_{-0.13}$} & \multirow{2}{*}{$\rm -18.89$}  & \multirow{2}{*}{$\rm 0.27$}\\
						 &BLR2& $\rm 7.5^{+0.31}_{-0.31}$ & $\rm 44.3$ & $\rm 0.06$ & &  &  &  &   & \\
8083 &   & $\rm 7.25^{+0.31}_{-0.31}$ & $\rm 44.6$ & $\rm 0.16$ & $\rm 8.45^{+0.03}_{-0.03}$ & $\rm 1.9^{+0.06}_{-0.07}$ & $\rm 0.11$ & $\rm 5.7$ & $\rm 8.84^{+0.11}_{-0.15}$ & $\rm -18.67$  & $\rm 0.64$\\
1093 &   & $\rm 7.36^{+0.32}_{-0.31}$ & $\rm 44.8$ & $\rm 0.2$ & $\rm 8.34^{+0.2}_{-0.2}$ & $\rm 1.95^{+0.05}_{-0.06}$ & -- & -- & -- & $\rm -17.48$ & $\rm 0.99$\\
3608 &   & $\rm 6.82^{+0.38}_{-0.33}$ & $\rm 44.0$ & $\rm 0.11$ & $\rm 8.38^{+0.11}_{-0.15}$ & $\rm 1.92^{+0.06}_{-0.07}$ & -- & -- & -- & $\rm -19.5$ & $\rm 0.48$\\
11836 &   & $\rm 7.13^{+0.31}_{-0.31}$ & $\rm 44.5$ & $\rm 0.2$ & $\rm 7.79^{+0.3}_{-0.3}$ & $\rm 1.96^{+0.05}_{-0.06}$ & $\rm 0.48$ & $\rm 0.8$ & $\rm 9.81^{+0.1}_{-0.14}$ & $\rm -18.75$ & $\rm 0.68$\\
20621 &   & $\rm 7.3^{+0.31}_{-0.31}$ & $\rm 44.7$ & $\rm 0.18$ & $\rm 8.06^{+0.7}_{-0.7}$ & $\rm 1.93^{+0.06}_{-0.07}$ & -- & -- & -- & $\rm -18.27$ & $\rm 0.67$\\
\multirow{2}{*}{73488} &BLR1& $\rm 6.18^{+0.3}_{-0.3}$ & $\rm 44.7$ & $\rm 2.48$ & \multirow{2}{*}{$\rm 9.78^{+0.2}_{-0.2}$} & \multirow{2}{*}{$\rm 1.64^{+0.11}_{-0.15}$} & \multirow{2}{*}{$\rm 0.59$} &
\multirow{2}{*}{$\rm 0.8$} & \multirow{2}{*}{$\rm 9.28^{+0.21}_{-0.41}$} & \multirow{2}{*}{$\rm -18.73$} & \multirow{2}{*}{$\rm 0.45$}\\
 &BLR2& $\rm 7.71^{+0.3}_{-0.3}$ & $\rm 45.0$ & $\rm 0.16$ &  &  &  &  &  & \\
77652 &   & $\rm 6.86^{+0.35}_{-0.34}$ & $\rm 44.5$ & $\rm 0.38$ & $\rm 7.87^{+0.16}_{-0.28}$ & $\rm 1.95^{+0.06}_{-0.07}$ & -- & -- & -- & $\rm -18.28$ & $\rm 0.39$\\
61888 &   & $\rm 7.22^{+0.31}_{-0.31}$ & $\rm 44.8$ & $\rm 0.32$ & $\rm 8.11^{+0.92}_{-0.92}$ & $\rm 1.85^{+0.07}_{-0.09}$ & $\rm 0.09$ & $\rm 0.9$ & $\rm 8.92^{+0.22}_{-0.46}$ & $\rm -19.0$ & $\rm 0.69$\\
62309 &   & $\rm 6.56^{+0.32}_{-0.31}$ & $\rm 44.2$ & $\rm 0.39$ & $\rm 8.12^{+0.12}_{-0.13}$ & $\rm 1.87^{+0.07}_{-0.08}$ & $\rm 0.21$ & $\rm 1.3$ & $\rm 9.27^{+0.14}_{-0.2}$ & $\rm -18.67$ & $\rm 0.74$\\
\multirow{2}{*}{53757} &BLR1& $\rm 6.29^{+0.33}_{-0.32}$ & $\rm 44.1$ & $\rm 0.56$ & \multirow{2}{*}{$\rm 10.18^{+0.13}_{-0.12}$} & \multirow{2}{*}{$\rm 1.77^{+0.09}_{-0.11}$} & \multirow{2}{*}{--} &
\multirow{2}{*}{--} & \multirow{2}{*}{--} & \multirow{2}{*}{$\rm -18.9$} & \multirow{2}{*}{$\rm 0.36$}\\
	 &BLR2& $\rm 7.69^{+0.32}_{-0.31}$ & $\rm 44.4$ & $\rm 0.05$ &  &  &  &  &  &  & \\
954$^c$ &   & $\rm 7.9^{+0.3}_{-0.3}$ & $\rm 45.6$ & $\rm 0.42$ & $\rm 10.66^{+0.09}_{-0.1}$ & $\rm 1.91^{+0.06}_{-0.06}$ & $\rm 0.35$ & $\rm 0.8$ & $\rm 9.56^{+0.13}_{-0.18}$ & $\rm -19.78$   & $\rm 0.64$\\
GN-z11$^d$ &   & $\rm 6.2^{+0.3}_{-0.3}$ & $\rm 45.0$ & $\rm 5.5$ & $\rm 8.9^{+0.2}_{-0.3}$ & -- & $\rm 0.20$ & $\rm 0.9$ &  & $-21.79$  & $\rm 0.0$\\
\hline
    \end{tabular}
    \label{tab:inferred_properties}
Notes: $^a$Central velocity dispersion of the host galaxy inferred from the narrow component of either H$\alpha$ or, if not available, [OIII], from the high resolution spectrum; the velocity dispersion is corrected for instrumental resolution and we have also applied a correction factor of 0.175~dex to take into account for the offset between gaseous and stellar velocity dispersion inferred by \cite{Bezanson18b}. $^b$Dust extinction toward the AGN from the low resolution spectral fitting. $^c$This is the most luminous AGN in the sample and the stellar and AGN light decomposition has been more challenging, and therefore the stellar mass might be significantly overestimated. $^d$ GN-z11 at z=10.6, is still part of the JADES survey but the derivation of the parameters is reported in \cite{maiolino_bh_2023} and \cite{tacchella_jades_2023}.
\end{table*}

\begin{table*}
    \centering
    \caption{Density of broad line AGN at 4$<$z$<$6 with $\rm L_{bol}> 10^{44}$ as a function of absolute UV magnitude, as inferred from the JADES survey. The last column gives the minimum density assuming the most conservative and extreme case that the AGN identified in the spectroscopic survey are the only AGN in the volume sampled.}
    \begin{tabular}{ccc}
    \hline
    $\rm M_{UV}$ &
    $\rm \log \Phi$ & $\rm \log \Phi _{min}$\\
    $\rm [mag]$ &
    $\rm [Mpc^{-3} mag^{-1}]$\\
    \hline
    $-18.0\pm0.75$ & $-3.06_{-0.26}^{+0.18}$ & $> -5.73$\\
    $-19.5\pm0.75$ & $-3.83_{-0.44}^{+0.23}$ & $> -5.98$\\
    \hline
\end{tabular}
    \label{tab:lf}
\end{table*}

\begin{acknowledgements}

We thank Marta Volonteri, Raffaella Schneider, Alessandro Trinca, Debora Sijiacki, Martin Haehnelt, Jake Bennett, Sophie Koudmani and the anonymous referee for useful comments.

FDE, JS, RM, TL, WB acknowledge support by the Science and Technology Facilities Council (STFC), by the ERC through Advanced Grant 695671 “QUENCH”, and by the
UKRI Frontier Research grant RISEandFALL. RM also acknowledges funding from a research professorship from the Royal Society.
ECL acknowledges support of an STFC Webb Fellowship (ST/W001438/1)
S.C \& G.V. acknowledge support by European Union’s HE ERC Starting Grant No. 101040227 - WINGS.
H{\"U} gratefully acknowledges support by the Isaac Newton Trust and by the Kavli Foundation through a Newton-Kavli Junior Fellowship.
S.A. \& M.P. acknowledge support from Grant PID2021-127718NB-I00 funded by the Spanish Ministry of Science and Innovation/State Agency of Research (MICIN/AEI/ 10.13039/501100011033). 
AJB \& GCJ acknowledge funding from the "FirstGalaxies" Advanced Grant from the European Research Council (ERC) under the European Union’s Horizon 2020 research and innovation programme (Grant agreement No. 789056)
E.E. \& DJE are supported as a Simons Investigator and by JWST/NIRCam contract to the University of Arizona, NAS5-02015
BER acknowledges support from the NIRCam Science Team contract to the University of Arizona, NAS5-02015. 
M.P. acknowledges support from the research project PID2021-127718NB-I00 of the Spanish Ministry of Science and Innovation/State Agency of Research (MICIN/AEI/ 10.13039/501100011033), and the Programa Atraccion de Talento de la Comunidad de Madrid via grant 2018-T2/TIC-11715
The research of CCW is supported by NOIRLab, which is managed by the Association of Universities for Research in Astronomy (AURA) under a cooperative agreement with the National Science Foundation.

The authors acknowledge use of the lux supercomputer at UC Santa Cruz, funded by NSF MRI grant AST 1828315.

\end{acknowledgements}

\bibliographystyle{aa}
\bibliography{Roberto,Roberto2,Roberto4}

\appendix

\section{[OIII]+H$\beta$ spectra}\label{app:oiii}

\begin{figure*}[h]%
\centering
\includegraphics[width=0.65\columnwidth]{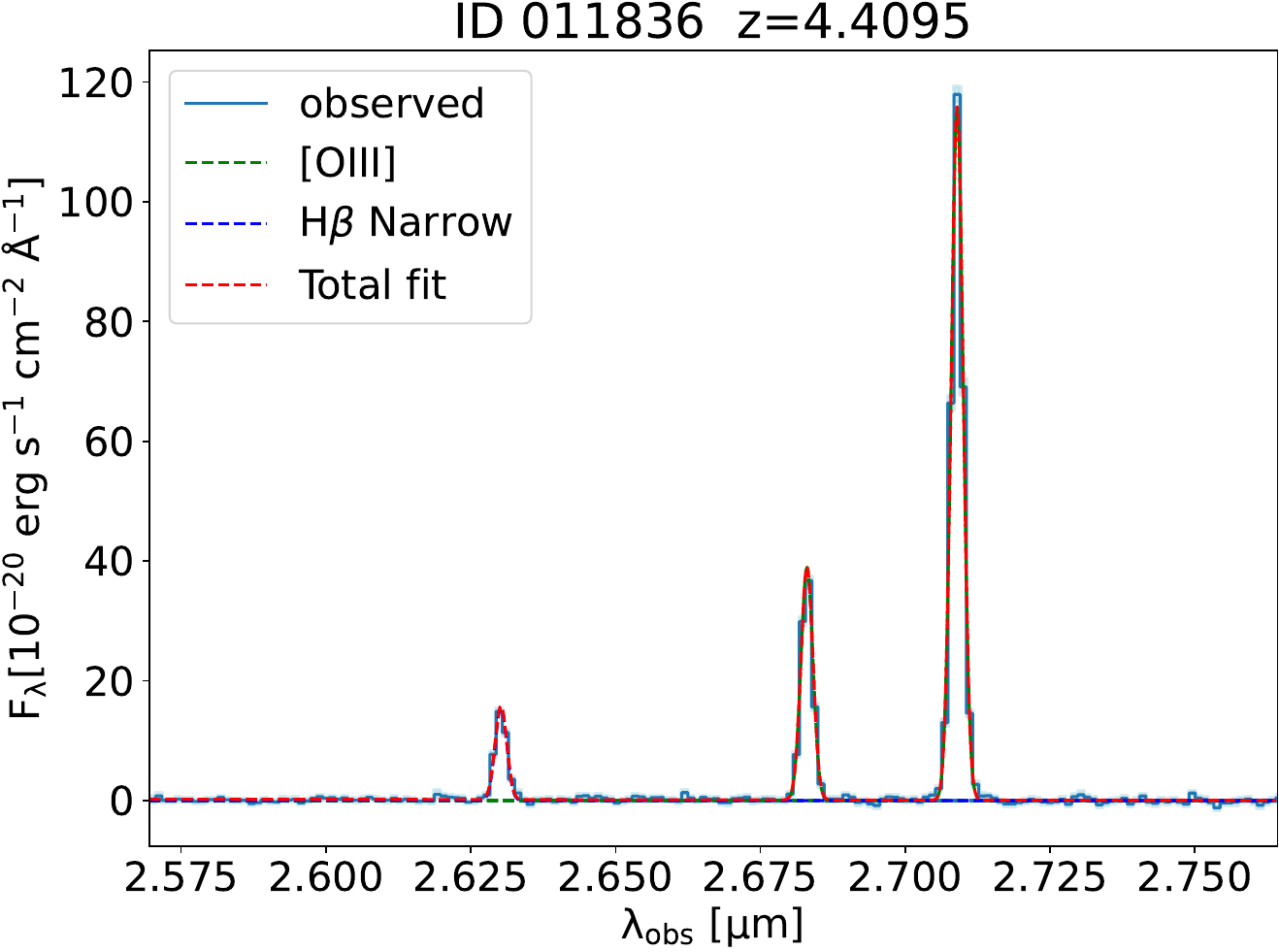}
\includegraphics[width=0.65\columnwidth]{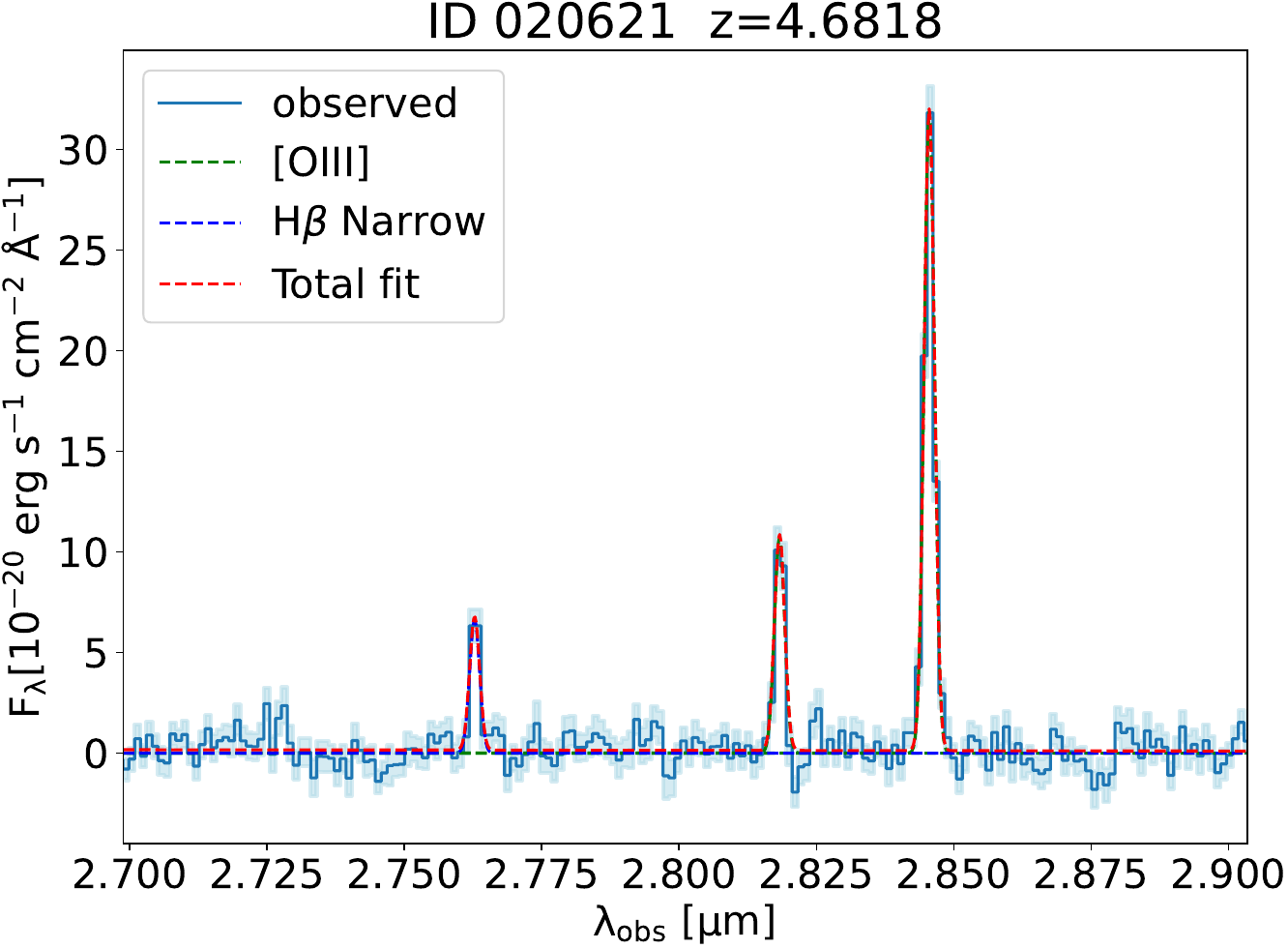}
\includegraphics[width=0.65\columnwidth]{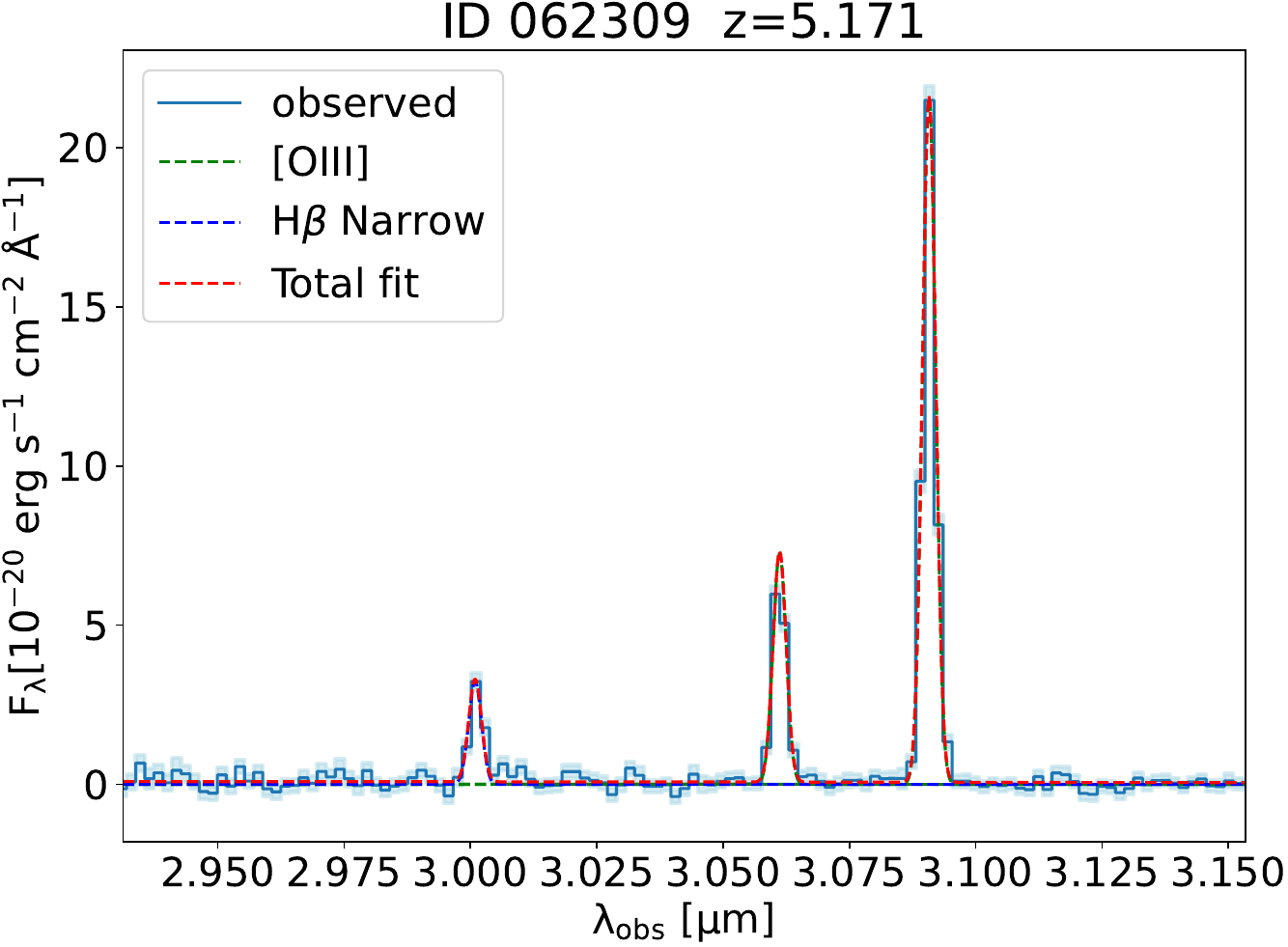}
\includegraphics[width=0.65\columnwidth]{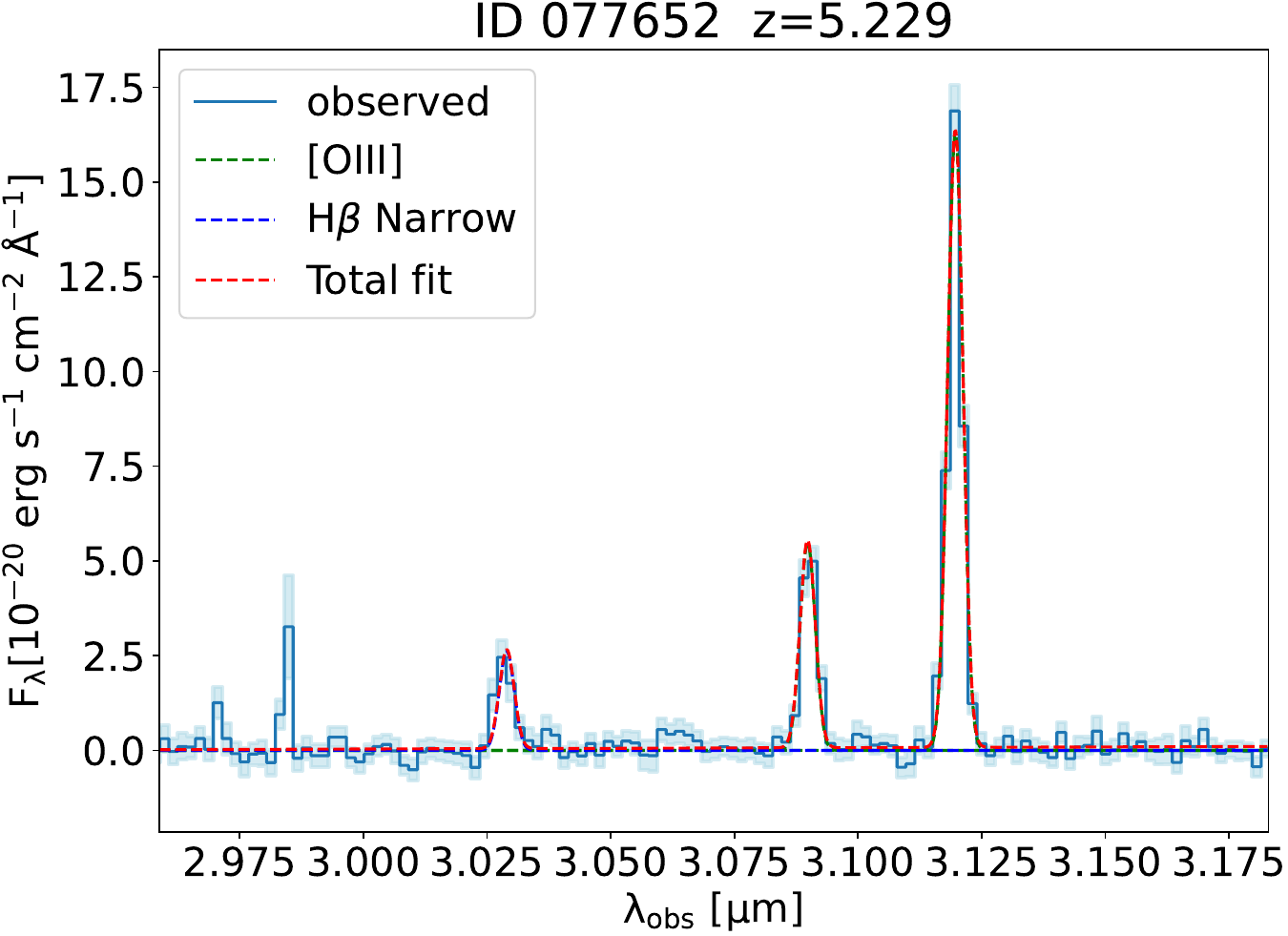}
\includegraphics[width=0.65
\columnwidth]{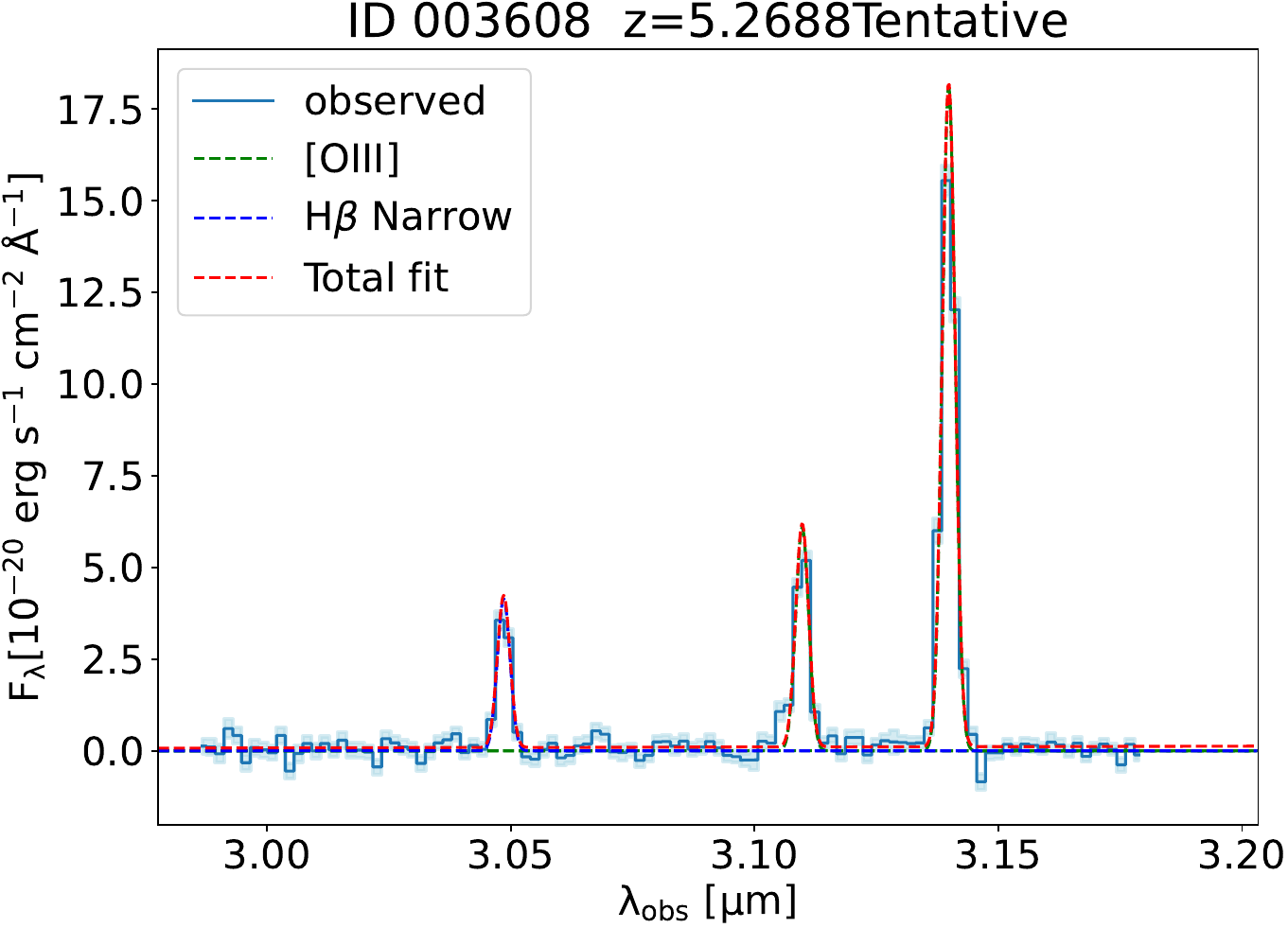}
\includegraphics[width=0.65
\columnwidth]{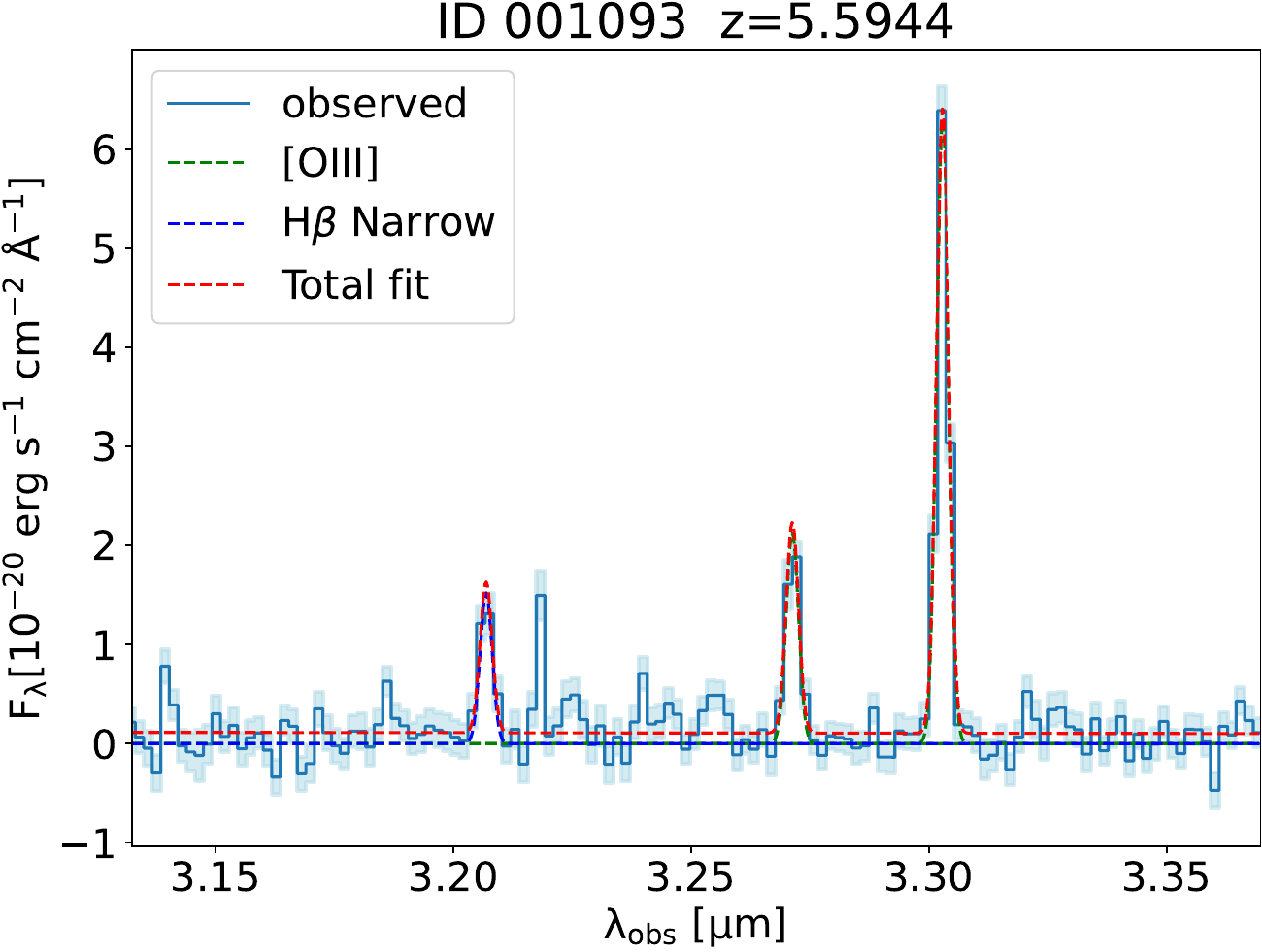}
\includegraphics[width=0.65\columnwidth]{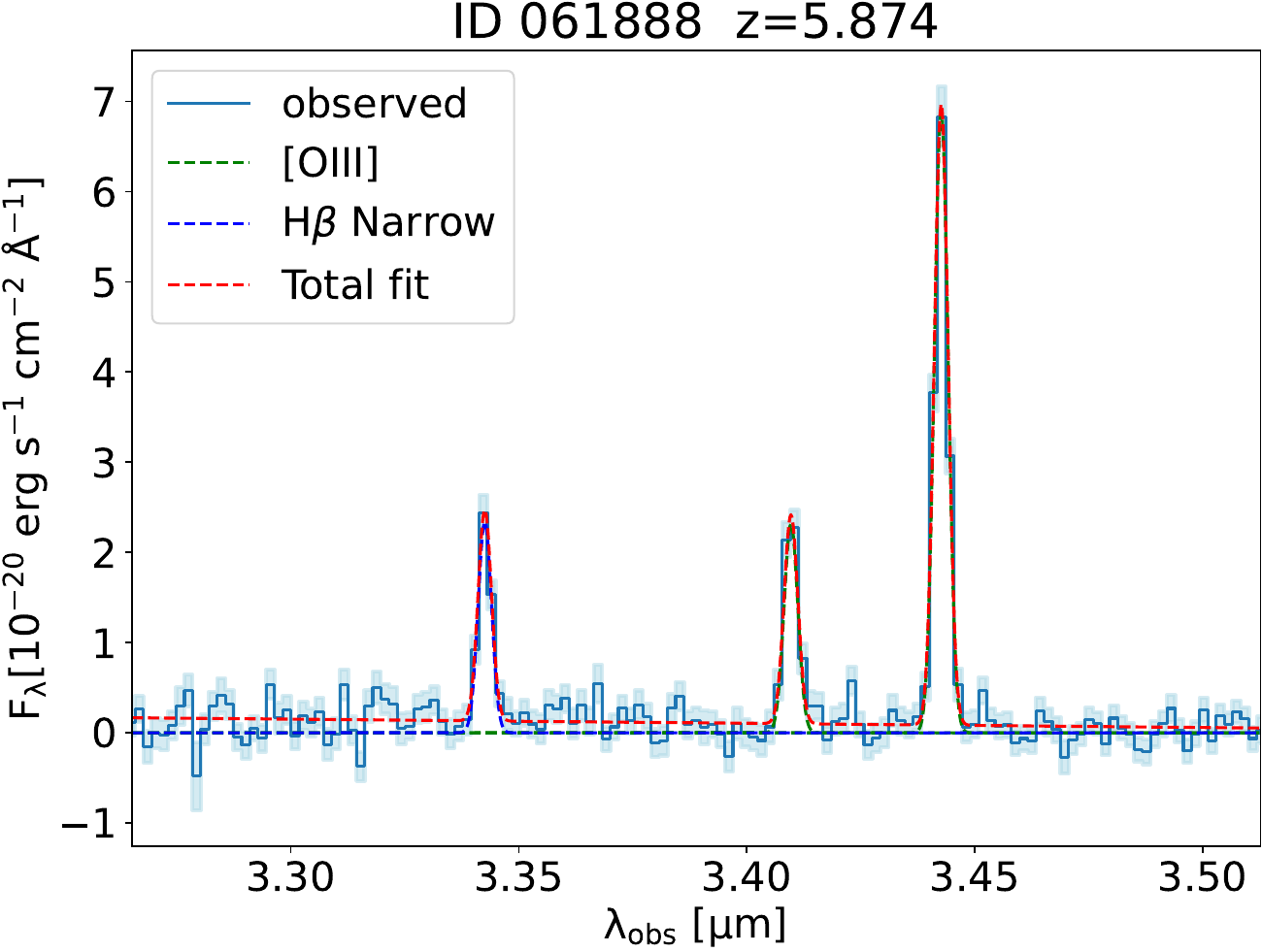}
\includegraphics[width=0.65\columnwidth]{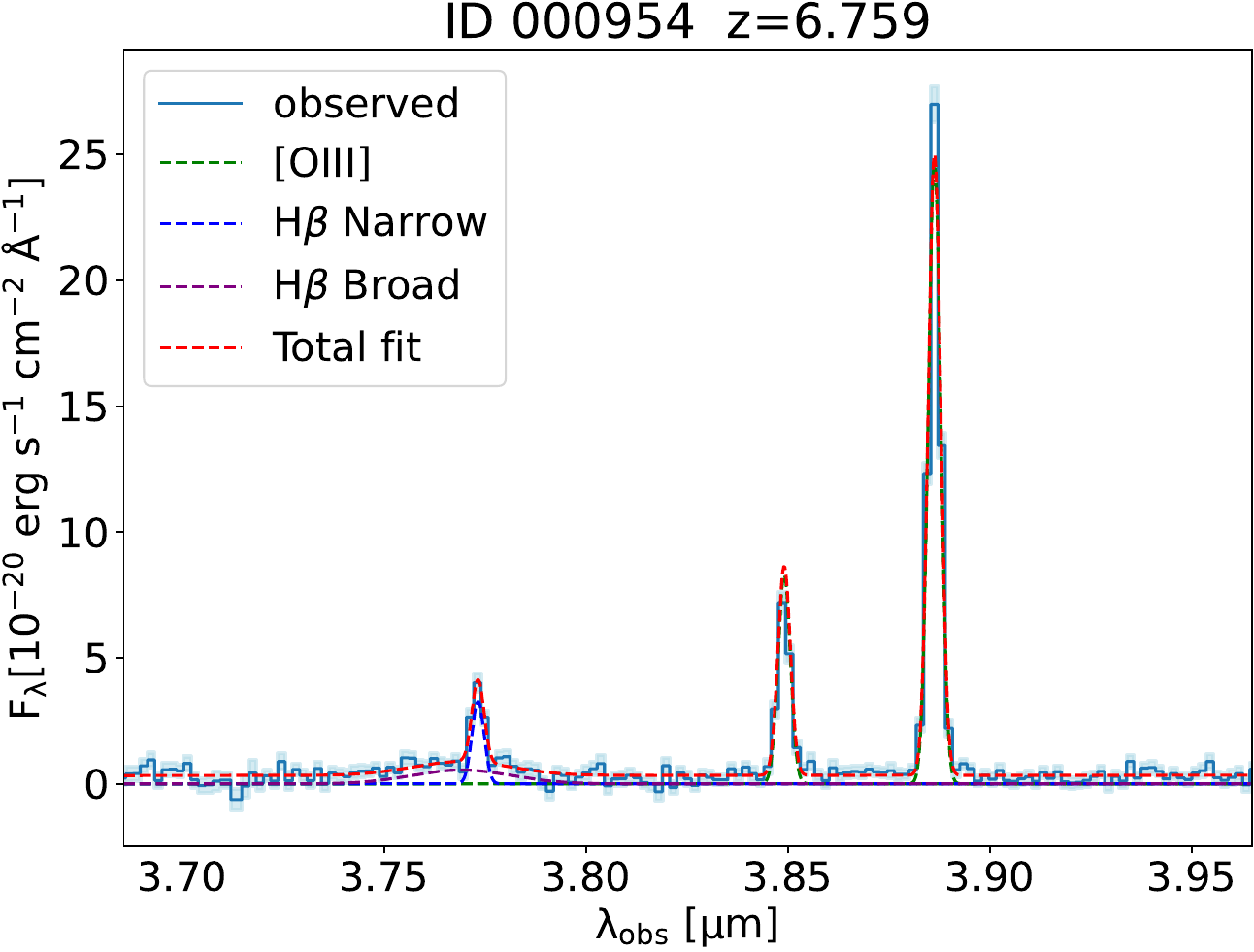}
\includegraphics[width=0.65\columnwidth]{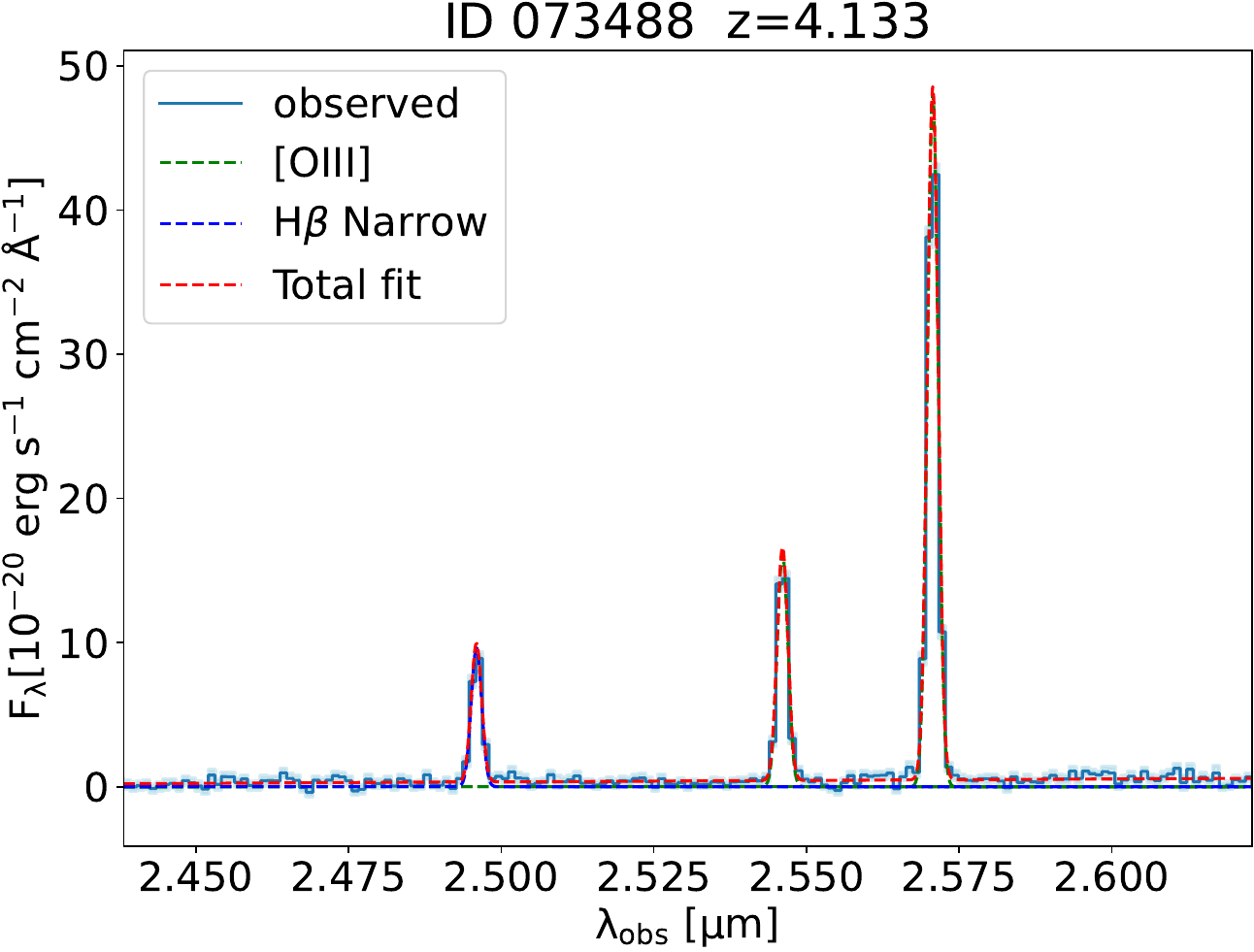}
\includegraphics[width=0.65\columnwidth]{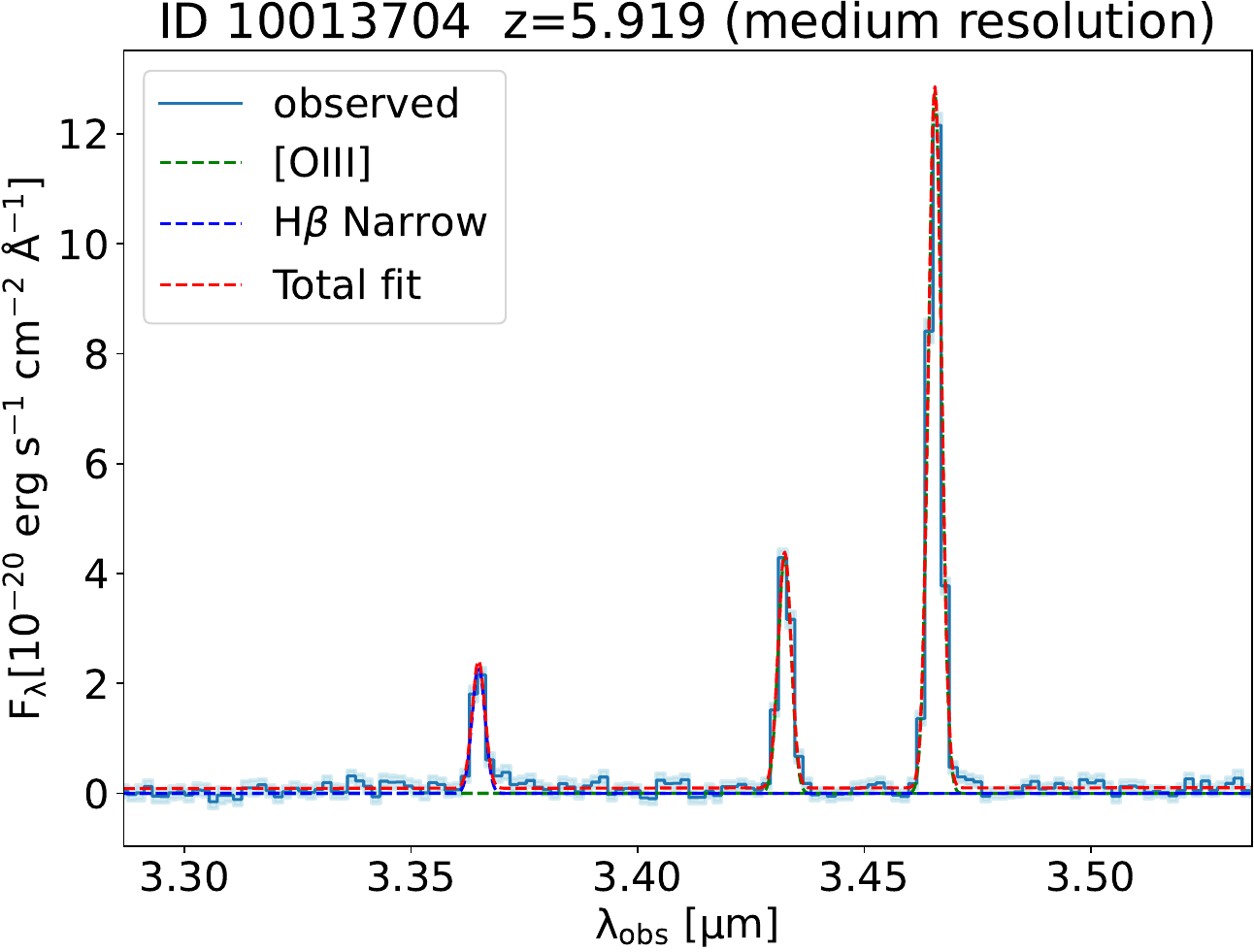}
\includegraphics[width=0.65\columnwidth]{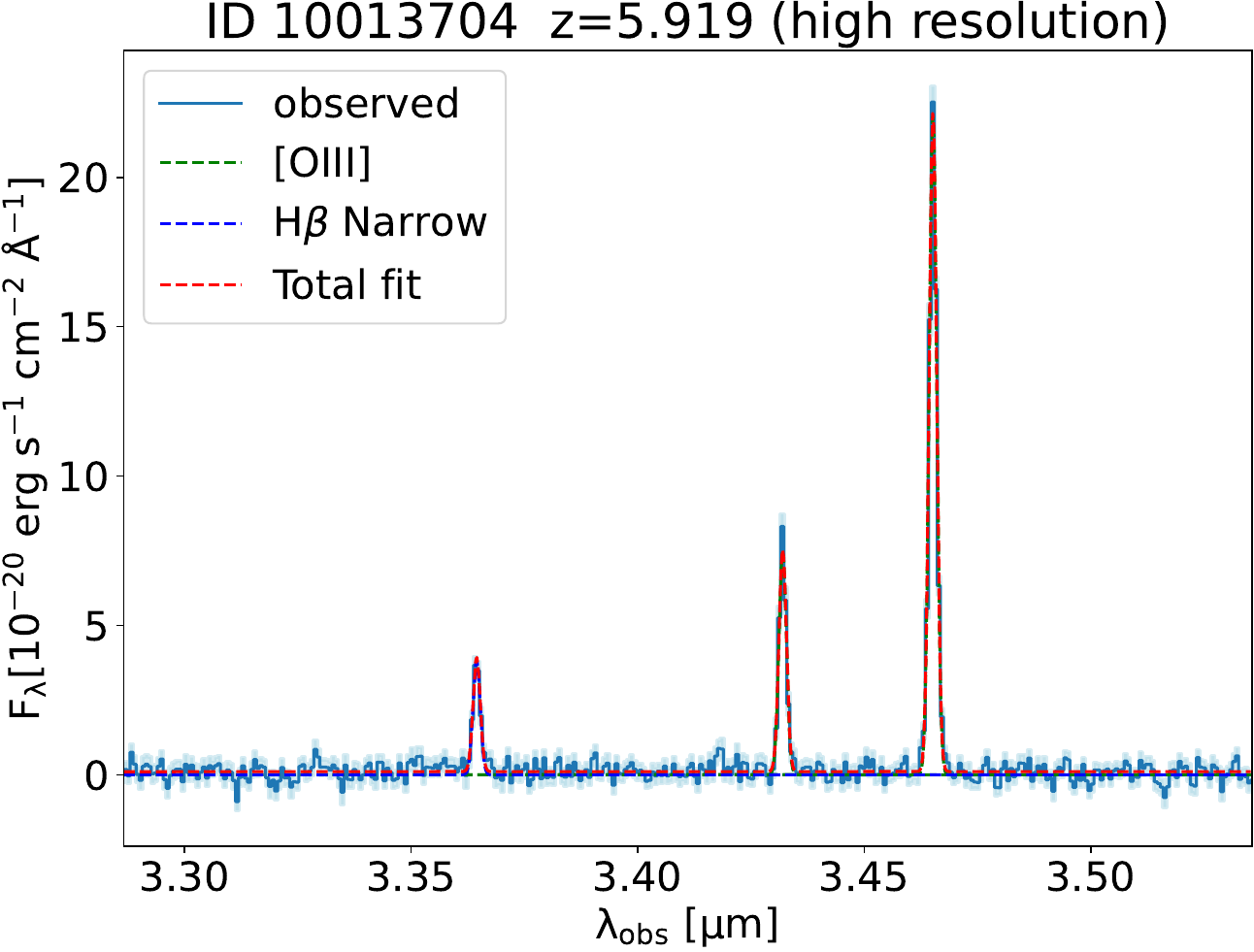}
\caption{Spectra around [OIII] and H$\beta$ for ten out of the twelve new JADES galaxies with evidence for a broad component of H$\alpha$. The blue solid line shows the spectrum (not continuum subtracted) along with the errors (light blue shaded area). The red dashed line shows the total multi-component fit; the blue and purple dashed lines show the narrow and broad components of H$\beta$, respectively (the broad component of H$\beta$ is detected only in ID 954), while the green dashed lines show the  components the [OIII]
doublets. The main purpose of these spectra is to illustrate that there no [OIII] counterpart of the H$\alpha$ broad component. For simplicity, for all galaxies in the sample we only show the medium resolution grating spectrum, with the exception of the candidate dual AGN ID 10013704, for which we also show the high resolution spectrum.
}\label{fig:oiii}
\end{figure*}

Fig.~\ref{fig:oiii} shows the zoom of the spectra around H$\beta$ and [OIII], primarily to illustrate that the broad component is not seen in the [OIII] line, despite being brighter than H$\alpha$, and the S/N being very high. In two cases, namely ID 8083 and ID 53757, [OIII] is not covered by grating spectra (because in the detectors gap); however, the H$\alpha$ line profile is so broad and symmetric that it is very unlikely that is due to an outflow. Moreover, in these cases the flux of the broad H$\alpha$ is higher than the narrow component, if the former was due to an outflow then it would imply that the mass of ionized gas in outflow is larger than in the ISM of the whole galaxy, which is very  unlikely.

We generally show only the medium resolution grating, with the exception of the candidate double AGN ID 10013704, for which we show both the low and high resolution spectra, to illustrate that neither of the two BLR components is present.

The fainter H$\beta$ generally does not show a broad component. The broad component seen in H$\alpha$ is not expected to be seen in the H$\beta$, especially with the modest absorption inferred towards the central region, as discussed in the text. The only exception is ID 954 (the most luminous AGN in the sample) for which the BLR is seen also for H$\beta$. 

\section{[OIII] profile of the dual BLR in 10013704}\label{app:clara_gemma_oiii}

\begin{figure*}[h]%
\centering
\includegraphics[width=1.9\columnwidth]{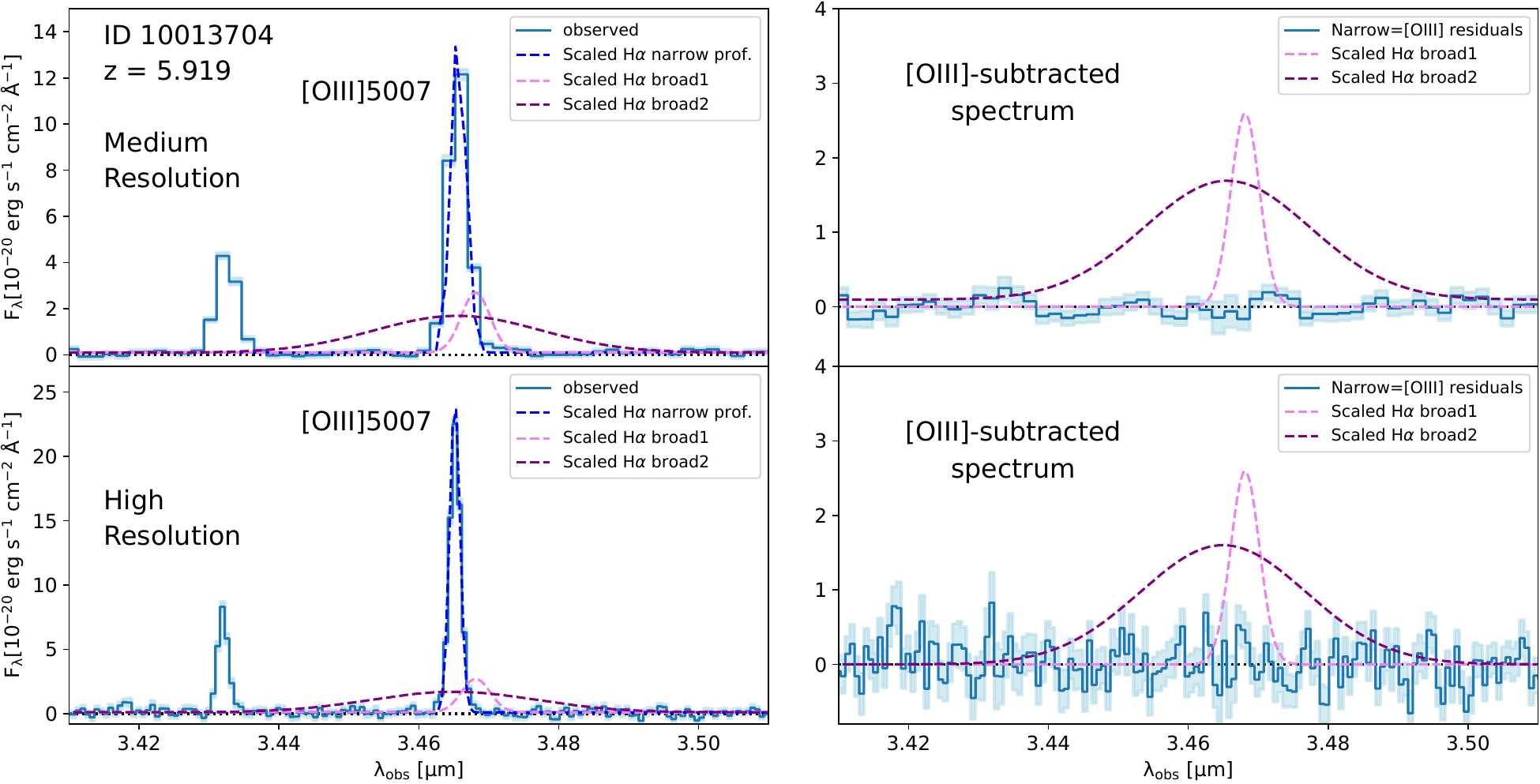}
\caption{Left panels: medium (top) and high (bottom) resolution spectra of ID 10013704 (candidate to host a dual BLR), zoomed around the [OIII]5007 line. The dashed lines indicate the Gaussian components used to fit the H$\alpha$ profile, rescaled so that the narrow component matches the flux of [OIII]. The right panels show the same spectra, where the [OIII] (narrow) profile has been subtracted. Clearly there is no [OIII] counterpart of the H$\alpha$ broad components.}
\label{fig:clara_gemma_oiii}
\end{figure*}

In this appendix we perform a closer analysis of the [OIII] profile of ID 10013704. The left panels of Fig.~\ref{fig:clara_gemma_oiii} show a  version of the medium (top) and high (bottom) resolution spectra zoomed around the [OIII]5007 line. We overplot the three Gaussian components used to reproduce the H$\alpha$ profile, rescaled so that the H$\alpha$ narrow component flux matches the [OIII] flux. Clearly, none of the two broad components has a counterpart in the [OIII] profile.

The medium resolution [OIII] profile has the hint of a redshifted component, but mostly consistent with the noise, not seen in the high-resolution spectrum, and anyway not matching the profile of either of the two H$\alpha$ broad components. This is more clearly seen in the right panels, where we show the same spectra after subtracting the narrow [OIII] component. The marginal redshifted residual is consistent with other features in the noise and not matching either of the two broad H$\alpha$ components. If confirmed with higher resolution data, this faint component may be associated with the remnant of the merging galaxy that hosted the smaller black hole.

\section{Attempt to fit a double powerlaw to the H$\alpha$ profile of ID  73488}\label{app:lara_emma_doublepowlaw}

\begin{figure*}[h]%
\centering
\includegraphics[width=1.9\columnwidth]{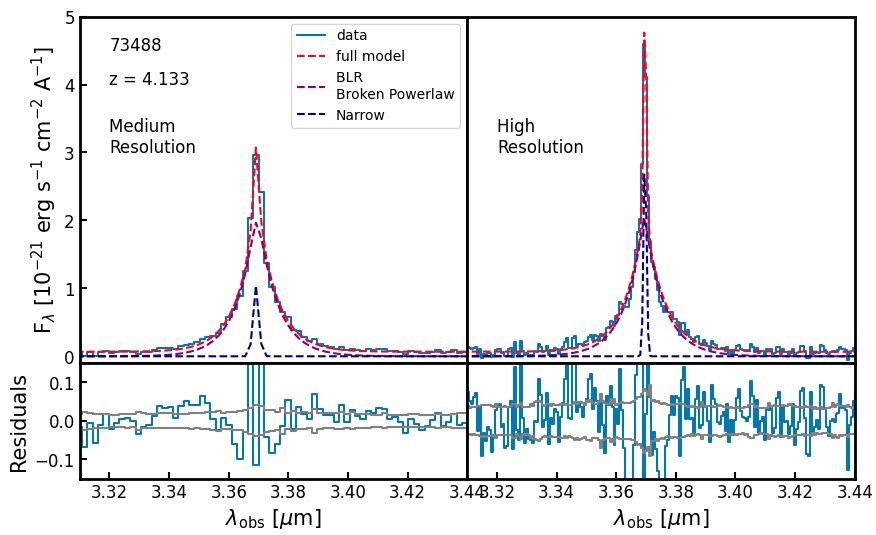}
\caption{Medium (left) and high (right) resolution spectra of ID 73488, zoomed around H$\alpha$ with overlaid the fit using a narrow (Gaussian) component (blue dashed line) and a double power-law profile (purple dashed line) to describe the broad component. The total fit is shown with the red dashed line. The double power-law profile fails to properly fit the broad component, as highlighted by the strong residuals, which are shown in the bottom panels.}
\label{fig:lara_emma_doublepowlaw}
\end{figure*}

In this Appendix we show that a double power-law profile \citep[sometimes used to fit the broad components of high-z quasars][]{Nagao06}, cannot reproduce the broad H$\alpha$ profile of ID 73488. The fit (in addition to the standard narrow component) is shown with dashed lines in Fig.~\ref{fig:lara_emma_doublepowlaw}, both for the low (left) and high (right) resolution spectra. The double powerlaw fails to properly fit the broad component of H$\alpha$, as highlighted by the strong systemic residuals, as shown in the bottom panels.

The $\Delta BIC$ of this fit, relative to the single Gaussian fit, is 273, much higher than the $\Delta BIC_{1B2B} =494$ obtained from the BIC difference between the fit with one Gaussian and two Gaussians. This indicates that the two-Gaussians models describes the data much better than the double-powerlaw model at very high level of confidence.

\section{Beagle spectral fits}\label{app:beagle_prism}

Fig.~\ref{app:beagle_prism} shows the low resolution prism spectra of the 12 AGN in our sample \citep[GN-z11 is discussed separately in ][]{maiolino_bh_2023}, with the Beagle best fit, adopting the composite model including both a stellar population and a reddened AGN power-law. The nebular emission lines are masked, so the fit is applied only to the continuum sections of the observed spectrum, shown in red. The Beagle best fit is shown with the blue line.

\begin{figure*}[h]%
\centering
 \begin{tabular}{ccc}
 ID 1093 & ID 3608 & ID 8083 \\
\includegraphics[width=0.65\columnwidth]{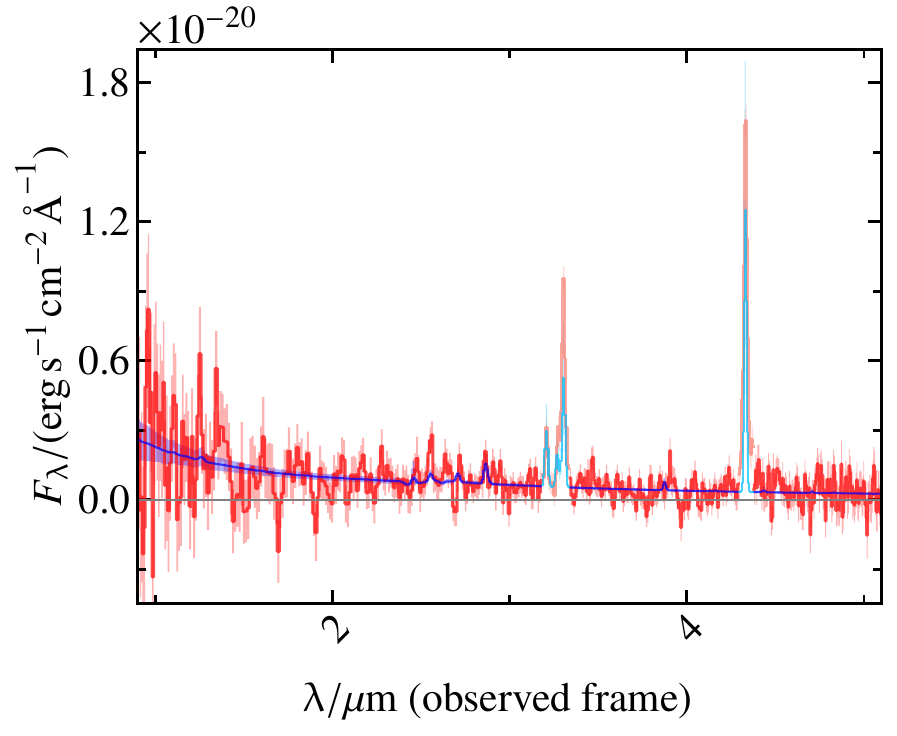} &
\includegraphics[width=0.65\columnwidth] {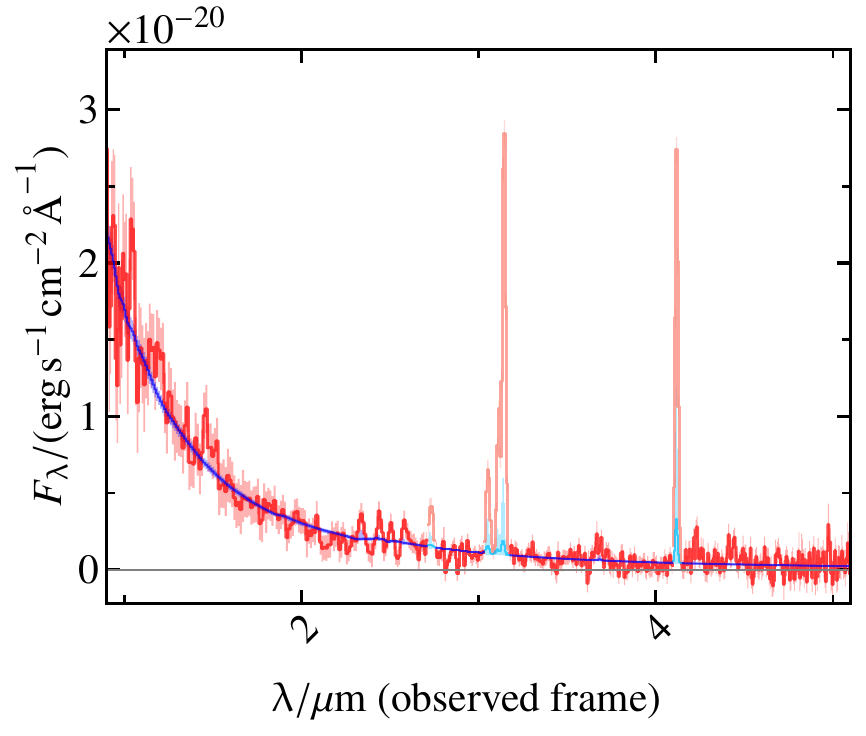} &
\includegraphics[width=0.65\columnwidth]{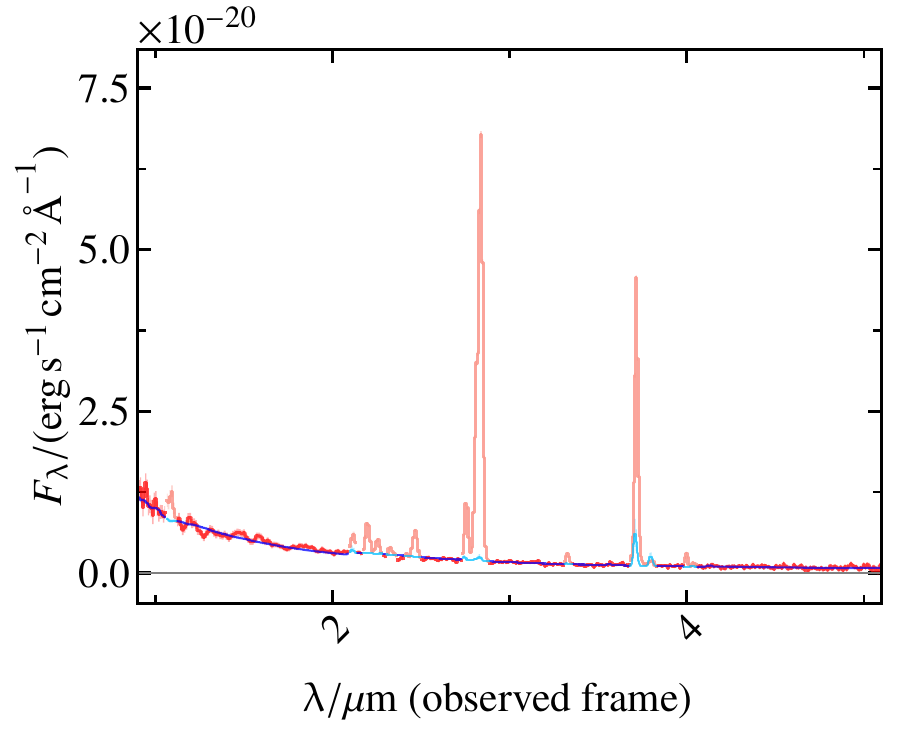} \\
ID 11836 & ID 53757 & ID 61888\\
 \includegraphics[width=0.65\columnwidth]{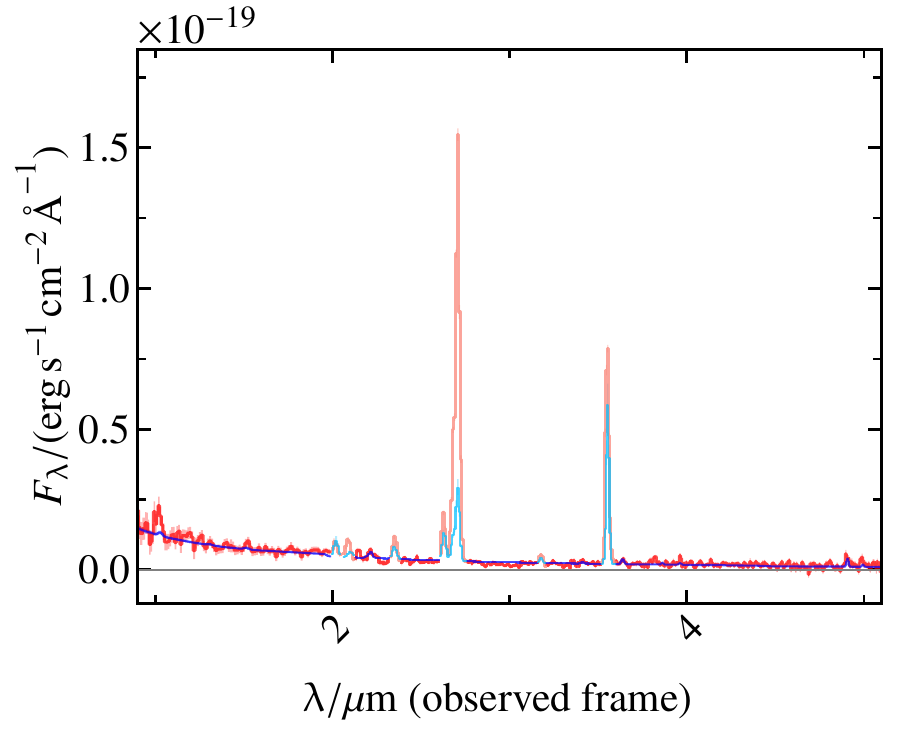} &
 \includegraphics[width=0.65\columnwidth]{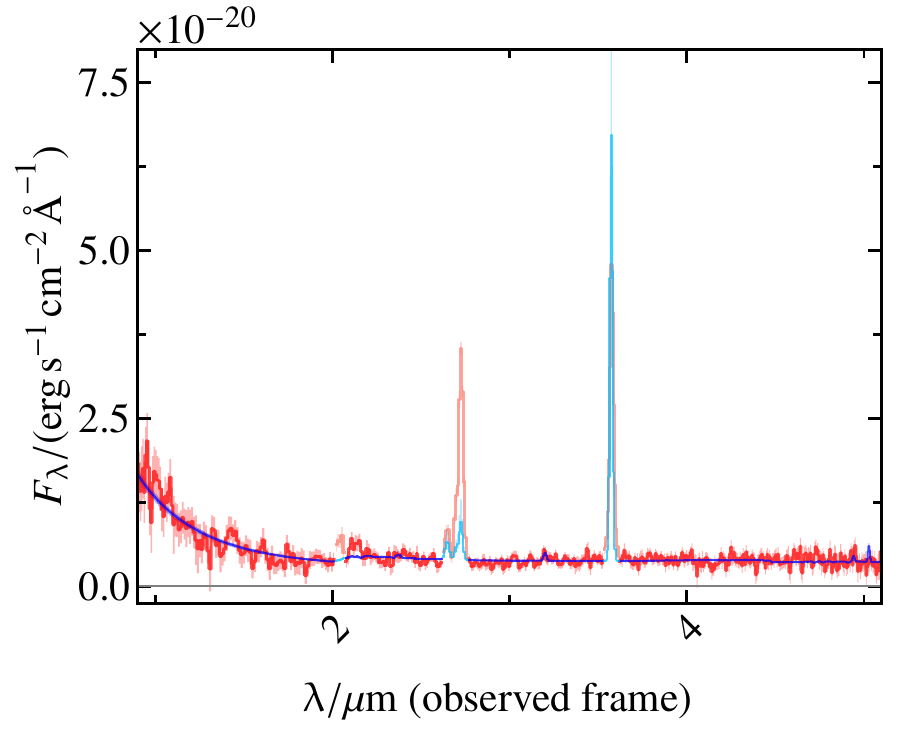} &
 \includegraphics[width=0.65\columnwidth]{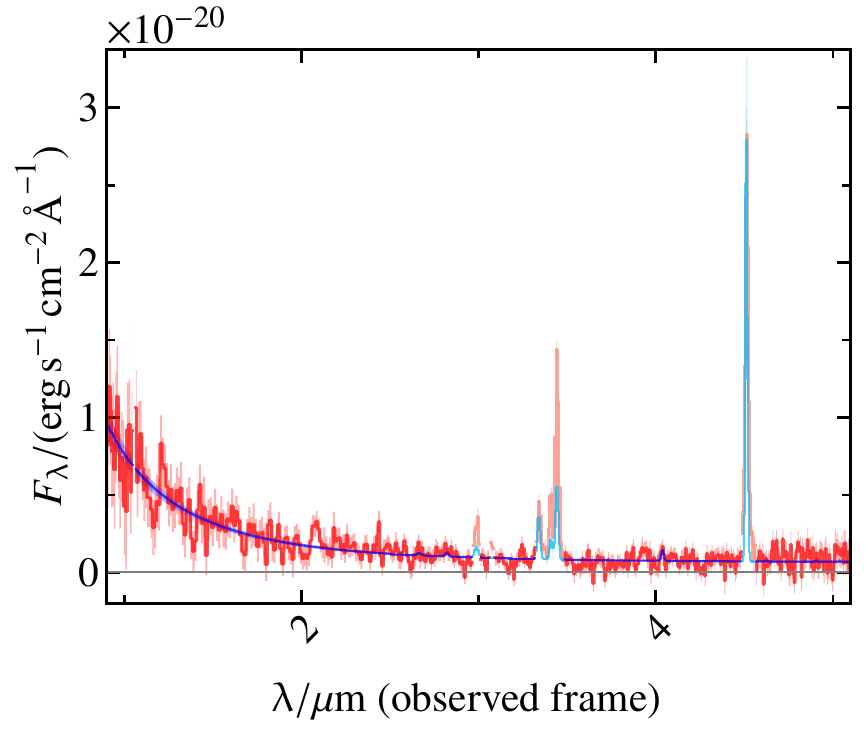} \\
ID 62309 & ID 73488 & ID 77652 \\
 \includegraphics[width=0.65\columnwidth]{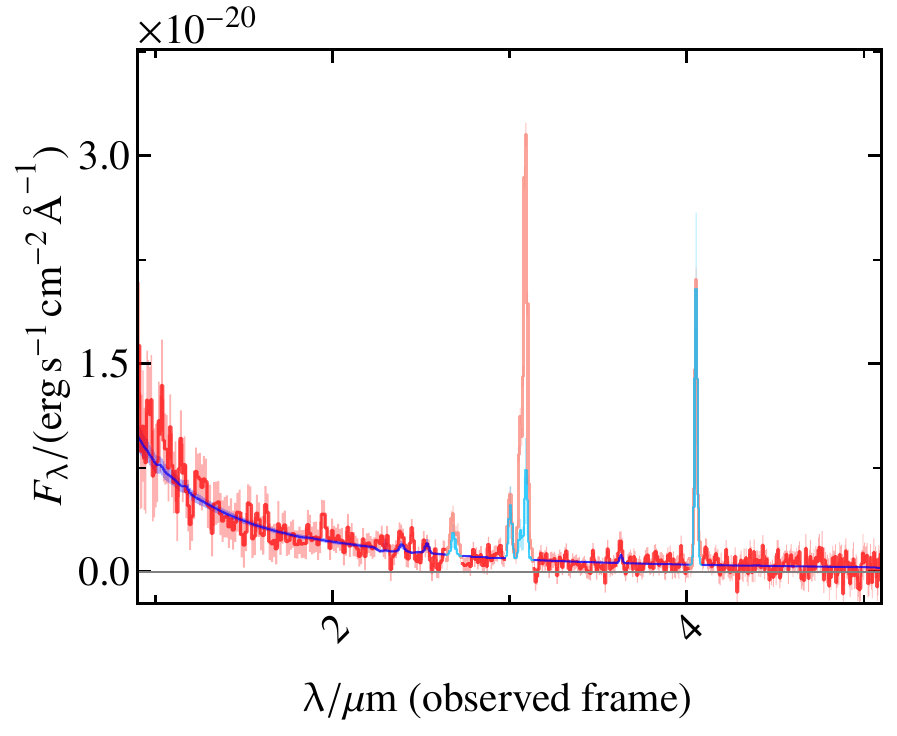} &
 \includegraphics[width=0.65\columnwidth]{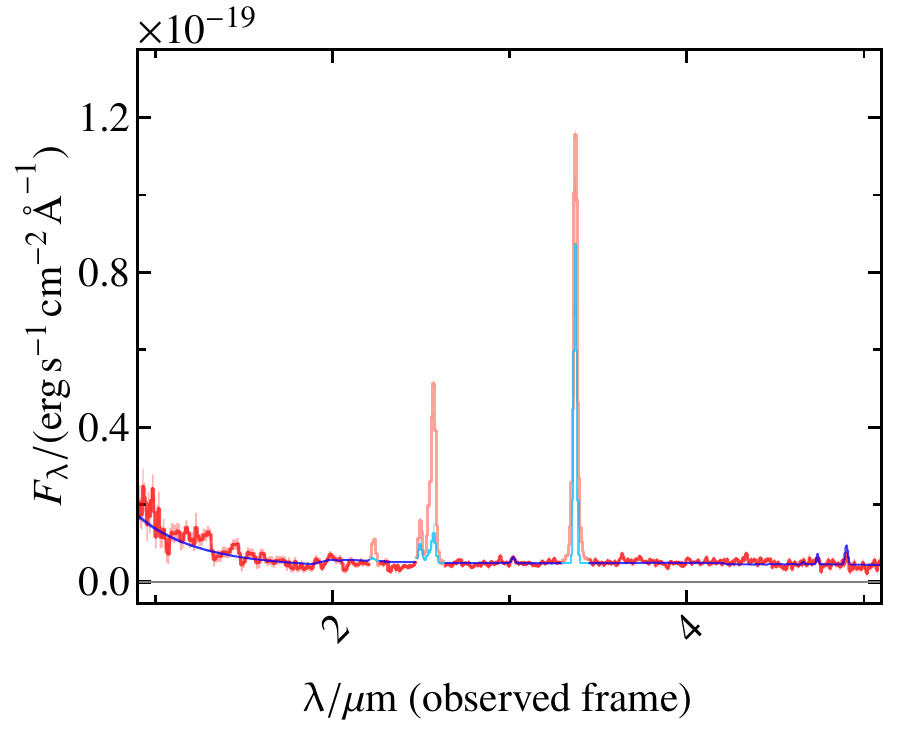} &
 \includegraphics[width=0.65\columnwidth]{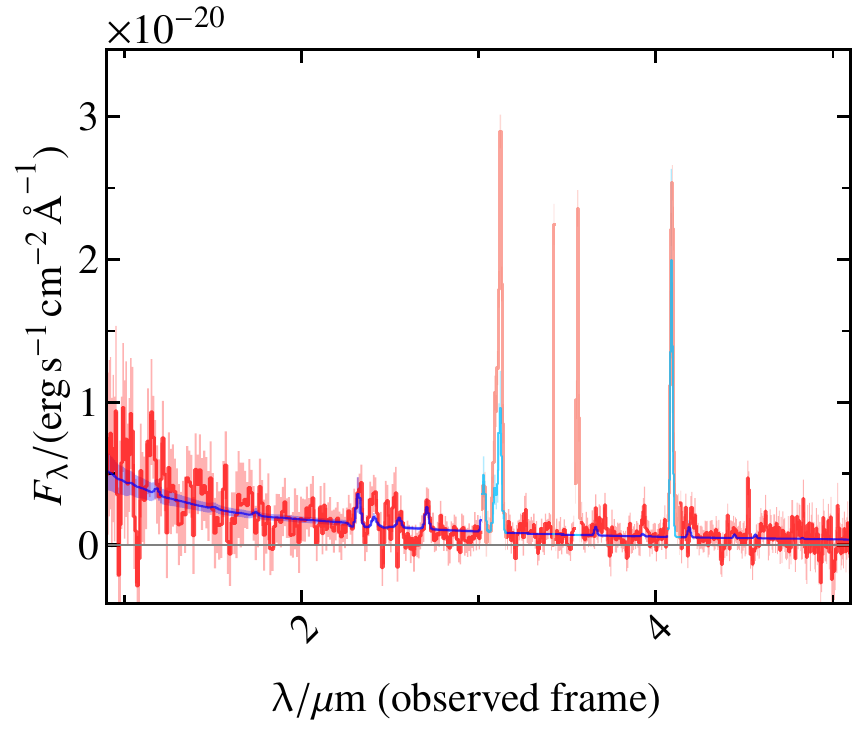} \\
ID 10013704 & & \\
 \includegraphics[width=0.65\columnwidth]{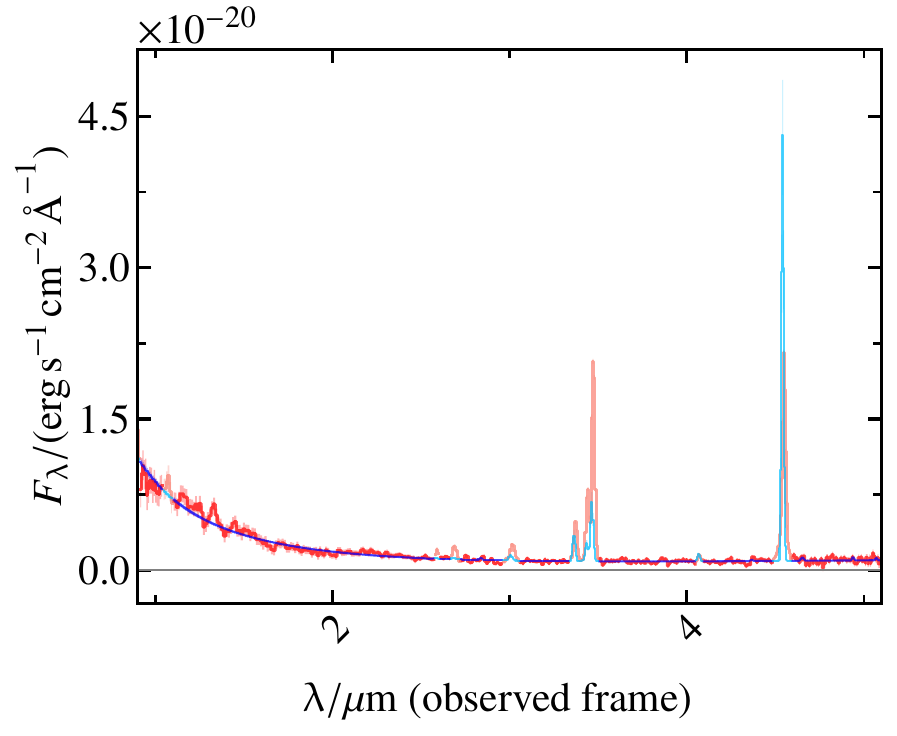} & & \\
\end{tabular}
\caption{Low-resolution prism spectra (dark and light red line) of the targets in our sample, with overlaid the Beagle best fit to the continuum (blue line).}
\label{fig:beagle_prism}
\end{figure*}

\end{document}

%% file: authors.tex
\author{
Roberto Maiolino
\inst{1,2,3}\fnmsep\thanks{rm665@cam.ac.uk}
\and
Jan Scholtz
\inst{1,2}
\and
Emma Curtis-Lake
\inst{4}
\and
Stefano Carniani
\inst{5}
\and
William Baker
\inst{1,2}
\and
Anna de Graaff
\inst{6}
\and
Sandro Tacchella
\inst{1,2}
\and
Hannah \"Ubler
\inst{1,2}
\and
Francesco D'Eugenio
\inst{1,2}
\and
Joris Witstok
\inst{1,2}
\and
Mirko Curti
\inst{7}
\and
Santiago Arribas
\inst{8}
Andrew J. Bunker 
\inst{9} 
\and
Stéphane Charlot
\inst{10} 
\and
Jacopo Chevallard
\inst{9} 
\and
Daniel J.\ Eisenstein
\inst{11}
\and
Eiichi Egami
\inst{12}
\and
Zhiyuan Ji
\inst{12}
\and
Gareth C. Jones
\inst{9}
\and
Jianwei Lyu
\inst{12}
\and
Tim Rawle
\inst{13}
\and
Brant Robertson
\inst{14}
\and
Wiphu Rujopakarn
\inst{15}
\and
Michele Perna
\inst{8}
\and
Fengwu Sun
\inst{12}
\and
Giacomo Venturi
\inst{5}
\and
Christina C. Williams
\inst{16}
\and
Chris Willott
\inst{17}
}

\institute{
Kavli Institute for Cosmology, University of Cambridge, Madingley Road, Cambridge, 
CB3 0HA, UK\\
\and
Cavendish Laboratory, University of Cambridge, 19 JJ Thomson Avenue, Cambridge CB3 0HE, UK\\
\and
Department of Physics and Astronomy, University College London, Gower Street, London WC1E 6BT, UK\\
\and
Centre for Astrophysics Research, Department of Physics, Astronomy and Mathematics, University of Hertfordshire, Hatfield AL10 9AB, UK\\
\and
Scuola Normale Superiore, Piazza dei Cavalieri 7, I-56126 Pisa, Italy\\
\and
Max-Planck-Institut f\"ur Astronomie, K\"onigstuhl 17, D-69117, Heidelberg, Germany
\and
European Southern Observatory, Karl-Schwarzschild-Strasse 2, 85748 Garching, Germany\\
\and
Centro de Astrobiolog\'ia (CAB), CSIC–INTA, Cra. de Ajalvir Km.~4, 28850- Torrej\'on de Ardoz, Madrid, Spain\\
\and
Department of Physics, University of Oxford, Denys Wilkinson Building, Keble Road, Oxford OX1 3RH, UK\\
\and
Sorbonne Universit\'e, CNRS, UMR 7095, Institut d'Astrophysique de Paris, 98 bis bd Arago, 75014 Paris, France\\
\and
Center for Astrophysics - Harvard \& Smithsonian, 60 Garden St., Cambridge MA 02138 USA\\
\and
Steward Observatory, University of Arizona, 933 N. Cherry Avenue, Tucson, AZ 85721, USA
\and
European Space Agency (ESA), European Space Astronomy Centre (ESAC), Camino Bajo del Castillo s/n, 28692 Villafranca del Castillo, Madrid, Spain
\and
Department of Astronomy and Astrophysics, University of California, Santa Cruz, 1156 High Street, Santa Cruz, CA 95064, USA\\
\and
1. National Astronomical Research Institute of Thailand, Don Kaeo, Mae Rim, Chiang Mai 50180, Thailand; 2. Department of Physics, Faculty of Science, Chulalongkorn University, 254 Phayathai Road, Pathumwan, Bangkok 10330, Thailand
\and
NSF’s National Optical-Infrared Astronomy Research Laboratory, 950 North Cherry Avenue, Tucson, AZ 85719, USA
\and
NRC Herzberg, 5071 West Saanich Rd, Victoria, BC V9E 2E7, Canada
}

   \authorrunning{Maiolino et al.}
   \date{}

%% file: Infant_BH_revised_clean2.bbl
\begin{thebibliography}{182}
\expandafter\ifx\csname natexlab\endcsname\relax\def\natexlab#1{#1}\fi

\bibitem[{{Agazie} {et~al.}(2023){Agazie}, {Anumarlapudi}, {Archibald},
  {Arzoumanian}, {Baker}, {B{\'e}csy}, {Blecha}, {Brazier}, {Brook},
  {Burke-Spolaor}, {Burnette}, {Case}, {Charisi}, {Chatterjee},
  {Chatziioannou}, {Cheeseboro}, {Chen}, {Cohen}, {Cordes}, {Cornish},
  {Crawford}, {Cromartie}, {Crowter}, {Cutler}, {Decesar}, {Degan}, {Demorest},
  {Deng}, {Dolch}, {Drachler}, {Ellis}, {Ferrara}, {Fiore}, {Fonseca},
  {Freedman}, {Garver-Daniels}, {Gentile}, {Gersbach}, {Glaser}, {Good},
  {G{\"u}ltekin}, {Hazboun}, {Hourihane}, {Islo}, {Jennings}, {Johnson},
  {Jones}, {Kaiser}, {Kaplan}, {Kelley}, {Kerr}, {Key}, {Klein}, {Laal}, {Lam},
  {Lamb}, {Lazio}, {Lewandowska}, {Littenberg}, {Liu}, {Lommen}, {Lorimer},
  {Luo}, {Lynch}, {Ma}, {Madison}, {Mattson}, {McEwen}, {McKee}, {McLaughlin},
  {McMann}, {Meyers}, {Meyers}, {Mingarelli}, {Mitridate}, {Natarajan}, {Ng},
  {Nice}, {Ocker}, {Olum}, {Pennucci}, {Perera}, {Petrov}, {Pol}, {Radovan},
  {Ransom}, {Ray}, {Romano}, {Sardesai}, {Schmiedekamp}, {Schmiedekamp},
  {Schmitz}, {Schult}, {Shapiro-Albert}, {Siemens}, {Simon}, {Siwek}, {Stairs},
  {Stinebring}, {Stovall}, {Sun}, {Susobhanan}, {Swiggum}, {Taylor}, {Taylor},
  {Turner}, {Unal}, {Vallisneri}, {van Haasteren}, {Vigeland}, {Wahl}, {Wang},
  {Witt}, {Young}, \& {Nanograv Collaboration}}]{Nanograv2023}
{Agazie}, G., {Anumarlapudi}, A., {Archibald}, A.~M., {et~al.} 2023, \apjl,
  951, L8

\bibitem[{{Amaro-Seoane} {et~al.}(2023){Amaro-Seoane}, {Andrews}, {Arca Sedda},
  {Askar}, {Baghi}, {Balasov}, {Bartos}, {Bavera}, {Bellovary}, {Berry},
  {Berti}, {Bianchi}, {Blecha}, {Blondin}, {Bogdanovi{\'c}}, {Boissier},
  {Bonetti}, {Bonoli}, {Bortolas}, {Breivik}, {Capelo}, {Caramete},
  {Cattorini}, {Charisi}, {Chaty}, {Chen}, {Chru{\'s}li{\'n}ska}, {Chua},
  {Church}, {Colpi}, {D'Orazio}, {Danielski}, {Davies}, {Dayal}, {De Rosa},
  {Derdzinski}, {Destounis}, {Dotti}, {Dutan}, {Dvorkin}, {Fabj}, {Foglizzo},
  {Ford}, {Fouvry}, {Franchini}, {Fragos}, {Fryer}, {Gaspari}, {Gerosa},
  {Graziani}, {Groot}, {Habouzit}, {Haggard}, {Haiman}, {Han}, {Istrate},
  {Johansson}, {Khan}, {Kimpson}, {Kokkotas}, {Kong}, {Korol}, {Kremer},
  {Kupfer}, {Lamberts}, {Larson}, {Lau}, {Liu}, {Lloyd-Ronning}, {Lodato},
  {Lupi}, {Ma}, {Maccarone}, {Mandel}, {Mangiagli}, {Mapelli}, {Mathis},
  {Mayer}, {McGee}, {McKernan}, {Miller}, {Mota}, {Mumpower}, {Nasim},
  {Nelemans}, {Noble}, {Pacucci}, {Panessa}, {Paschalidis}, {Pfister},
  {Porquet}, {Quenby}, {Ricarte}, {R{\"o}pke}, {Regan}, {Rosswog}, {Ruiter},
  {Ruiz}, {Runnoe}, {Schneider}, {Schnittman}, {Secunda}, {Sesana}, {Seto},
  {Shao}, {Shapiro}, {Sopuerta}, {Stone}, {Suvorov}, {Tamanini}, {Tamfal},
  {Tauris}, {Temmink}, {Tomsick}, {Toonen}, {Torres-Orjuela}, {Toscani},
  {Tsokaros}, {Unal}, {V{\'a}zquez-Aceves}, {Valiante}, {van Putten}, {van
  Roestel}, {Vignali}, {Volonteri}, {Wu}, {Younsi}, {Yu}, {Zane}, {Zwick},
  {Antonini}, {Baibhav}, {Barausse}, {Bonilla Rivera}, {Branchesi},
  {Branduardi-Raymont}, {Burdge}, {Chakraborty}, {Cuadra}, {Dage}, {Davis}, {de
  Mink}, {Decarli}, {Doneva}, {Escoffier}, {Gandhi}, {Haardt}, {Lousto},
  {Nissanke}, {Nordhaus}, {O'Shaughnessy}, {Portegies Zwart}, {Pound},
  {Schussler}, {Sergijenko}, {Spallicci}, {Vernieri}, \&
  {Vigna-G{\'o}mez}}]{Amaro2023}
{Amaro-Seoane}, P., {Andrews}, J., {Arca Sedda}, M., {et~al.} 2023, Living
  Reviews in Relativity, 26, 2

\bibitem[{{Amaro-Seoane} {et~al.}(2012){Amaro-Seoane}, {Aoudia}, {Babak},
  {Bin{\'e}truy}, {Berti}, {Boh{\'e}}, {Caprini}, {Colpi}, {Cornish},
  {Danzmann}, {Dufaux}, {Gair}, {Jennrich}, {Jetzer}, {Klein}, {Lang}, {Lobo},
  {Littenberg}, {McWilliams}, {Nelemans}, {Petiteau}, {Porter}, {Schutz},
  {Sesana}, {Stebbins}, {Sumner}, {Vallisneri}, {Vitale}, {Volonteri}, \&
  {Ward}}]{Amaro2012}
{Amaro-Seoane}, P., {Aoudia}, S., {Babak}, S., {et~al.} 2012, Classical and
  Quantum Gravity, 29, 124016

\bibitem[{{Ba{\~n}ados} {et~al.}(2018){Ba{\~n}ados}, {Venemans},
  {Mazzucchelli}, {Farina}, {Walter}, {Wang}, {Decarli}, {Stern}, {Fan},
  {Davies}, {Hennawi}, {Simcoe}, {Turner}, {Rix}, {Yang}, {Kelson}, {Rudie}, \&
  {Winters}}]{Banados18}
{Ba{\~n}ados}, E., {Venemans}, B.~P., {Mazzucchelli}, C., {et~al.} 2018, \nat,
  553, 473

\bibitem[{{Baker} {et~al.}(2023){Baker}, {Tacchella}, {Johnson}, {Nelson},
  {Suess}, {D'Eugenio}, {Curti}, {de Graaff}, {Ji}, {Maiolino}, {Robertson},
  {Scholtz}, {Alberts}, {Arribas}, {Boyett}, {Bunker}, {Carniani}, {Charlot},
  {Chen}, {Chevallard}, {Curtis-Lake}, {Danhaive}, {DeCoursey}, {Egami},
  {Eisenstein}, {Endsley}, {Hausen}, {Helton}, {Kumari}, {Looser}, {Maseda},
  {Pusk{\'a}s}, {Rieke}, {Sandles}, {Sun}, {{\"U}bler}, {Williams}, {Willmer},
  \& {Witstok}}]{Baker23bulge}
{Baker}, W.~M., {Tacchella}, S., {Johnson}, B.~D., {et~al.} 2023, arXiv
  e-prints, arXiv:2306.02472

\bibitem[{{Baldwin} {et~al.}(1981){Baldwin}, {Phillips}, \&
  {Terlevich}}]{Baldwin81}
{Baldwin}, J.~A., {Phillips}, M.~M., \& {Terlevich}, R. 1981, \pasp, 93, 5

\bibitem[{{Banik} {et~al.}(2019){Banik}, {Tan}, \& {Monaco}}]{Banik19}
{Banik}, N., {Tan}, J.~C., \& {Monaco}, P. 2019, \mnras, 483, 3592

\bibitem[{{Barai} {et~al.}(2018){Barai}, {Gallerani}, {Pallottini}, {Ferrara},
  {Marconi}, {Cicone}, {Maiolino}, \& {Carniani}}]{Barai2018}
{Barai}, P., {Gallerani}, S., {Pallottini}, A., {et~al.} 2018, \mnras, 473,
  4003

\bibitem[{{Barausse} {et~al.}(2020){Barausse}, {Dvorkin}, {Tremmel},
  {Volonteri}, \& {Bonetti}}]{Barausse20}
{Barausse}, E., {Dvorkin}, I., {Tremmel}, M., {Volonteri}, M., \& {Bonetti}, M.
  2020, \apj, 904, 16

\bibitem[{{Barro} {et~al.}(2023){Barro}, {Perez-Gonzalez}, {Kocevski},
  {McGrath}, {Trump}, {Simons}, {Somerville}, {Yung}, {Arrabal Haro}, {Bagley},
  {Cleri}, {Costantin}, {Davis}, {Dickinson}, {Finkelstein}, {Giavalisco},
  {Gomez-Guijarro}, {Hathi}, {Hirschmann}, {Akins}, {Holwerda},
  {Huertas-Company}, {Lucas}, {Papovich}, {Seille}, {Tacchella}, {Wilkins}, {de
  la Vega}, {Yang}, \& {Zavala}}]{Barro2023}
{Barro}, G., {Perez-Gonzalez}, P.~G., {Kocevski}, D.~D., {et~al.} 2023, arXiv
  e-prints, arXiv:2305.14418

\bibitem[{{Beckmann} {et~al.}(2023){Beckmann}, {Dubois}, {Volonteri},
  {Dong-P{\'a}ez}, {Trebitsch}, {Devriendt}, {Kaviraj}, {Kimm}, \&
  {Peirani}}]{beckman2023}
{Beckmann}, R.~S., {Dubois}, Y., {Volonteri}, M., {et~al.} 2023, \mnras, 523,
  5610

\bibitem[{{Beifiori} {et~al.}(2012){Beifiori}, {Courteau}, {Corsini}, \&
  {Zhu}}]{Beifiori12}
{Beifiori}, A., {Courteau}, S., {Corsini}, E.~M., \& {Zhu}, Y. 2012, \mnras,
  419, 2497

\bibitem[{{Bennert} {et~al.}(2021){Bennert}, {Treu}, {Ding}, {Stomberg},
  {Birrer}, {Snyder}, {Malkan}, {Stephens}, \& {Auger}}]{Bennert21}
{Bennert}, V.~N., {Treu}, T., {Ding}, X., {et~al.} 2021, \apj, 921, 36

\bibitem[{{Bennett} {et~al.}(2023){Bennett}, {Sijacki}, {Costa}, {Laporte}, \&
  {Witten}}]{Bennett23}
{Bennett}, J.~S., {Sijacki}, D., {Costa}, T., {Laporte}, N., \& {Witten}, C.
  2023, arXiv e-prints, arXiv:2305.11932

\bibitem[{{Bezanson} {et~al.}(2018){Bezanson}, {van der Wel}, {Straatman},
  {Pacifici}, {Wu}, {Bari{\v{s}}i{\'c}}, {Bell}, {Conroy}, {D'Eugenio},
  {Franx}, {Gallazzi}, {van Houdt}, {Maseda}, {Muzzin}, {van de Sande},
  {Sobral}, \& {Spilker}}]{Bezanson18b}
{Bezanson}, R., {van der Wel}, A., {Straatman}, C., {et~al.} 2018, \apjl, 868,
  L36

\bibitem[{{Blecha} {et~al.}(2011){Blecha}, {Cox}, {Loeb}, \&
  {Hernquist}}]{Blecha2011}
{Blecha}, L., {Cox}, T.~J., {Loeb}, A., \& {Hernquist}, L. 2011, \mnras, 412,
  2154

\bibitem[{{Blecha} {et~al.}(2016){Blecha}, {Sijacki}, {Kelley}, {Torrey},
  {Vogelsberger}, {Nelson}, {Springel}, {Snyder}, \& {Hernquist}}]{Blecha2016}
{Blecha}, L., {Sijacki}, D., {Kelley}, L.~Z., {et~al.} 2016, \mnras, 456, 961

\bibitem[{{Bogdan} {et~al.}(2023){Bogdan}, {Goulding}, {Natarajan}, {Kovacs},
  {Tremblay}, {Chadayammuri}, {Volonteri}, {Kraft}, {Forman}, {Jones},
  {Churazov}, \& {Zhuravleva}}]{Bogdan2023}
{Bogdan}, A., {Goulding}, A., {Natarajan}, P., {et~al.} 2023, arXiv e-prints,
  arXiv:2305.15458

\bibitem[{{B{\"o}ker} {et~al.}(2023){B{\"o}ker}, {Beck}, {Birkmann},
  {Giardino}, {Keyes}, {Kumari}, {Muzerolle}, {Rawle}, {Zeidler}, {Abul-Huda},
  {de Oliveira}, {Arribas}, {Bechtold}, {Bhatawdekar}, {Bonaventura}, {Bunker},
  {Cameron}, {Carniani}, {Charlot}, {Curti}, {Espinoza}, {Ferruit}, {Franx},
  {Jakobsen}, {Karakla}, {L{\'o}pez-Caniego}, {L{\"u}tzgendorf}, {Maiolino},
  {Manjavacas}, {Marston}, {Moseley}, {Ogle}, {Perna}, {Pe{\~n}a-Guerrero},
  {Pirzkal}, {Plesha}, {Proffitt}, {Rauscher}, {Rix}, {Rodr{\'\i}guez del
  Pino}, {Rustamkulov}, {Sabbi}, {Sing}, {Sirianni}, {te Plate}, {{\'U}beda},
  {Wahlgren}, {Wislowski}, {Wu}, \& {Willott}}]{boker23}
{B{\"o}ker}, T., {Beck}, T.~L., {Birkmann}, S.~M., {et~al.} 2023, \pasp, 135,
  038001

\bibitem[{{Bongiorno} {et~al.}(2014){Bongiorno}, {Maiolino}, {Brusa},
  {Marconi}, {Piconcelli}, {Lamastra}, {Cano-D{\'\i}az}, {Schulze}, {Magnelli},
  {Vignali}, {Fiore}, {Menci}, {Cresci}, {La Franca}, \&
  {Merloni}}]{Bongiorno14}
{Bongiorno}, A., {Maiolino}, R., {Brusa}, M., {et~al.} 2014, \mnras, 443, 2077

\bibitem[{{Bosman} {et~al.}(2023){Bosman}, {{\'A}lvarez-M{\'a}rquez}, {Colina},
  {Walter}, {Alonso-Herrero}, {Ward}, {{\"O}stlin}, {Greve}, {Wright}, {Bik},
  {Boogaard}, {Caputi}, {Costantin}, {Eckart}, {Garc{\'\i}a-Mar{\'\i}n},
  {Gillman}, {G{\"u}del}, {Henning}, {Hjorth}, {Iani}, {Ilbert}, {Jermann},
  {Labiano}, {Lagage}, {Langeroodi}, {Pei{\ss}ker}, {Ray}, {Rinaldi},
  {Topinka}, {van Dishoeck}, {van der Werf}, \& {Vandenbussche}}]{Bosman2023}
{Bosman}, S. E.~I., {{\'A}lvarez-M{\'a}rquez}, J., {Colina}, L., {et~al.} 2023,
  arXiv e-prints, arXiv:2307.14414

\bibitem[{{Bouwens} {et~al.}(2021){Bouwens}, {Oesch}, {Stefanon},
  {Illingworth}, {Labb{\'e}}, {Reddy}, {Atek}, {Montes}, {Naidu},
  {Nanayakkara}, {Nelson}, \& {Wilkins}}]{Bouwens21}
{Bouwens}, R.~J., {Oesch}, P.~A., {Stefanon}, M., {et~al.} 2021, \aj, 162, 47

\bibitem[{{Bunker} {et~al.}(2023{\natexlab{a}}){Bunker}, {Saxena}, {Cameron},
  {Willott}, {Curtis-Lake}, {Jakobsen}, {Carniani}, {Smit}, {Maiolino},
  {Witstok}, {Curti}, {D'Eugenio}, {Jones}, {Ferruit}, {Arribas}, {Charlot},
  {Chevallard}, {Giardino}, {de Graaff}, {Looser}, {Luetzgendorf}, {Maseda},
  {Rawle}, {Rix}, {Rodriguez Del Pino}, {Alberts}, {Egami}, {Eisenstein},
  {Endsley}, {Hainline}, {Hausen}, {Johnson}, {Rieke}, {Rieke}, {Robertson},
  {Shivaei}, {Stark}, {Sun}, {Tacchella}, {Tang}, {Williams}, {Willmer},
  {Baker}, {Baum}, {Bhatawdekar}, {Bowler}, {Boyett}, {Chen}, {Circosta},
  {Helton}, {Ji}, {Lyu}, {Nelson}, {Parlanti}, {Perna}, {Sandles}, {Scholtz},
  {Suess}, {Topping}, {Uebler}, {Wallace}, \& {Whitler}}]{bunker_jades_2023}
{Bunker}, A.~J., {Saxena}, A., {Cameron}, A.~J., {et~al.} 2023{\natexlab{a}},
  arXiv e-prints, arXiv:2302.07256

\bibitem[{{Bunker} {et~al.}(2023{\natexlab{b}}){Bunker}, {Saxena}, {Cameron},
  {Willott}, {Curtis-Lake}, {Jakobsen}, {Carniani}, {Smit}, {Maiolino},
  {Witstok}, {Curti}, {D'Eugenio}, {Jones}, {Ferruit}, {Arribas}, {Charlot},
  {Chevallard}, {Giardino}, {de Graaff}, {Looser}, {Luetzgendorf}, {Maseda},
  {Rawle}, {Rix}, {Rodriguez Del Pino}, {Alberts}, {Egami}, {Eisenstein},
  {Endsley}, {Hainline}, {Hausen}, {Johnson}, {Rieke}, {Rieke}, {Robertson},
  {Shivaei}, {Stark}, {Sun}, {Tacchella}, {Tang}, {Williams}, {Willmer},
  {Baker}, {Baum}, {Bhatawdekar}, {Bowler}, {Boyett}, {Chen}, {Circosta},
  {Helton}, {Ji}, {Lyu}, {Nelson}, {Parlanti}, {Perna}, {Sandles}, {Scholtz},
  {Suess}, {Topping}, {Uebler}, {Wallace}, \& {Whitler}}]{Bunker2023GNz11}
{Bunker}, A.~J., {Saxena}, A., {Cameron}, A.~J., {et~al.} 2023{\natexlab{b}},
  arXiv e-prints, arXiv:2302.07256

\bibitem[{{Cappellari} {et~al.}(2006{\natexlab{a}}){Cappellari}, {Bacon},
  {Bureau}, {Damen}, {Davies}, {de Zeeuw}, {Emsellem}, {Falc{\'o}n-Barroso},
  {Krajnovi{\'c}}, {Kuntschner}, {McDermid}, {Peletier}, {Sarzi}, {van den
  Bosch}, \& {van de Ven}}]{Cappellari2006}
{Cappellari}, M., {Bacon}, R., {Bureau}, M., {et~al.} 2006{\natexlab{a}},
  \mnras, 366, 1126

\bibitem[{{Cappellari} {et~al.}(2006{\natexlab{b}}){Cappellari}, {Bacon},
  {Bureau}, {Damen}, {Davies}, {de Zeeuw}, {Emsellem}, {Falc{\'o}n-Barroso},
  {Krajnovi{\'c}}, {Kuntschner}, {McDermid}, {Peletier}, {Sarzi}, {van den
  Bosch}, \& {van de Ven}}]{Cappellari06}
{Cappellari}, M., {Bacon}, R., {Bureau}, M., {et~al.} 2006{\natexlab{b}},
  \mnras, 366, 1126

\bibitem[{{Cappellari} {et~al.}(2013){Cappellari}, {Scott}, {Alatalo}, {Blitz},
  {Bois}, {Bournaud}, {Bureau}, {Crocker}, {Davies}, {Davis}, {de Zeeuw},
  {Duc}, {Emsellem}, {Khochfar}, {Krajnovi{\'c}}, {Kuntschner}, {McDermid},
  {Morganti}, {Naab}, {Oosterloo}, {Sarzi}, {Serra}, {Weijmans}, \&
  {Young}}]{cappellari2013a}
{Cappellari}, M., {Scott}, N., {Alatalo}, K., {et~al.} 2013, \mnras, 432, 1709

\bibitem[{{Carnall} {et~al.}(2023{\natexlab{a}}){Carnall}, {McLeod}, {McLure},
  {Dunlop}, {Begley}, {Cullen}, {Donnan}, {Hamadouche}, {Jewell}, {Jones},
  {Pollock}, \& {Wild}}]{Carnall23b}
{Carnall}, A.~C., {McLeod}, D.~J., {McLure}, R.~J., {et~al.}
  2023{\natexlab{a}}, \mnras, 520, 3974

\bibitem[{{Carnall} {et~al.}(2023{\natexlab{b}}){Carnall}, {McLure}, {Dunlop},
  {McLeod}, {Wild}, {Cullen}, {Magee}, {Begley}, {Cimatti}, {Donnan},
  {Hamadouche}, {Jewell}, \& {Walker}}]{Carnall23a}
{Carnall}, A.~C., {McLure}, R.~J., {Dunlop}, J.~S., {et~al.}
  2023{\natexlab{b}}, arXiv e-prints, arXiv:2301.11413

\bibitem[{{Carniani} {et~al.}(2015){Carniani}, {Marconi}, {Maiolino},
  {Balmaverde}, {Brusa}, {Cano-D{\'\i}az}, {Cicone}, {Comastri}, {Cresci},
  {Fiore}, {Feruglio}, {La Franca}, {Mainieri}, {Mannucci}, {Nagao}, {Netzer},
  {Piconcelli}, {Risaliti}, {Schneider}, \& {Shemmer}}]{Carniani15}
{Carniani}, S., {Marconi}, A., {Maiolino}, R., {et~al.} 2015, \aap, 580, A102

\bibitem[{{Carniani} {et~al.}(2023){Carniani}, {Venturi}, {Parlanti}, {de
  Graaff}, {Maiolino}, {Arribas}, {Bonaventura}, {Boyett}, {Bunker}, {Cameron},
  {Charlot}, {Chevallard}, {Curti}, {Curtis-Lake}, {Eisenstein}, {Giardino},
  {Hausen}, {Kumari}, {Maseda}, {Nelson}, {Perna}, {Rix}, {Robertson},
  {Rodr{\'\i}guez Del Pino}, {Sandles}, {Scholtz}, {Simmonds}, {Smit},
  {Tacchella}, {{\"U}bler}, {Williams}, {Willott}, \& {Witstok}}]{Carniani23}
{Carniani}, S., {Venturi}, G., {Parlanti}, E., {et~al.} 2023, arXiv e-prints,
  arXiv:2306.11801

\bibitem[{{Charlot} \& {Fall}(2000)}]{CF00}
{Charlot}, S. \& {Fall}, S.~M. 2000, \apj, 539, 718

\bibitem[{{Chen} {et~al.}(2023){Chen}, {Di Matteo}, {Ni}, {Tremmel}, {DeGraf},
  {Shen}, {Holgado}, {Bird}, {Croft}, \& {Feng}}]{Chen23_dualbh_astrid}
{Chen}, N., {Di Matteo}, T., {Ni}, Y., {et~al.} 2023, \mnras, 522, 1895

\bibitem[{{Chevallard} \& {Charlot}(2016)}]{Chevallard2016}
{Chevallard}, J. \& {Charlot}, S. 2016, \mnras, 462, 1415

\bibitem[{{Chiaberge} {et~al.}(2018){Chiaberge}, {Tremblay}, {Capetti}, \&
  {Norman}}]{Chiaberge2018}
{Chiaberge}, M., {Tremblay}, G.~R., {Capetti}, A., \& {Norman}, C. 2018, \apj,
  861, 56

\bibitem[{{Ciurlo} {et~al.}(2023){Ciurlo}, {Mannucci}, {Yeh}, {Amiri},
  {Carniani}, {Cicone}, {Cresci}, {Lusso}, {Marasco}, {Marconcini}, {Marconi},
  {Nardini}, {Pancino}, {Rosati}, {Rubinur}, {Severgnini}, {Scialpi}, {Tozzi},
  {Venturi}, {Vignali}, \& {Volonteri}}]{Ciurlo2023}
{Ciurlo}, A., {Mannucci}, F., {Yeh}, S., {et~al.} 2023, \aap, 671, L4

\bibitem[{{Civano} {et~al.}(2010){Civano}, {Elvis}, {Lanzuisi}, {Jahnke},
  {Zamorani}, {Blecha}, {Bongiorno}, {Brusa}, {Comastri}, {Hao}, {Leauthaud},
  {Loeb}, {Mainieri}, {Piconcelli}, {Salvato}, {Scoville}, {Trump}, {Vignali},
  {Aldcroft}, {Bolzonella}, {Bressert}, {Finoguenov}, {Fruscione}, {Koekemoer},
  {Cappelluti}, {Fiore}, {Giodini}, {Gilli}, {Impey}, {Lilly}, {Lusso},
  {Puccetti}, {Silverman}, {Aussel}, {Capak}, {Frayer}, {Le Floch},
  {McCracken}, {Sanders}, {Schiminovich}, \& {Taniguchi}}]{Civano2010}
{Civano}, F., {Elvis}, M., {Lanzuisi}, G., {et~al.} 2010, \apj, 717, 209

\bibitem[{{Curti} {et~al.}(2023){Curti}, {Maiolino}, {Carniani}, {D'Eugenio},
  {Chevallard}, {Curtis-Lake}, {Looser}, {Scholtz}, {{\"U}bler}, {Witstok},
  {Cameron}, {Charlot}, {Laseter}, {Sandles}, {Arribas}, {Bunker}, {Giardino},
  {Maseda}, {Rawle}, {Rodr{\'\i}guez Del Pino}, {Smit}, {Willott},
  {Eisenstein}, {Hausen}, {Johnson}, {Rieke}, {Robertson}, {Tacchella},
  {Williams}, {Willmer}, {Baker}, {Bhatawdekar}, {Boyett}, {Egami}, {Helton},
  {Ji}, {Kumari}, {Shivaei}, \& {Sun}}]{Curti23}
{Curti}, M., {Maiolino}, R., {Carniani}, S., {et~al.} 2023, arXiv e-prints,
  arXiv:2304.08516

\bibitem[{{Curti} {et~al.}(2020){Curti}, {Mannucci}, {Cresci}, \&
  {Maiolino}}]{Curti2023}
{Curti}, M., {Mannucci}, F., {Cresci}, G., \& {Maiolino}, R. 2020, \mnras, 491,
  944

\bibitem[{{de Graaff} {et~al.}(2024){de Graaff}, {Rix}, {Carniani}, {Suess},
  {Charlot}, {Curtis-Lake}, {Arribas}, {Baker}, {Boyett}, {Bunker}, {Cameron},
  {Chevallard}, {Curti}, {Eisenstein}, {Franx}, {Hainline}, {Hausen}, {Ji},
  {Johnson}, {Jones}, {Maiolino}, {Maseda}, {Nelson}, {Parlanti}, {Rawle},
  {Robertson}, {Tacchella}, {{\"U}bler}, {Williams}, {Willmer}, \&
  {Willott}}]{de_Graaff2024}
{de Graaff}, A., {Rix}, H.-W., {Carniani}, S., {et~al.} 2024, \aap, 684, A87

\bibitem[{{DeGraf} \& {Sijacki}(2020)}]{degraf2020}
{DeGraf}, C. \& {Sijacki}, D. 2020, \mnras, 491, 4973

\bibitem[{{Dessauges-Zavadsky} {et~al.}(2020){Dessauges-Zavadsky}, {Ginolfi},
  {Pozzi}, {B{\'e}thermin}, {Le F{\`e}vre}, {Fujimoto}, {Silverman}, {Jones},
  {Vallini}, {Schaerer}, {Faisst}, {Khusanova}, {Fudamoto}, {Cassata},
  {Loiacono}, {Capak}, {Yan}, {Amorin}, {Bardelli}, {Boquien}, {Cimatti},
  {Gruppioni}, {Hathi}, {Ibar}, {Koekemoer}, {Lemaux}, {Narayanan}, {Oesch},
  {Rodighiero}, {Romano}, {Talia}, {Toft}, {Vergani}, {Zamorani}, \&
  {Zucca}}]{Dessauges2020}
{Dessauges-Zavadsky}, M., {Ginolfi}, M., {Pozzi}, F., {et~al.} 2020, \aap, 643,
  A5

\bibitem[{{Di Mascia} {et~al.}(2021){Di Mascia}, {Gallerani}, {Behrens},
  {Pallottini}, {Carniani}, {Ferrara}, {Barai}, {Vito}, \&
  {Zana}}]{Dimascia2021}
{Di Mascia}, F., {Gallerani}, S., {Behrens}, C., {et~al.} 2021, \mnras, 503,
  2349

\bibitem[{{Di Matteo} {et~al.}(2023){Di Matteo}, {Angles-Alcazar}, \&
  {Shankar}}]{Dimatteo2022}
{Di Matteo}, T., {Angles-Alcazar}, D., \& {Shankar}, F. 2023, arXiv e-prints,
  arXiv:2304.11541

\bibitem[{{Di Matteo} {et~al.}(2022){Di Matteo}, {Ni}, {Chen}, {Croft}, {Bird},
  {Pacucci}, {Ricarte}, \& {Tremmel}}]{Dimatteo2022a}
{Di Matteo}, T., {Ni}, Y., {Chen}, N., {et~al.} 2022, arXiv e-prints,
  arXiv:2210.14960

\bibitem[{{Ding} {et~al.}(2022){Ding}, {Onoue}, {Silverman}, {Matsuoka},
  {Izumi}, {Strauss}, {Jahnke}, {Phillips}, {Li}, {Volonteri}, {Haiman},
  {Taufik Andika}, {Aoki}, {Baba}, {Bieri}, {Bosman}, {Bottrell}, {Eilers},
  {Fujimoto}, {Habouzit}, {Imanishi}, {Inayoshi}, {Iwasawa}, {Kashikawa},
  {Kawaguchi}, {Kohno}, {Lee}, {Lupi}, {Lyu}, {Nagao}, {Overzier}, {Schindler},
  {Schramm}, {Shimasaku}, {Toba}, {Trakhtenbrot}, {Trebitsch}, {Treu},
  {Umehata}, {Venemans}, {Vestergaard}, {Walter}, {Wang}, \& {Yang}}]{Ding2022}
{Ding}, X., {Onoue}, M., {Silverman}, J.~D., {et~al.} 2022, arXiv e-prints,
  arXiv:2211.14329

\bibitem[{{Dome} {et~al.}(2023){Dome}, {Tacchella}, {Fialkov}, {Dekel},
  {Ginzburg}, {Lapiner}, \& {Looser}}]{Dome23}
{Dome}, T., {Tacchella}, S., {Fialkov}, A., {et~al.} 2023, arXiv e-prints,
  arXiv:2305.07066

\bibitem[{{Duras} {et~al.}(2020){Duras}, {Bongiorno}, {Ricci}, {Piconcelli},
  {Shankar}, {Lusso}, {Bianchi}, {Fiore}, {Maiolino}, {Marconi}, {Onori},
  {Sani}, {Schneider}, {Vignali}, \& {La Franca}}]{Duras20}
{Duras}, F., {Bongiorno}, A., {Ricci}, F., {et~al.} 2020, \aap, 636, A73

\bibitem[{{Eisenstein} {et~al.}(2023){Eisenstein}, {Willott}, {Alberts},
  {Arribas}, {Bonaventura}, {Bunker}, {Cameron}, {Carniani}, {Charlot},
  {Curtis-Lake}, {D'Eugenio}, {Endsley}, {Ferruit}, {Giardino}, {Hainline},
  {Hausen}, {Jakobsen}, {Johnson}, {Maiolino}, {Rieke}, {Rieke}, {Rix},
  {Robertson}, {Stark}, {Tacchella}, {Williams}, {Willmer}, {Baker}, {Baum},
  {Bhatawdekar}, {Boyett}, {Chen}, {Chevallard}, {Circosta}, {Curti},
  {Danhaive}, {DeCoursey}, {de Graaff}, {Dressler}, {Egami}, {Helton},
  {Hviding}, {Ji}, {Jones}, {Kumari}, {L{\"u}tzgendorf}, {Laseter}, {Looser},
  {Lyu}, {Maseda}, {Nelson}, {Parlanti}, {Perna}, {Pusk{\'a}s}, {Rawle},
  {Rodr{\'\i}guez Del Pino}, {Sandles}, {Saxena}, {Scholtz}, {Sharpe},
  {Shivaei}, {Silcock}, {Simmonds}, {Skarbinski}, {Smit}, {Stone}, {Suess},
  {Sun}, {Tang}, {Topping}, {{\"U}bler}, {Villanueva}, {Wallace}, {Whitler},
  {Witstok}, \& {Woodrum}}]{Eisenstein23}
{Eisenstein}, D.~J., {Willott}, C., {Alberts}, S., {et~al.} 2023, arXiv
  e-prints, arXiv:2306.02465

\bibitem[{{Eracleous} \& {Halpern}(2003)}]{Eracleous03}
{Eracleous}, M. \& {Halpern}, J.~P. 2003, \apj, 599, 886

\bibitem[{{Eracleous} {et~al.}(1997){Eracleous}, {Halpern}, {M. Gilbert},
  {Newman}, \& {Filippenko}}]{Eracleous97}
{Eracleous}, M., {Halpern}, J.~P., {M. Gilbert}, A., {Newman}, J.~A., \&
  {Filippenko}, A.~V. 1997, \apj, 490, 216

\bibitem[{{Fan} {et~al.}(2022){Fan}, {Banados}, \& {Simcoe}}]{Fan22}
{Fan}, X., {Banados}, E., \& {Simcoe}, R.~A. 2022, arXiv e-prints,
  arXiv:2212.06907

\bibitem[{{Ferrara} {et~al.}(2014){Ferrara}, {Salvadori}, {Yue}, \&
  {Schleicher}}]{Ferrara2014}
{Ferrara}, A., {Salvadori}, S., {Yue}, B., \& {Schleicher}, D. 2014, \mnras,
  443, 2410

\bibitem[{{Ferruit} {et~al.}(2022){Ferruit}, {Jakobsen}, {Giardino}, {Rawle},
  {Alves de Oliveira}, {Arribas}, {Beck}, {Birkmann}, {B{\"o}ker}, {Bunker},
  {Charlot}, {de Marchi}, {Franx}, {Henry}, {Karakla}, {Kassin}, {Kumari},
  {L{\'o}pez-Caniego}, {L{\"u}tzgendorf}, {Maiolino}, {Manjavacas}, {Marston},
  {Moseley}, {Muzerolle}, {Pirzkal}, {Rauscher}, {Rix}, {Sabbi}, {Sirianni},
  {te Plate}, {Valenti}, {Willott}, \& {Zeidler}}]{Ferruit22}
{Ferruit}, P., {Jakobsen}, P., {Giardino}, G., {et~al.} 2022, \aap, 661, A81

\bibitem[{{Fiore} {et~al.}(2023){Fiore}, {Ferrara}, {Bischetti}, {Feruglio}, \&
  {Travascio}}]{fiore+2023}
{Fiore}, F., {Ferrara}, A., {Bischetti}, M., {Feruglio}, C., \& {Travascio}, A.
  2023, \apjl, 943, L27

\bibitem[{{Furtak} {et~al.}(2022){Furtak}, {Zitrin}, {Plat}, {Fujimoto},
  {Wang}, {Nelson}, {Labb{\'e}}, {Bezanson}, {Brammer}, {van Dokkum},
  {Endsley}, {Glazebrook}, {Greene}, {Leja}, {Price}, {Smit}, {Stark},
  {Weaver}, {Whitaker}, {Atek}, {Chevallard}, {Curtis-Lake}, {Dayal}, {Feltre},
  {Franx}, {Fudamoto}, {Marchesini}, {Mowla}, {Pan}, {Suess},
  {Vidal-Garc{\'\i}a}, \& {Williams}}]{Furtak2022}
{Furtak}, L.~J., {Zitrin}, A., {Plat}, A., {et~al.} 2022, arXiv e-prints,
  arXiv:2212.10531

\bibitem[{{Gallerani} {et~al.}(2010){Gallerani}, {Maiolino}, {Juarez}, {Nagao},
  {Marconi}, {Bianchi}, {Schneider}, {Mannucci}, {Oliva}, {Willott}, {Jiang},
  \& {Fan}}]{Gallerani2010}
{Gallerani}, S., {Maiolino}, R., {Juarez}, Y., {et~al.} 2010, \aap, 523, A85

\bibitem[{{Gardner} {et~al.}(2023){Gardner}, {Mather}, {Abbott}, {Abell},
  {Abernathy}, {Abney}, {Abraham}, {Abraham}, {Abul-Huda}, {Acton}, {Adams},
  {Adams}, {Adler}, {Adriaensen}, {Aguilar}, {Ahmed}, {Ahmed}, {Ahmed},
  {Albat}, {Albert}, {Alberts}, {Aldridge}, {Allen}, {Allen}, {Altenburg},
  {Altunc}, {Alvarez}, {{\'A}lvarez-M{\'a}rquez}, {de Oliveira}, {Ambrose},
  {Anandakrishnan}, {Andersen}, {Anderson}, {Anderson}, {Anderson}, {Anderson},
  {Aprea}, {Archer}, {Arenberg}, {Argyriou}, {Arribas}, {Artigau}, {Arvai},
  {Atcheson}, {Atkinson}, {Averbukh}, {Aymergen}, {Bacinski}, {Baggett},
  {Bagnasco}, {Baker}, {Balzano}, {Banks}, {Baran}, {Barker}, {Barrett},
  {Barringer}, {Barto}, {Bast}, {Baudoz}, {Baum}, {Beatty}, {Beaulieu},
  {Bechtold}, {Beck}, {Beddard}, {Beichman}, {Bellagama}, {Bely}, {Berger},
  {Bergeron}, {Bernier}, {Bertch}, {Beskow}, {Betz}, {Biagetti}, {Birkmann},
  {Bjorklund}, {Blackwood}, {Blazek}, {Blossfeld}, {Bluth}, {Boccaletti},
  {Boegner}, {Bohlin}, {Boia}, {B{\"o}ker}, {Bonaventura}, {Bond}, {Bosley},
  {Boucarut}, {Bouchet}, {Bouwman}, {Bower}, {Bowers}, {Bowers}, {Boyce},
  {Boyer}, {Boyer}, {Boyer}, {Boyer}, {Bradley}, {Brady}, {Brandl}, {Brannen},
  {Breda}, {Bremmer}, {Brennan}, {Bresnahan}, {Bright}, {Broiles},
  {Bromenschenkel}, {Brooks}, {Brooks}, {Brown}, {Brown}, {Brown}, {Bruce},
  {Bryson}, {Bujanda}, {Bullock}, {Bunker}, {Bureo}, {Burt}, {Bush},
  {Bushouse}, {Bussman}, {Cabaud}, {Cale}, {Calhoon}, {Calvani}, {Canipe},
  {Caputo}, {Cara}, {Carey}, {Case}, {Cesari}, {Cetorelli}, {Chance},
  {Chandler}, {Chaney}, {Chapman}, {Charlot}, {Chayer}, {Cheezum}, {Chen},
  {Chen}, {Cherinka}, {Chichester}, {Chilton}, {Chittiraibalan}, {Clampin},
  {Clark}, {Clark}, {Clark}, {Claybrooks}, {Cleveland}, {Cohen}, {Cohen},
  {Col{\'o}n}, {Coleman}, {Colina}, {Comber}, {Comeau}, {Comer}, {Reis},
  {Connolly}, {Conroy}, {Contos}, {Contreras}, {Cook}, {Cooper}, {Cooper},
  {Correia}, {Correnti}, {Cossou}, {Costanza}, {Coulais}, {Cox}, {Coyle},
  {Cracraft}, {Crew}, {Curtis}, {Cusveller}, {Maciel}, {Dailey}, {Daugeron},
  {Davidson}, {Davies}, {Davis}, {Davis}, {Day}, {de Chambure}, {de Jong}, {De
  Marchi}, {Dean}, {Decker}, {Delisa}, {Dell}, {Dellagatta}, {Dembinska},
  {Demosthenes}, {Dencheva}, {Deneu}, {DePriest}, {Deschenes}, {Dethienne},
  {Detre}, {Diaz}, {Dicken}, {DiFelice}, {Dillman}, {Disharoon}, {Dixon},
  {Doggett}, {Dominguez}, {Donaldson}, {Doria-Warner}, {Santos}, {Doty},
  {Douglas}, {Doyon}, {Dressler}, {Driggers}, {Driggers}, {Dunn}, {DuPrie},
  {Dupuis}, {Durning}, {Dutta}, {Earl}, {Eccleston}, {Ecobichon}, {Egami},
  {Ehrenwinkler}, {Eisenhamer}, {Eisenhower}, {Eisenstein}, {El Hamel}, {Elie},
  {Elliott}, {Elliott}, {Engesser}, {Espinoza}, {Etienne}, {Etxaluze}, {Evans},
  {Fabreguettes}, {Falcolini}, {Falini}, {Fatig}, {Feeney}, {Feinberg}, {Fels},
  {Ferdous}, {Ferguson}, {Ferrarese}, {Ferreira}, {Ferruit}, {Ferry},
  {Filippazzo}, {Firre}, {Fix}, {Flagey}, {Flanagan}, {Fleming}, {Florian},
  {Flynn}, {Foiadelli}, {Fontaine}, {Fontanella}, {Forshay}, {Fortner}, {Fox},
  {Framarini}, {Francisco}, {Franck}, {Franx}, {Franz}, {Friedman}, {Friend},
  {Frost}, {Fu}, {Fullerton}, {Gaillard}, {Galkin}, {Gallagher}, {Galyer},
  {Garc{\'\i}a Mar{\'\i}n}, {Gardner}, {Garland}, {Garrett}, {Gasman},
  {G{\'a}sp{\'a}r}, {Gastaud}, {Gaudreau}, {Gauthier}, {Geers}, {Geithner},
  {Gennaro}, {Gerber}, {Gereau}, {Giampaoli}, {Giardino}, {Gibbons}, {Gilbert},
  {Gilman}, {Girard}, {Giuliano}, {Gkountis}, {Glasse}, {Glassmire}, {Glauser},
  {Glazer}, {Goldberg}, {Golimowski}, {Gonzaga}, {Gordon}, {Gordon},
  {Goudfrooij}, {Gough}, {Graham}, {Grau}, {Green}, {Greene}, {Greene},
  {Greenfield}, {Greenhouse}, {Greve}, {Greville}, {Grimaldi}, {Groe},
  {Groebner}, {Grumm}, {Grundy}, {G{\"u}del}, {Guillard}, {Guldalian}, {Gunn},
  {Gurule}, {Gutman}, {Guy}, {Guyot}, {Hack}, {Haderlein}, {Hagan}, {Hagedorn},
  {Hainline}, {Haley}, {Hami}, {Hamilton}, {Hammann}, {Hammel}, {Hanley},
  {Hansen}, {Hardy}, {Harnisch}, {Harr}, {Harris}, {Hart}, {Hartig}, {Hasan},
  {Hashim}, {Hashimoto}, {Haskins}, {Hawkins}, {Hayden}, {Hayden}, {Healy},
  {Hecht}, {Heeg}, {Hejal}, {Helm}, {Hengemihle}, {Henning}, {Henry}, {Henry},
  {Henshaw}, {Hernandez}, {Herrington}, {Heske}, {Hesman}, {Hickey}, {Hilbert},
  {Hines}, {Hinz}, {Hirsch}, {Hitcho}, {Hodapp}, {Hodge}, {Hoffman},
  {Holfeltz}, {Holler}, {Hoppa}, {Horner}, {Howard}, {Howard}, {Huber},
  {Hunkeler}, {Hunter}, {Hunter}, {Hurd}, {Hurst}, {Hutchings}, {Hylan},
  {Ignat}, {Illingworth}, {Irish}, {Isaacs}, {Jackson}, {Jaffe}, {Jahic},
  {Jahromi}, {Jakobsen}, {James}, {James}, {James}, {Jamieson}, {Jandra},
  {Jayawardhana}, {Jedrzejewski}, {Jeffers}, {Jensen}, {Joanne}, {Johns},
  {Johnson}, {Johnson}, {Johnson}, {Johnson}, {Johnson}, {Johnson},
  {Johnstone}, {Jollet}, {Jones}, {Jones}, {Jones}, {Jones}, {Jones}, {Jordan},
  {Jordan}, {Jue}, {Jurkowski}, {Justis}, {Justtanont}, {Kaleida}, {Kalirai},
  {Kalmanson}, {Kaltenegger}, {Kammerer}, {Kan}, {Kanarek}, {Kao}, {Karakla},
  {Karl}, {Kassin}, {Kauffman}, {Kavanagh}, {Kelley}, {Kelly}, {Kendrew},
  {Kennedy}, {Kenny}, {Keski-Kuha}, {Keyes}, {Khan}, {Kidwell}, {Kimble},
  {King}, {King}, {Kinzel}, {Kirk}, {Kirkpatrick}, {Klaassen}, {Klingemann},
  {Klintworth}, {Knapp}, {Knight}, {Knollenberg}, {Knutsen}, {Koehler},
  {Koekemoer}, {Kofler}, {Kontson}, {Kovacs}, {Kozhurina-Platais}, {Krause},
  {Kriss}, {Krist}, {Kristoffersen}, {Krogel}, {Krueger}, {Kulp}, {Kumari},
  {Kwan}, {Kyprianou}, {Labador}, {Labiano}, {Lafreni{\`e}re}, {Lagage},
  {Laidler}, {Laine}, {Laird}, {Lajoie}, {Lallo}, {Lam}, {LaMassa}, {Lambros},
  {Lampenfield}, {Lander}, {Langston}, {Larson}, {Larson}, {LaVerghetta},
  {Law}, {Lawrence}, {Lee}, {Lee}, {Lee}, {Leisenring}, {Leveille}, {Levenson},
  {Levi}, {Levine}, {Lewis}, {Lewis}, {Lewis}, {Libralato}, {Lidon},
  {Liebrecht}, {Lightsey}, {Lilly}, {Lim}, {Lim}, {Ling}, {Link}, {Link},
  {Lipinski}, {Liu}, {Lo}, {Lobmeyer}, {Logue}, {Long}, {Long}, {Long}, {Long},
  {L{\'o}pez-Caniego}, {Lotz}, {Love-Pruitt}, {Lubskiy}, {Luers}, {Luetgens},
  {Luevano}, {Lui}, {Lund}, {Lundquist}, {Lunine}, {L{\"u}tzgendorf}, {Lynch},
  {MacDonald}, {MacDonald}, {Macias}, {Macklis}, {Maghami}, {Maharaja},
  {Maiolino}, {Makrygiannis}, {Malla}, {Malumuth}, {Manjavacas}, {Marini},
  {Marrione}, {Marston}, {Martel}, {Martin}, {Martin}, {Martinez}, {Maschmann},
  {Masci}, {Masetti}, {Maszkiewicz}, {Matthews}, {Matuskey}, {McBrayer},
  {McCarthy}, {McCaughrean}, {McClare}, {McClare}, {McCloskey}, {McClurg},
  {McCoy}, {McElwain}, {McGregor}, {McGuffey}, {McKay}, {McKenzie}, {McLean},
  {McMaster}, {McNeil}, {De Meester}, {Mehalick}, {Meixner}, {Mel{\'e}ndez},
  {Menzel}, {Menzel}, {Merz}, {Mesterharm}, {Meyer}, {Meyett}, {Meza},
  {Midwinter}, {Milam}, {Miller}, {Miller}, {Miskey}, {Misselt}, {Mitchell},
  {Mohan}, {Montoya}, {Moran}, {Morishita}, {Moro-Mart{\'\i}n}, {Morrison},
  {Morrison}, {Morse}, {Moschos}, {Moseley}, {Mosier}, {Mosner}, {Mountain},
  {Muckenthaler}, {Mueller}, {Mueller}, {Muhiem}, {M{\"u}hlmann}, {Mullally},
  {Mullen}, {Munger}, {Murphy}, {Murray}, {Muzerolle}, {Mycroft}, {Myers},
  {Myers}, {Myers}, {Myers}, {Myrick}, {Nagle}, {Nayak}, {Naylor}, {Neff},
  {Nelan}, {Nella}, {Nguyen}, {Nguyen}, {Nickson}, {Nidhiry}, {Niedner},
  {Nieto-Santisteban}, {Nikolov}, {Nishisaka}, {Noriega-Crespo}, {Nota},
  {O'Mara}, {Oboryshko}, {O'Brien}, {Ochs}, {Offenberg}, {Ogle}, {Ohl},
  {Olmsted}, {Osborne}, {O'Shaughnessy}, {{\"O}stlin}, {O'Sullivan}, {Otor},
  {Ottens}, {Ouellette}, {Outlaw}, {Owens}, {Pacifici}, {Page}, {Paranilam},
  {Park}, {Parrish}, {Paschal}, {Patapis}, {Patel}, {Patrick}, {Pattishall},
  {Paul}, {Paul}, {Pauly}, {Pavlovsky}, {Pe{\~n}a-Guerrero}, {Pedder}, {Peek},
  {Pelham}, {Penanen}, {Perriello}, {Perrin}, {Perrine}, {Perrygo}, {Peslier},
  {Petach}, {Peterson}, {Pfarr}, {Pierson}, {Pietraszkiewicz}, {Pilchen},
  {Pipher}, {Pirzkal}, {Pitman}, {Player}, {Plesha}, {Plitzke}, {Pohner},
  {Poletis}, {Pollizzi}, {Polster}, {Pontius}, {Pontoppidan}, {Porges},
  {Potter}, {Prescott}, {Proffitt}, {Pueyo}, {Quispe Neira}, {Radich}, {Rager},
  {Rameau}, {Ramey}, {Alarcon}, {Rampini}, {Rapp}, {Rashford}, {Rauscher},
  {Ravindranath}, {Rawle}, {Rawlings}, {Ray}, {Regan}, {Rehm}, {Rehm}, {Reid},
  {Reis}, {Renk}, {Reoch}, {Ressler}, {Rest}, {Reynolds}, {Richon}, {Richon},
  {Ridgaway}, {Riedel}, {Rieke}, {Rieke}, {Rifelli}, {Rigby}, {Riggs},
  {Ringel}, {Ritchie}, {Rix}, {Robberto}, {Robinson}, {Robinson}, {Robinson},
  {Rock}, {Rodriguez}, {Rodr{\'\i}guez del Pino}, {Roellig}, {Rohrbach},
  {Roman}, {Romelfanger}, {Romo}, {Rosales}, {Rose}, {Roteliuk}, {Roth},
  {Rothwell}, {Rouzaud}, {Rowe}, {Rowlands}, {Roy}, {Royer}, {Rui}, {Rumler},
  {Rumpl}, {Russ}, {Ryan}, {Ryan}, {Saad}, {Sabata}, {Sabatino}, {Sabbi},
  {Sabelhaus}, {Sabia}, {Sahu}, {Saif}, {Salvignol}, {Samara-Ratna},
  {Samuelson}, {Sanders}, {Sappington}, {Sargent}, {Sauer}, {Savadkin},
  {Sawicki}, {Schappell}, {Scheffer}, {Scheithauer}, {Scherer}, {Schiff},
  {Schlawin}, {Schmeitzky}, {Schmitz}, {Schmude}, {Schneider}, {Schreiber},
  {Schroeven-Deceuninck}, {Schultz}, {Schwab}, {Schwartz}, {Scoccimarro},
  {Scott}, {Scott}, {Seaton}, {Seely}, {Seery}, {Seidleck}, {Sembach},
  {Shanahan}, {Shaughnessy}, {Shaw}, {Shay}, {Sheehan}, {Sheth}, {Shih},
  {Shivaei}, {Siegel}, {Sienkiewicz}, {Simmons}, {Simon}, {Sirianni},
  {Sivaramakrishnan}, {Slade}, {Sloan}, {Slocum}, {Slowinski}, {Smith},
  {Smith}, {Smith}, {Smith}, {Smith}, {Smith}, {Smolik}, {Soderblom}, {Sohn},
  {Sokol}, {Sonneborn}, {Sontag}, {Sooy}, {Soummer}, {Southwood}, {Spain},
  {Sparmo}, {Speer}, {Spencer}, {Sprofera}, {Stallcup}, {Stanley},
  {Stansberry}, {Stark}, {Starr}, {Stassi}, {Steck}, {Steeley}, {Stephens},
  {Stephenson}, {Stewart}, {Stiavelli}, {}, {Strada}, {Straughn}, {Streetman},
  {Strickland}, {Strobele}, {Stuhlinger}, {Stys}, {Such}, {Sukhatme},
  {Sullivan}, {Sullivan}, {Sumner}, {Sun}, {Sunnquist}, {Swade}, {Swam},
  {Swenton}, {Swoish}, {Tam Litten}, {Tamas}, {Tao}, {Taylor}, {Taylor},
  {Plate}, {Van Tea}, {Teague}, {Telfer}, {Temim}, {Texter}, {Thatte},
  {Thompson}, {Thompson}, {Thomson}, {Thronson}, {Tierney}, {Tikkanen},
  {Tinnin}, {Tippet}, {Todd}, {Tran}, {Trauger}, {Trejo}, {Vinh Truong},
  {Tsukamoto}, {Tufail}, {Tumlinson}, {Tustain}, {Tyra}, {Ubeda}, {Underwood},
  {Uzzo}, {Vaclavik}, {Valenduc}, {Valenti}, {Van Campen}, {van de Wetering},
  {Van Der Marel}, {van Haarlem}, {Vandenbussche}, {van Dishoeck},
  {Vanterpool}, {Vernoy}, {Vila Costas}, {Volk}, {Voorzaat}, {Voyton}, {Vydra},
  {Waddy}, {Waelkens}, {Wahlgren}, {Walker}, {Wander}, {Warfield}, {Warner},
  {Wasiak}, {Wasiak}, {Wehner}, {Weiler}, {Weilert}, {Weiss}, {Wells}, {Welty},
  {Wheate}, {Wheeler}, {White}, {Whitehouse}, {Whiteleather}, {Whitman},
  {Williams}, {Willmer}, {Willott}, {Willoughby}, {Wilson}, {Wilson}, {Wilson},
  {Windhorst}, {Wislowski}, {Wolfe}, {Wolfe}, {Wolff}, {Wondel}, {Woo},
  {Woods}, {Worden}, {Workman}, {Wright}, {Wu}, {Wu}, {Wun}, {Wymer},
  {Yadetie}, {Yan}, {Yang}, {Yates}, {Yeager}, {Yerger}, {Young}, {Young},
  {Yu}, {Yu}, {Zak}, {Zeidler}, {Zepp}, {Zhou}, {Zincke}, {Zonak}, \&
  {Zondag}}]{Gardner2023}
{Gardner}, J.~P., {Mather}, J.~C., {Abbott}, R., {et~al.} 2023, \pasp, 135,
  068001

\bibitem[{{Giallongo} {et~al.}(2019){Giallongo}, {Grazian}, {Fiore}, {Kodra},
  {Urrutia}, {Castellano}, {Cristiani}, {Dickinson}, {Fontana}, {Menci},
  {Pentericci}, {Boutsia}, {Newman}, \& {Puccetti}}]{Giallongo19}
{Giallongo}, E., {Grazian}, A., {Fiore}, F., {et~al.} 2019, \apj, 884, 19

\bibitem[{{Goulding} {et~al.}(2023){Goulding}, {Greene}, {Setton}, {Labbe},
  {Bezanson}, {Miller}, {Atek}, {Bogdan}, {Brammer}, {Chemerynska}, {Cutler},
  {Dayal}, {Fudamoto}, {Fujimoto}, {Furtak}, {Kokorev}, {Khullar}, {Leja},
  {Marchesini}, {Natarajan}, {Nelson}, {Oesch}, {Pan}, {Papovich}, {Price},
  {van Dokkum}, {Wang}, {Weaver}, {Whitaker}, \& {Zitrin}}]{Goulding2023}
{Goulding}, A.~D., {Greene}, J.~E., {Setton}, D.~J., {et~al.} 2023, arXiv
  e-prints, arXiv:2308.02750

\bibitem[{{Greene} \& {Ho}(2004)}]{Greene2004}
{Greene}, J.~E. \& {Ho}, L.~C. 2004, \apj, 610, 722

\bibitem[{{Greene} {et~al.}(2020{\natexlab{a}}){Greene}, {Strader}, \&
  {Ho}}]{Greene20}
{Greene}, J.~E., {Strader}, J., \& {Ho}, L.~C. 2020{\natexlab{a}}, \araa, 58,
  257

\bibitem[{{Greene} {et~al.}(2020{\natexlab{b}}){Greene}, {Strader}, \&
  {Ho}}]{greene+2020}
{Greene}, J.~E., {Strader}, J., \& {Ho}, L.~C. 2020{\natexlab{b}}, \araa, 58,
  257

\bibitem[{{Habouzit} {et~al.}(2022){Habouzit}, {Onoue}, {Ba{\~n}ados},
  {Neeleman}, {Angl{\'e}s-Alc{\'a}zar}, {Walter}, {Pillepich}, {Dav{\'e}},
  {Jahnke}, \& {Dubois}}]{Habouzit22}
{Habouzit}, M., {Onoue}, M., {Ba{\~n}ados}, E., {et~al.} 2022, \mnras, 511,
  3751

\bibitem[{{Haidar} {et~al.}(2022){Haidar}, {Habouzit}, {Volonteri}, {Mezcua},
  {Greene}, {Neumayer}, {Angl{\'e}s-Alc{\'a}zar}, {Martin-Navarro}, {Hoyer},
  {Dubois}, \& {Dav{\'e}}}]{haidar2022}
{Haidar}, H., {Habouzit}, M., {Volonteri}, M., {et~al.} 2022, \mnras, 514, 4912

\bibitem[{{Harikane} {et~al.}(2023){Harikane}, {Zhang}, {Nakajima}, {Ouchi},
  {Isobe}, {Ono}, {Hatano}, {Xu}, \& {Umeda}}]{Harikane23BH}
{Harikane}, Y., {Zhang}, Y., {Nakajima}, K., {et~al.} 2023, arXiv e-prints,
  arXiv:2303.11946

\bibitem[{{Holden} \& {Tadhunter}(2023)}]{Holden23}
{Holden}, L.~R. \& {Tadhunter}, C.~N. 2023, \mnras, 524, 886

\bibitem[{{Inayoshi} {et~al.}(2020){Inayoshi}, {Visbal}, \&
  {Haiman}}]{inayoshi+2020}
{Inayoshi}, K., {Visbal}, E., \& {Haiman}, Z. 2020, \araa, 58, 27

\bibitem[{{Izumi} {et~al.}(2019){Izumi}, {Onoue}, {Matsuoka}, {Nagao},
  {Strauss}, {Imanishi}, {Kashikawa}, {Fujimoto}, {Kohno}, {Toba}, {Umehata},
  {Goto}, {Ueda}, {Shirakata}, {Silverman}, {Greene}, {Harikane}, {Hashimoto},
  {Ikarashi}, {Iono}, {Iwasawa}, {Lee}, {Minezaki}, {Nakanishi}, {Tamura},
  {Tang}, \& {Taniguchi}}]{Izumi19}
{Izumi}, T., {Onoue}, M., {Matsuoka}, Y., {et~al.} 2019, \pasj, 71, 111

\bibitem[{{Jakobsen} {et~al.}(2022){Jakobsen}, {Ferruit}, {Alves de Oliveira},
  {Arribas}, {Bagnasco}, {Barho}, {Beck}, {Birkmann}, {B{\"o}ker}, {Bunker},
  {Charlot}, {de Jong}, {de Marchi}, {Ehrenwinkler}, {Falcolini}, {Fels},
  {Franx}, {Franz}, {Funke}, {Giardino}, {Gnata}, {Holota}, {Honnen}, {Jensen},
  {Jentsch}, {Johnson}, {Jollet}, {Karl}, {Kling}, {K{\"o}hler}, {Kolm},
  {Kumari}, {Lander}, {Lemke}, {L{\'o}pez-Caniego}, {L{\"u}tzgendorf},
  {Maiolino}, {Manjavacas}, {Marston}, {Maschmann}, {Maurer}, {Messerschmidt},
  {Moseley}, {Mosner}, {Mott}, {Muzerolle}, {Pirzkal}, {Pittet}, {Plitzke},
  {Posselt}, {Rapp}, {Rauscher}, {Rawle}, {Rix}, {R{\"o}del}, {Rumler},
  {Sabbi}, {Salvignol}, {Schmid}, {Sirianni}, {Smith}, {Strada}, {te Plate},
  {Valenti}, {Wettemann}, {Wiehe}, {Wiesmayer}, {Willott}, {Wright}, {Zeidler},
  \& {Zincke}}]{Jakobsen22}
{Jakobsen}, P., {Ferruit}, P., {Alves de Oliveira}, C., {et~al.} 2022, \aap,
  661, A80

\bibitem[{{Juod{\v{z}}balis} {et~al.}(2023){Juod{\v{z}}balis}, {Conselice},
  {Singh}, {Adams}, {Ormerod}, {Harvey}, {Austin}, {Volonteri}, {Cohen},
  {Jansen}, {Summers}, {Windhorst}, {D'Silva}, {Koekemoer}, {Coe}, {Driver},
  {Frye}, {Grogin}, {Marshall}, {Nonino}, {Pirzkal}, {Robotham}, {Ryan},
  {Ortiz}, {Tompkins}, {Willmer}, \& {Yan}}]{Ignas2023J}
{Juod{\v{z}}balis}, I., {Conselice}, C.~J., {Singh}, M., {et~al.} 2023, arXiv
  e-prints, arXiv:2307.07535

\bibitem[{{Juod{\v{z}}balis} {et~al.}(2024{\natexlab{a}}){Juod{\v{z}}balis},
  {Ji}, {Maiolino}, {D'Eugenio}, {Scholtz}, {Risaliti}, {Fabian}, {Mazzolari},
  {Gilli}, {Prandoni}, {Arribas}, {Bunker}, {Carniani}, {Charlot},
  {Curtis-Lake}, {de Graaff}, {Hainline}, {Parlanti}, {Perna},
  {P{\'e}rez-Gonz{\'a}lez}, {Robertson}, {Tacchella}, {{\"U}bler}, {Williams},
  {Willott}, \& {Witstok}}]{Juodzbalis2024b}
{Juod{\v{z}}balis}, I., {Ji}, X., {Maiolino}, R., {et~al.} 2024{\natexlab{a}},
  arXiv e-prints, arXiv:2407.08643

\bibitem[{{Juod{\v{z}}balis} {et~al.}(2024{\natexlab{b}}){Juod{\v{z}}balis},
  {Maiolino}, {Baker}, {Tacchella}, {Scholtz}, {D'Eugenio}, {Schneider},
  {Trinca}, {Valiante}, {DeCoursey}, {Curti}, {Carniani}, {Chevallard}, {de
  Graaff}, {Arribas}, {Bennett}, {Bourne}, {Bunker}, {Charlot}, {Jiang},
  {Koudmani}, {Perna}, {Robertson}, {Sijacki}, {{\"U}bler}, {Williams},
  {Willott}, \& {Witstok}}]{Juodzbalis2024}
{Juod{\v{z}}balis}, I., {Maiolino}, R., {Baker}, W.~M., {et~al.}
  2024{\natexlab{b}}, arXiv e-prints, arXiv:2403.03872

\bibitem[{{Kauffmann} {et~al.}(2003){Kauffmann}, {Heckman}, {Tremonti},
  {Brinchmann}, {Charlot}, {White}, {Ridgway}, {Brinkmann}, {Fukugita}, {Hall},
  {Ivezi{\'c}}, {Richards}, \& {Schneider}}]{Kauffmann03}
{Kauffmann}, G., {Heckman}, T.~M., {Tremonti}, C., {et~al.} 2003, \mnras, 346,
  1055

\bibitem[{{Kewley} {et~al.}(2001){Kewley}, {Dopita}, {Sutherland}, {Heisler},
  \& {Trevena}}]{Kewley01}
{Kewley}, L.~J., {Dopita}, M.~A., {Sutherland}, R.~S., {Heisler}, C.~A., \&
  {Trevena}, J. 2001, \apj, 556, 121

\bibitem[{{Kocevski} {et~al.}(2023){Kocevski}, {Onoue}, {Inayoshi}, {Trump},
  {Arrabal Haro}, {Grazian}, {Dickinson}, {Finkelstein}, {Kartaltepe},
  {Hirschmann}, {Fujimoto}, {Juneau}, {Amorin}, {Bagley}, {Barro}, {Bell},
  {Bisigello}, {Calabro}, {Cleri}, {Cooper}, {Ding}, {Grogin}, {Ho}, {Inoue},
  {Jiang}, {Jones}, {Koekemoer}, {Li}, {Li}, {McGrath}, {Molina}, {Papovich},
  {Perez-Gonzalez}, {Pirzkal}, {Wilkins}, {Yang}, \& {Yung}}]{Kocevski23}
{Kocevski}, D.~D., {Onoue}, M., {Inayoshi}, K., {et~al.} 2023, arXiv e-prints,
  arXiv:2302.00012

\bibitem[{{Kormendy} \& {Ho}(2013)}]{Kormendy13}
{Kormendy}, J. \& {Ho}, L.~C. 2013, \araa, 51, 511

\bibitem[{{Koudmani} {et~al.}(2022{\natexlab{a}}){Koudmani}, {Sijacki}, \&
  {Smith}}]{Koudmani2022}
{Koudmani}, S., {Sijacki}, D., \& {Smith}, M.~C. 2022{\natexlab{a}}, \mnras,
  516, 2112

\bibitem[{{Koudmani} {et~al.}(2022{\natexlab{b}}){Koudmani}, {Sijacki}, \&
  {Smith}}]{koudmani+2022}
{Koudmani}, S., {Sijacki}, D., \& {Smith}, M.~C. 2022{\natexlab{b}}, \mnras,
  516, 2112

\bibitem[{{Krolik} {et~al.}(2019){Krolik}, {Volonteri}, {Dubois}, \&
  {Devriendt}}]{Krolik2019}
{Krolik}, J.~H., {Volonteri}, M., {Dubois}, Y., \& {Devriendt}, J. 2019, \apj,
  879, 110

\bibitem[{{Laseter} {et~al.}(2023){Laseter}, {Maseda}, {Curti}, {Maiolino},
  {D'Eugenio}, {Cameron}, {Looser}, {Arribas}, {Baker}, {Bhatawdekar},
  {Boyett}, {Bunker}, {Carniani}, {Charlot}, {Chevallard}, {Curtis-lake},
  {Egami}, {Eisenstein}, {Hainline}, {Hausen}, {Ji}, {Kumari}, {Perna},
  {Rawle}, {Rix}, {Robertson}, {Rodr{\'\i}guez Del Pino}, {Sandles}, {Scholtz},
  {Smit}, {Tacchella}, {{\"U}bler}, {Williams}, {Willott}, \&
  {Witstok}}]{Laseter23}
{Laseter}, I.~H., {Maseda}, M.~V., {Curti}, M., {et~al.} 2023, arXiv e-prints,
  arXiv:2306.03120

\bibitem[{{Lauer} {et~al.}(2007){Lauer}, {Tremaine}, {Richstone}, \&
  {Faber}}]{Lauer2007}
{Lauer}, T.~R., {Tremaine}, S., {Richstone}, D., \& {Faber}, S.~M. 2007, \apj,
  670, 249

\bibitem[{{Li} {et~al.}(2024){Li}, {Silverman}, {Shen}, {Volonteri}, {Jahnke},
  {Zhuang}, {Scoggins}, {Ding}, {Harikane}, {Onoue}, \& {Tanaka}}]{Li2024bias}
{Li}, J., {Silverman}, J.~D., {Shen}, Y., {et~al.} 2024, arXiv e-prints,
  arXiv:2403.00074

\bibitem[{{Liddle}(2007)}]{Liddle2007}
{Liddle}, A.~R. 2007, \mnras, 377, L74

\bibitem[{{Liu} {et~al.}(2019){Liu}, {Schinnerer}, {Groves}, {Magnelli},
  {Lang}, {Leslie}, {Jim{\'e}nez-Andrade}, {Riechers}, {Popping}, {Magdis},
  {Daddi}, {Sargent}, {Gao}, {Fudamoto}, {Oesch}, \& {Bertoldi}}]{Liu2019}
{Liu}, D., {Schinnerer}, E., {Groves}, B., {et~al.} 2019, \apj, 887, 235

\bibitem[{{Looser} {et~al.}(2023{\natexlab{a}}){Looser}, {D'Eugenio},
  {Maiolino}, {Tacchella}, {Curti}, {Arribas}, {Baker}, {Baum}, {Bonaventura},
  {Boyett}, {Bunker}, {Carniani}, {Charlot}, {Chevallard}, {Curtis-Lake},
  {Danhaive}, {Eisenstein}, {de Graaff}, {Hainline}, {Ji}, {Johnson}, {Kumari},
  {Nelson}, {Parlanti}, {Rix}, {Robertson}, {Rodr{\'\i}guez Del Pino},
  {Sandles}, {Scholtz}, {Smit}, {Stark}, {{\"U}bler}, {Williams}, {Willott}, \&
  {Witstok}}]{Looser23b}
{Looser}, T.~J., {D'Eugenio}, F., {Maiolino}, R., {et~al.} 2023{\natexlab{a}},
  arXiv e-prints, arXiv:2306.02470

\bibitem[{{Looser} {et~al.}(2023{\natexlab{b}}){Looser}, {D'Eugenio},
  {Maiolino}, {Witstok}, {Sandles}, {Curtis-Lake}, {Chevallard}, {Tacchella},
  {Johnson}, {Baker}, {Suess}, {Carniani}, {Ferruit}, {Arribas}, {Bonaventura},
  {Bunker}, {Cameron}, {Charlot}, {Curti}, {de Graaff}, {Maseda}, {Rawle},
  {Rix}, {Rodriguez Del Pino}, {Smit}, {{\"U}bler}, {Willott}, {Alberts},
  {Egami}, {Eisenstein}, {Endsley}, {Hausen}, {Rieke}, {Robertson}, {Shivaei},
  {Williams}, {Boyett}, {Chen}, {Ji}, {Jones}, {Kumari}, {Nelson}, {Perna},
  {Saxena}, \& {Scholtz}}]{Looser23a}
{Looser}, T.~J., {D'Eugenio}, F., {Maiolino}, R., {et~al.} 2023{\natexlab{b}},
  arXiv e-prints, arXiv:2302.14155

\bibitem[{{Lyu} {et~al.}(2022){Lyu}, {Alberts}, {Rieke}, \&
  {Rujopakarn}}]{Lyu2022}
{Lyu}, J., {Alberts}, S., {Rieke}, G.~H., \& {Rujopakarn}, W. 2022, \apj, 941,
  191

\bibitem[{{Madau} {et~al.}(2024){Madau}, {Giallongo}, {Grazian}, \&
  {Haardt}}]{Madau2024}
{Madau}, P., {Giallongo}, E., {Grazian}, A., \& {Haardt}, F. 2024, \apj, 971,
  75

\bibitem[{{Maiolino} \& {Mannucci}(2019)}]{Maiolino2019}
{Maiolino}, R. \& {Mannucci}, F. 2019, \aapr, 27, 3

\bibitem[{{Maiolino} \& {Rieke}(1995)}]{Maiolino95}
{Maiolino}, R. \& {Rieke}, G.~H. 1995, \apj, 454, 95

\bibitem[{{Maiolino} {et~al.}(2024{\natexlab{a}}){Maiolino}, {Risaliti},
  {Signorini}, {Trefoloni}, {Juodzbalis}, {Scholtz}, {Uebler}, {D'Eugenio},
  {Carniani}, {Fabian}, {Ji}, {Mazzolari}, {Bertola}, {Brusa}, {Bunker},
  {Charlot}, {Comastri}, {Cresci}, {DeCoursey}, {Egami}, {Fiore}, {Gilli},
  {Perna}, {Tacchella}, \& {Venturi}}]{Maiolino24X}
{Maiolino}, R., {Risaliti}, G., {Signorini}, M., {et~al.} 2024{\natexlab{a}},
  arXiv e-prints, arXiv:2405.00504

\bibitem[{{Maiolino} {et~al.}(2004){Maiolino}, {Schneider}, {Oliva}, {Bianchi},
  {Ferrara}, {Mannucci}, {Pedani}, \& {Roca Sogorb}}]{Maiolino2004}
{Maiolino}, R., {Schneider}, R., {Oliva}, E., {et~al.} 2004, \nat, 431, 533

\bibitem[{{Maiolino} {et~al.}(2024{\natexlab{b}}){Maiolino}, {Scholtz},
  {Witstok}, {Carniani}, {D'Eugenio}, {de Graaff}, {{\"U}bler}, {Tacchella},
  {Curtis-Lake}, {Arribas}, {Bunker}, {Charlot}, {Chevallard}, {Curti},
  {Looser}, {Maseda}, {Rawle}, {Rodr{\'\i}guez del Pino}, {Willott}, {Egami},
  {Eisenstein}, {Hainline}, {Robertson}, {Williams}, {Willmer}, {Baker},
  {Boyett}, {DeCoursey}, {Fabian}, {Helton}, {Ji}, {Jones}, {Kumari},
  {Laporte}, {Nelson}, {Perna}, {Sandles}, {Shivaei}, \&
  {Sun}}]{maiolino_bh_2023}
{Maiolino}, R., {Scholtz}, J., {Witstok}, J., {et~al.} 2024{\natexlab{b}},
  \nat, 627, 59

\bibitem[{{Maiolino} {et~al.}(2007){Maiolino}, {Shemmer}, {Imanishi}, {Netzer},
  {Oliva}, {Lutz}, \& {Sturm}}]{Maiolino2007}
{Maiolino}, R., {Shemmer}, O., {Imanishi}, M., {et~al.} 2007, \aap, 468, 979

\bibitem[{{Maiolino} {et~al.}(2023){Maiolino}, {Uebler}, {Perna}, {Scholtz},
  {D'Eugenio}, {Witten}, {Laporte}, {Witstok}, {Carniani}, {Tacchella},
  {Baker}, {Arribas}, {Nakajima}, {Eisenstein}, {Bunker}, {Charlot}, {Cresci},
  {Curti}, {Curtis-Lake}, {de Graaff}, {Ji}, {Johnson}, {Kumari}, {Looser},
  {Maseda}, {Robertson}, {Rodriguez Del Pino}, {Sandles}, {Simmonds}, {Smit},
  {Sun}, {Venturi}, {Williams}, \& {Willmer}}]{Maiolino23heii}
{Maiolino}, R., {Uebler}, H., {Perna}, M., {et~al.} 2023, arXiv e-prints,
  arXiv:2306.00953

\bibitem[{{Mannerkoski} {et~al.}(2022){Mannerkoski}, {Johansson}, {Rantala},
  {Naab}, {Liao}, \& {Rawlings}}]{Mannerkoski2022}
{Mannerkoski}, M., {Johansson}, P.~H., {Rantala}, A., {et~al.} 2022, \apj, 929,
  167

\bibitem[{{Mannucci} {et~al.}(2022){Mannucci}, {Pancino}, {Belfiore}, {Cicone},
  {Ciurlo}, {Cresci}, {Lusso}, {Marasco}, {Marconi}, {Nardini}, {Pinna},
  {Severgnini}, {Saracco}, {Tozzi}, \& {Yeh}}]{Mannucci22}
{Mannucci}, F., {Pancino}, E., {Belfiore}, F., {et~al.} 2022, Nature Astronomy,
  6, 1185

\bibitem[{{Mannucci} {et~al.}(2023){Mannucci}, {Scialpi}, {Ciurlo}, {Yeh},
  {Marconcini}, {Tozzi}, {Cresci}, {Marconi}, {Amiri}, {Belfiore}, {Carniani},
  {Cicone}, {Nardini}, {Pancino}, {Rubinur}, {Severgnini}, {Ulivi}, {Venturi},
  {Vignali}, {Volonteri}, {Pinna}, {Rossi}, {Puglisi}, {Agapito}, {Plantet},
  {Ghose}, {Carbonaro}, {Xompero}, {Grani}, {Esposito}, {Power}, {Guerra
  Ramon}, {Lefebvre}, {Cavallaro}, {Davies}, {Riccardi}, {Macintosh}, {Taylor},
  {Dolci}, {Baruffolo}, {Feuchtgruber}, {Kravchenko}, {Rau}, {Sturm},
  {Wiezorrek}, {Dallilar}, \& {Kenworthy}}]{Mannucci2023}
{Mannucci}, F., {Scialpi}, M., {Ciurlo}, A., {et~al.} 2023, arXiv e-prints,
  arXiv:2305.07396

\bibitem[{{Marshall} {et~al.}(2023){Marshall}, {Perna}, {Willott}, {Maiolino},
  {Scholtz}, {{\"U}bler}, {Carniani}, {Arribas}, {L{\"u}tzgendorf}, {Bunker},
  {Charlot}, {Ferruit}, {Jakobsen}, {Rodr{\'\i}guez Del Pino}, {B{\"o}ker},
  {Cameron}, {Cresci}, {Curtis-Lake}, {Jones}, {Kumari}, \&
  {P{\'e}rez-Gonz{\'a}lez}}]{Marshall2023}
{Marshall}, M.~A., {Perna}, M., {Willott}, C.~J., {et~al.} 2023, arXiv
  e-prints, arXiv:2302.04795

\bibitem[{{Marziani} {et~al.}(2019){Marziani}, {del Olmo},
  {Mart{\'\i}nez-Carballo}, {Mart{\'\i}nez-Aldama}, {Stirpe}, {Negrete},
  {Dultzin}, {D'Onofrio}, {Bon}, \& {Bon}}]{Marziani2019}
{Marziani}, P., {del Olmo}, A., {Mart{\'\i}nez-Carballo}, M.~A., {et~al.} 2019,
  \aap, 627, A88

\bibitem[{{Mathur} {et~al.}(2012){Mathur}, {Fields}, {Peterson}, \&
  {Grupe}}]{Mathur2012}
{Mathur}, S., {Fields}, D., {Peterson}, B.~M., \& {Grupe}, D. 2012, \apj, 754,
  146

\bibitem[{{Matthee} {et~al.}(2023){Matthee}, {Naidu}, {Brammer}, {Chisholm},
  {Eilers}, {Goulding}, {Greene}, {Kashino}, {Labbe}, {Lilly}, {Mackenzie},
  {Oesch}, {Weibel}, {Wuyts}, {Xiao}, {Bordoloi}, {Bouwens}, {van Dokkum},
  {Illingworth}, {Kramarenko}, {Maseda}, {Mason}, {Meyer}, {Nelson}, {Reddy},
  {Shivaei}, {Simcoe}, \& {Yue}}]{Matthee23}
{Matthee}, J., {Naidu}, R.~P., {Brammer}, G., {et~al.} 2023, arXiv e-prints,
  arXiv:2306.05448

\bibitem[{{McKee} \& {Tan}(2008)}]{Mckee08}
{McKee}, C.~F. \& {Tan}, J.~C. 2008, \apj, 681, 771

\bibitem[{{Merloni} {et~al.}(2010){Merloni}, {Bongiorno}, {Bolzonella},
  {Brusa}, {Civano}, {Comastri}, {Elvis}, {Fiore}, {Gilli}, {Hao}, {Jahnke},
  {Koekemoer}, {Lusso}, {Mainieri}, {Mignoli}, {Miyaji}, {Renzini}, {Salvato},
  {Silverman}, {Trump}, {Vignali}, {Zamorani}, {Capak}, {Lilly}, {Sanders},
  {Taniguchi}, {Bardelli}, {Carollo}, {Caputi}, {Contini}, {Coppa}, {Cucciati},
  {de la Torre}, {de Ravel}, {Franzetti}, {Garilli}, {Hasinger}, {Impey},
  {Iovino}, {Iwasawa}, {Kampczyk}, {Kneib}, {Knobel}, {Kova{\v{c}}},
  {Lamareille}, {Le Borgne}, {Le Brun}, {Le F{\`e}vre}, {Maier}, {Pello},
  {Peng}, {Perez Montero}, {Ricciardelli}, {Scodeggio}, {Tanaka}, {Tasca},
  {Tresse}, {Vergani}, \& {Zucca}}]{Merloni10}
{Merloni}, A., {Bongiorno}, A., {Bolzonella}, M., {et~al.} 2010, \apj, 708, 137

\bibitem[{{Merloni} {et~al.}(2014){Merloni}, {Bongiorno}, {Brusa}, {Iwasawa},
  {Mainieri}, {Magnelli}, {Salvato}, {Berta}, {Cappelluti}, {Comastri},
  {Fiore}, {Gilli}, {Koekemoer}, {Le Floc'h}, {Lusso}, {Lutz}, {Miyaji},
  {Pozzi}, {Riguccini}, {Rosario}, {Silverman}, {Symeonidis}, {Treister},
  {Vignali}, \& {Zamorani}}]{Merloni2014}
{Merloni}, A., {Bongiorno}, A., {Brusa}, M., {et~al.} 2014, \mnras, 437, 3550

\bibitem[{{Mezcua} {et~al.}(2018){Mezcua}, {Civano}, {Marchesi}, {Suh},
  {Fabbiano}, \& {Volonteri}}]{Mezcua2018}
{Mezcua}, M., {Civano}, F., {Marchesi}, S., {et~al.} 2018, \mnras, 478, 2576

\bibitem[{{Morishita} {et~al.}(2022){Morishita}, {Chiaberge}, {Hilbert},
  {Lambrides}, {Blecha}, {Baum}, {Bianchi}, {Capetti}, {Castignani},
  {Macchetto}, {Miley}, {O'Dea}, \& {Norman}}]{Morishita2022}
{Morishita}, T., {Chiaberge}, M., {Hilbert}, B., {et~al.} 2022, \apj, 931, 165

\bibitem[{{Nagao} {et~al.}(2006){Nagao}, {Marconi}, \& {Maiolino}}]{Nagao06}
{Nagao}, T., {Marconi}, A., \& {Maiolino}, R. 2006, \aap, 447, 157

\bibitem[{{Nakajima} \& {Maiolino}(2022)}]{nakajima_diagnostics_2022}
{Nakajima}, K. \& {Maiolino}, R. 2022, \mnras, 513, 5134

\bibitem[{{Netzer}(2019)}]{Netzer19}
{Netzer}, H. 2019, \mnras, 488, 5185

\bibitem[{{Netzer} {et~al.}(2016){Netzer}, {Lani}, {Nordon}, {Trakhtenbrot},
  {Lira}, \& {Shemmer}}]{Netzer2016}
{Netzer}, H., {Lani}, C., {Nordon}, R., {et~al.} 2016, \apj, 819, 123

\bibitem[{{Ni} {et~al.}(2022){Ni}, {Di Matteo}, {Bird}, {Croft}, {Feng},
  {Chen}, {Tremmel}, {DeGraf}, \& {Li}}]{Ni2022}
{Ni}, Y., {Di Matteo}, T., {Bird}, S., {et~al.} 2022, \mnras, 513, 670

\bibitem[{{Niida} {et~al.}(2020){Niida}, {Nagao}, {Ikeda}, {Akiyama},
  {Matsuoka}, {He}, {Matsuoka}, {Toba}, {Onoue}, {Kobayashi}, {Taniguchi},
  {Furusawa}, {Harikane}, {Imanishi}, {Kashikawa}, {Kawaguchi}, {Komiyama},
  {Shirakata}, {Terashima}, \& {Ueda}}]{Niida2020}
{Niida}, M., {Nagao}, T., {Ikeda}, H., {et~al.} 2020, \apj, 904, 89

\bibitem[{{Oesch} {et~al.}(2016){Oesch}, {Brammer}, {van Dokkum},
  {Illingworth}, {Bouwens}, {Labb{\'e}}, {Franx}, {Momcheva}, {Ashby}, {Fazio},
  {Gonzalez}, {Holden}, {Magee}, {Skelton}, {Smit}, {Spitler}, {Trenti}, \&
  {Willner}}]{Oesch16}
{Oesch}, P.~A., {Brammer}, G., {van Dokkum}, P.~G., {et~al.} 2016, \apj, 819,
  129

\bibitem[{{Onoue} {et~al.}(2023){Onoue}, {Inayoshi}, {Ding}, {Li}, {Li},
  {Molina}, {Inoue}, {Jiang}, \& {Ho}}]{Onoue2023}
{Onoue}, M., {Inayoshi}, K., {Ding}, X., {et~al.} 2023, \apjl, 942, L17

\bibitem[{{Pei}(1992)}]{Pei1992}
{Pei}, Y.~C. 1992, \apj, 395, 130

\bibitem[{{Piotrowska} {et~al.}(2022){Piotrowska}, {Bluck}, {Maiolino}, \&
  {Peng}}]{Piotrowska22}
{Piotrowska}, J.~M., {Bluck}, A. F.~L., {Maiolino}, R., \& {Peng}, Y. 2022,
  \mnras, 512, 1052

\bibitem[{{Planck Collaboration} {et~al.}(2020){Planck Collaboration},
  {Aghanim}, {Akrami}, {Ashdown}, {Aumont}, {Baccigalupi}, {Ballardini},
  {Banday}, {Barreiro}, {Bartolo}, {Basak}, {Battye}, {Benabed}, {Bernard},
  {Bersanelli}, {Bielewicz}, {Bock}, {Bond}, {Borrill}, {Bouchet}, {Boulanger},
  {Bucher}, {Burigana}, {Butler}, {Calabrese}, {Cardoso}, {Carron},
  {Challinor}, {Chiang}, {Chluba}, {Colombo}, {Combet}, {Contreras}, {Crill},
  {Cuttaia}, {de Bernardis}, {de Zotti}, {Delabrouille}, {Delouis}, {Di
  Valentino}, {Diego}, {Dor{\'e}}, {Douspis}, {Ducout}, {Dupac}, {Dusini},
  {Efstathiou}, {Elsner}, {En{\ss}lin}, {Eriksen}, {Fantaye}, {Farhang},
  {Fergusson}, {Fernandez-Cobos}, {Finelli}, {Forastieri}, {Frailis},
  {Fraisse}, {Franceschi}, {Frolov}, {Galeotta}, {Galli}, {Ganga},
  {G{\'e}nova-Santos}, {Gerbino}, {Ghosh}, {Gonz{\'a}lez-Nuevo}, {G{\'o}rski},
  {Gratton}, {Gruppuso}, {Gudmundsson}, {Hamann}, {Handley}, {Hansen},
  {Herranz}, {Hildebrandt}, {Hivon}, {Huang}, {Jaffe}, {Jones}, {Karakci},
  {Keih{\"a}nen}, {Keskitalo}, {Kiiveri}, {Kim}, {Kisner}, {Knox},
  {Krachmalnicoff}, {Kunz}, {Kurki-Suonio}, {Lagache}, {Lamarre}, {Lasenby},
  {Lattanzi}, {Lawrence}, {Le Jeune}, {Lemos}, {Lesgourgues}, {Levrier},
  {Lewis}, {Liguori}, {Lilje}, {Lilley}, {Lindholm}, {L{\'o}pez-Caniego},
  {Lubin}, {Ma}, {Mac{\'\i}as-P{\'e}rez}, {Maggio}, {Maino}, {Mandolesi},
  {Mangilli}, {Marcos-Caballero}, {Maris}, {Martin}, {Martinelli},
  {Mart{\'\i}nez-Gonz{\'a}lez}, {Matarrese}, {Mauri}, {McEwen}, {Meinhold},
  {Melchiorri}, {Mennella}, {Migliaccio}, {Millea}, {Mitra},
  {Miville-Desch{\^e}nes}, {Molinari}, {Montier}, {Morgante}, {Moss}, {Natoli},
  {N{\o}rgaard-Nielsen}, {Pagano}, {Paoletti}, {Partridge}, {Patanchon},
  {Peiris}, {Perrotta}, {Pettorino}, {Piacentini}, {Polastri}, {Polenta},
  {Puget}, {Rachen}, {Reinecke}, {Remazeilles}, {Renzi}, {Rocha}, {Rosset},
  {Roudier}, {Rubi{\~n}o-Mart{\'\i}n}, {Ruiz-Granados}, {Salvati}, {Sandri},
  {Savelainen}, {Scott}, {Shellard}, {Sirignano}, {Sirri}, {Spencer},
  {Sunyaev}, {Suur-Uski}, {Tauber}, {Tavagnacco}, {Tenti}, {Toffolatti},
  {Tomasi}, {Trombetti}, {Valenziano}, {Valiviita}, {Van Tent}, {Vibert},
  {Vielva}, {Villa}, {Vittorio}, {Wandelt}, {Wehus}, {White}, {White},
  {Zacchei}, \& {Zonca}}]{Planck20}
{Planck Collaboration}, {Aghanim}, N., {Akrami}, Y., {et~al.} 2020, \aap, 641,
  A6

\bibitem[{{Reichard} {et~al.}(2003){Reichard}, {Richards}, {Hall}, {Schneider},
  {Vanden Berk}, {Fan}, {York}, {Knapp}, \& {Brinkmann}}]{Reichard2003A}
{Reichard}, T.~A., {Richards}, G.~T., {Hall}, P.~B., {et~al.} 2003, \aj, 126,
  2594

\bibitem[{{Reines} {et~al.}(2013){Reines}, {Greene}, \& {Geha}}]{Reines13}
{Reines}, A.~E., {Greene}, J.~E., \& {Geha}, M. 2013, \apj, 775, 116

\bibitem[{{Reines} \& {Volonteri}(2015)}]{Reines15}
{Reines}, A.~E. \& {Volonteri}, M. 2015, \apj, 813, 82

\bibitem[{{Richards} {et~al.}(2003){Richards}, {Hall}, {Vanden Berk},
  {Strauss}, {Schneider}, {Weinstein}, {Reichard}, {York}, {Knapp}, {Fan},
  {Ivezi{\'c}}, {Brinkmann}, {Budav{\'a}ri}, {Csabai}, \&
  {Nichol}}]{Richards2003}
{Richards}, G.~T., {Hall}, P.~B., {Vanden Berk}, D.~E., {et~al.} 2003, \aj,
  126, 1131

\bibitem[{{Rieke} \& {the JADES Collaboration}(2023)}]{Rieke2023}
{Rieke}, M. \& {the JADES Collaboration}. 2023, arXiv e-prints,
  arXiv:2306.02466

\bibitem[{{Rigby} {et~al.}(2023){Rigby}, {Perrin}, {McElwain}, {Kimble},
  {Friedman}, {Lallo}, {Doyon}, {Feinberg}, {Ferruit}, {Glasse}, {Rieke},
  {Rieke}, {Wright}, {Willott}, {Colon}, {Milam}, {Neff}, {Stark}, {Valenti},
  {Abell}, {Abney}, {Abul-Huda}, {Acton}, {Adams}, {Adler}, {Aguilar}, {Ahmed},
  {Albert}, {Alberts}, {Aldridge}, {Allen}, {Altenburg},
  {{\'A}lvarez-M{\'a}rquez}, {Alves de Oliveira}, {Andersen}, {Anderson},
  {Anderson}, {Argyriou}, {Armstrong}, {Arribas}, {Artigau}, {Arvai},
  {Atkinson}, {Bacon}, {Bair}, {Banks}, {Barrientes}, {Barringer}, {Bartosik},
  {Bast}, {Baudoz}, {Beatty}, {Bechtold}, {Beck}, {Bergeron}, {Bergkoetter},
  {Bhatawdekar}, {Birkmann}, {Blazek}, {Blome}, {Boccaletti}, {B{\"o}ker},
  {Boia}, {Bonaventura}, {Bond}, {Bosley}, {Boucarut}, {Bourque}, {Bouwman},
  {Bower}, {Bowers}, {Boyer}, {Bradley}, {Brady}, {Braun}, {Breda},
  {Bresnahan}, {Bright}, {Britt}, {Bromenschenkel}, {Brooks}, {Brooks},
  {Brown}, {Brown}, {Brown}, {Bunker}, {Burger}, {Bushouse}, {Cale}, {Cameron},
  {Cameron}, {Canipe}, {Caplinger}, {Caputo}, {Cara}, {Carey}, {Carniani},
  {Carrasquilla}, {Carruthers}, {Case}, {Catherine}, {Chance}, {Chapman},
  {Charlot}, {Charlow}, {Chayer}, {Chen}, {Cherinka}, {Chichester}, {Chilton},
  {Chonis}, {Clampin}, {Clark}, {Clark}, {Coe}, {Coleman}, {Comber}, {Comeau},
  {Connolly}, {Cooper}, {Cooper}, {Coppock}, {Correnti}, {Cossou}, {Coulais},
  {Coyle}, {Cracraft}, {Curti}, {Cuturic}, {Davis}, {Davis}, {Dean}, {DeLisa},
  {deMeester}, {Dencheva}, {Dencheva}, {DePasquale}, {Deschenes}, {Hunor
  Detre}, {Diaz}, {Dicken}, {DiFelice}, {Dillman}, {Dixon}, {Doggett},
  {Donaldson}, {Douglas}, {DuPrie}, {Dupuis}, {Durning}, {Easmin}, {Eck},
  {Edeani}, {Egami}, {Ehrenwinkler}, {Eisenhamer}, {Eisenhower}, {Elie},
  {Elliott}, {Elliott}, {Ellis}, {Engesser}, {Espinoza}, {Etienne}, {Etxaluze},
  {Falini}, {Feeney}, {Ferry}, {Filippazzo}, {Fincham}, {Fix}, {Flagey},
  {Florian}, {Flynn}, {Fontanella}, {Ford}, {Forshay}, {Fox}, {Franz}, {Fu},
  {Fullerton}, {Galkin}, {Galyer}, {Garc{\'\i}a Mar{\'\i}n}, {Gardner},
  {Gardner}, {Garland}, {Garrett}, {Gasman}, {Gaspar}, {Gaudreau}, {Gauthier},
  {Geers}, {Geithner}, {Gennaro}, {Giardino}, {Girard}, {Giuliano},
  {Glassmire}, {Glauser}, {Glazer}, {Godfrey}, {Golimowski}, {Gollnitz},
  {Gong}, {Gonzaga}, {Gordon}, {Gordon}, {Goudfrooij}, {Greene}, {Greenhouse},
  {Grimaldi}, {Groebner}, {Grundy}, {Guillard}, {Gutman}, {Ha}, {Haderlein},
  {Hagedorn}, {Hainline}, {Haley}, {Hami}, {Hamilton}, {Hammel}, {Hansen},
  {Harkins}, {Harr}, {Hart}, {Hart}, {Hartig}, {Hashimoto}, {Haskins},
  {Hathaway}, {Havey}, {Hayden}, {Hecht}, {Heller-Boyer}, {Henriques}, {Henry},
  {Hermann}, {Hernandez}, {Hesman}, {Hicks}, {Hilbert}, {Hines}, {Hoffman},
  {Holfeltz}, {Holler}, {Hoppa}, {Hott}, {Howard}, {Howard}, {Hunter},
  {Hunter}, {Hurst}, {Husemann}, {Hustak}, {Ilinca Ignat}, {Illingworth},
  {Irish}, {Jackson}, {Jahromi}, {Jakobsen}, {James}, {James}, {Januszewski},
  {Jenkins}, {Jirdeh}, {Johnson}, {Johnson}, {Jones}, {Jones}, {Jones},
  {Jones}, {Jordan}, {Jordan}, {Jurczyk}, {Jurling}, {Kaleida}, {Kalmanson},
  {Kammerer}, {Kang}, {Kao}, {Karakla}, {Kavanagh}, {Kelly}, {Kendrew},
  {Kennedy}, {Kenny}, {Keski-kuha}, {Keyes}, {Kidwell}, {Kinzel}, {Kirk},
  {Kirkpatrick}, {Kirshenblat}, {Klaassen}, {Knapp}, {Knight}, {Knollenberg},
  {Koehler}, {Koekemoer}, {Kovacs}, {Kulp}, {Kumari}, {Kyprianou}, {La Massa},
  {Labador}, {Labiano}, {Lagage}, {Lajoie}, {Lallo}, {Lam}, {Lamb}, {Lambros},
  {Lampenfield}, {Langston}, {Larson}, {Law}, {Lawrence}, {Lee}, {Leisenring},
  {Lepo}, {Leveille}, {Levenson}, {Levine}, {Levy}, {Lewis}, {Lewis},
  {Libralato}, {Lightsey}, {Link}, {Liu}, {Lo}, {Lockwood}, {Logue}, {Long},
  {Long}, {Loomis}, {Lopez-Caniego}, {Lorenzo Alvarez}, {Love-Pruitt}, {Lucy},
  {Luetzgendorf}, {Maghami}, {Maiolino}, {Major}, {Malla}, {Malumuth},
  {Manjavacas}, {Mannfolk}, {Marrione}, {Marston}, {Martel}, {Maschmann},
  {Masci}, {Masciarelli}, {Maszkiewicz}, {Mather}, {McKenzie}, {McLean},
  {McMaster}, {Melbourne}, {Mel{\'e}ndez}, {Menzel}, {Merz}, {Meyett}, {Meza},
  {Miskey}, {Misselt}, {Moller}, {Morrison}, {Morse}, {Moseley}, {Mosier},
  {Mountain}, {Mueckay}, {Mueller}, {Mullally}, {Murphy}, {Murray}, {Murray},
  {Mustelier}, {Muzerolle}, {Mycroft}, {Myers}, {Myrick}, {Nanavati}, {Nance},
  {Nayak}, {Naylor}, {Nelan}, {Nickson}, {Nielson}, {Nieto-Santisteban},
  {Nikolov}, {Noriega-Crespo}, {O'Shaughnessy}, {O'Sullivan}, {Ochs}, {Ogle},
  {Oleszczuk}, {Olmsted}, {Osborne}, {Ottens}, {Owens}, {Pacifici}, {Pagan},
  {Page}, {Park}, {Parrish}, {Patapis}, {Paul}, {Pauly}, {Pavlovsky}, {Pedder},
  {Peek}, {Pena-Guerrero}, {Penanen}, {Perez}, {Perna}, {Perriello},
  {Phillips}, {Pietraszkiewicz}, {Pinaud}, {Pirzkal}, {Pitman}, {Piwowar},
  {Platais}, {Player}, {Plesha}, {Pollizi}, {Polster}, {Pontoppidan},
  {Porterfield}, {Proffitt}, {Pueyo}, {Pulliam}, {Quirt}, {Quispe Neira},
  {Ramos Alarcon}, {Ramsay}, {Rapp}, {Rapp}, {Rauscher}, {Ravindranath},
  {Rawle}, {Regan}, {Reichard}, {Reis}, {Ressler}, {Rest}, {Reynolds}, {Rhue},
  {Richon}, {Rickman}, {Ridgaway}, {Ritchie}, {Rix}, {Robberto}, {Robinson},
  {Robinson}, {Robinson}, {Rock}, {Rodriguez}, {Rodriguez Del Pino}, {Roellig},
  {Rohrbach}, {Roman}, {Romelfanger}, {Rose}, {Roteliuk}, {Roth}, {Rothwell},
  {Rowlands}, {Roy}, {Royer}, {Royle}, {Rui}, {Rumler}, {Runnels}, {Russ},
  {Rustamkulov}, {Ryden}, {Ryer}, {Sabata}, {Sabatke}, {Sabbi}, {Samuelson},
  {Sapp}, {Sappington}, {Sargent}, {Sauer}, {Scheithauer}, {Schlawin},
  {Schlitz}, {Schmitz}, {Schneider}, {Schreiber}, {Schulze}, {Schwab}, {Scott},
  {Sembach}, {Shanahan}, {Shaughnessy}, {Shaw}, {Shawger}, {Shay}, {Sheehan},
  {Shen}, {Sherman}, {Shiao}, {Shih}, {Shivaei}, {Sienkiewicz}, {Sing},
  {Sirianni}, {Sivaramakrishnan}, {Skipper}, {Sloan}, {Slocum}, {Slowinski},
  {Smith}, {Smith}, {Smith}, {Smith}, {Snyder}, {Soh}, {Sohn}, {Soto},
  {Spencer}, {Stallcup}, {Stansberry}, {Starr}, {Starr}, {Stewart},
  {Stiavelli}, {Straughn}, {Strickland}, {Stys}, {Summers}, {Sun}, {Sunnquist},
  {Swade}, {Swam}, {Swaters}, {Swoish}, {Taylor}, {Taylor}, {Te Plate}, {Tea},
  {Teague}, {Telfer}, {Temim}, {Thatte}, {Thompson}, {Thompson}, {Thomson},
  {Tikkanen}, {Tippet}, {Todd}, {Toolan}, {Tran}, {Trejo}, {Truong},
  {Tsukamoto}, {Tustain}, {Tyra}, {Ubeda}, {Underwood}, {Uzzo}, {Van Campen},
  {Vandal}, {Vandenbussche}, {Vila}, {Volk}, {Wahlgren}, {Waldman}, {Walker},
  {Wander}, {Warfield}, {Warner}, {Wasiak}, {Watkins}, {Weaver}, {Weilert},
  {Weiser}, {Weiss}, {Weissman}, {Welty}, {West}, {Wheate}, {Wheatley},
  {Wheeler}, {White}, {Whiteaker}, {Whitehouse}, {Whiteleather}, {Whitman},
  {Williams}, {Willmer}, {Willoughby}, {Wilson}, {Wirth}, {Wislowski}, {Wolf},
  {Wolfe}, {Wolff}, {Workman}, {Wright}, {Wu}, {Wu}, {Wymer}, {Yates},
  {Yeager}, {Yeates}, {Yerger}, {Yoon}, {Young}, {Yu}, {Zak}, {Zeidler},
  {Zhou}, {Zielinski}, {Zincke}, \& {Zonak}}]{Rigby2023}
{Rigby}, J., {Perrin}, M., {McElwain}, M., {et~al.} 2023, \pasp, 135, 048001

\bibitem[{{Risaliti} {et~al.}(1999){Risaliti}, {Maiolino}, \&
  {Salvati}}]{Risaliti1999}
{Risaliti}, G., {Maiolino}, R., \& {Salvati}, M. 1999, \apj, 522, 157

\bibitem[{{Robertson} {et~al.}(2023){Robertson}, {Tacchella}, {Johnson},
  {Hainline}, {Whitler}, {Eisenstein}, {Endsley}, {Rieke}, {Stark}, {Alberts},
  {Dressler}, {Egami}, {Hausen}, {Rieke}, {Shivaei}, {Williams}, {Willmer},
  {Arribas}, {Bonaventura}, {Bunker}, {Cameron}, {Carniani}, {Charlot},
  {Chevallard}, {Curti}, {Curtis-Lake}, {D'Eugenio}, {Jakobsen}, {Looser},
  {L{\"u}tzgendorf}, {Maiolino}, {Maseda}, {Rawle}, {Rix}, {Smit}, {{\"U}bler},
  {Willott}, {Witstok}, {Baum}, {Bhatawdekar}, {Boyett}, {Chen}, {de Graaff},
  {Florian}, {Helton}, {Hviding}, {Ji}, {Kumari}, {Lyu}, {Nelson}, {Sandles},
  {Saxena}, {Suess}, {Sun}, {Topping}, \& {Wallace}}]{Robertson2023}
{Robertson}, B.~E., {Tacchella}, S., {Johnson}, B.~D., {et~al.} 2023, Nature
  Astronomy, 7, 611

\bibitem[{{Saccheo} {et~al.}(2023){Saccheo}, {Bongiorno}, {Piconcelli},
  {Testa}, {Bischetti}, {Bisogni}, {Bruni}, {Cresci}, {Feruglio}, {Fiore},
  {Grazian}, {Luminari}, {Lusso}, {Mainieri}, {Maiolino}, {Marconi}, {Ricci},
  {Tombesi}, {Travascio}, {Vietri}, {Vignali}, {Zappacosta}, \& {La
  Franca}}]{Saccheo23}
{Saccheo}, I., {Bongiorno}, A., {Piconcelli}, E., {et~al.} 2023, \aap, 671, A34

\bibitem[{{Sandles} {et~al.}(2023){Sandles}, {D'Eugenio}, {Maiolino}, {Looser},
  {Arribas}, {Baker}, {Bonaventura}, {Bunker}, {Cameron}, {Carniani},
  {Charlot}, {Chevallard}, {Curti}, {Curtis-Lake}, {de Graaff}, {Eisenstein},
  {Hainline}, {Ji}, {Johnson}, {Jones}, {Kumari}, {Nelson}, {Perna}, {Rawle},
  {Rix}, {Robertson}, {Rodriguez Del Pino}, {Scholtz}, {Shivaei}, {Smit},
  {Sun}, {Tacchella}, {Uebler}, {Williams}, {Willott}, \&
  {Witstok}}]{Sandles23a}
{Sandles}, L., {D'Eugenio}, F., {Maiolino}, R., {et~al.} 2023, arXiv e-prints,
  arXiv:2306.03931

\bibitem[{{Santini} {et~al.}(2014){Santini}, {Maiolino}, {Magnelli}, {Lutz},
  {Lamastra}, {Li Causi}, {Eales}, {Andreani}, {Berta}, {Buat}, {Cooray},
  {Cresci}, {Daddi}, {Farrah}, {Fontana}, {Franceschini}, {Genzel}, {Granato},
  {Grazian}, {Le Floc'h}, {Magdis}, {Magliocchetti}, {Mannucci}, {Menci},
  {Nordon}, {Oliver}, {Popesso}, {Pozzi}, {Riguccini}, {Rodighiero}, {Rosario},
  {Salvato}, {Scott}, {Silva}, {Tacconi}, {Viero}, {Wang}, {Wuyts}, \&
  {Xu}}]{Santini2014}
{Santini}, P., {Maiolino}, R., {Magnelli}, B., {et~al.} 2014, \aap, 562, A30

\bibitem[{{Sassano} {et~al.}(2023){Sassano}, {Capelo}, {Mayer}, {Schneider}, \&
  {Valiante}}]{Sassano2023}
{Sassano}, F., {Capelo}, P.~R., {Mayer}, L., {Schneider}, R., \& {Valiante}, R.
  2023, \mnras, 519, 1837

\bibitem[{{Sassano} {et~al.}(2021){Sassano}, {Schneider}, {Valiante},
  {Inayoshi}, {Chon}, {Omukai}, {Mayer}, \& {Capelo}}]{Sassano2021}
{Sassano}, F., {Schneider}, R., {Valiante}, R., {et~al.} 2021, \mnras, 506, 613

\bibitem[{{Schneider} {et~al.}(2023){Schneider}, {Valiante}, {Trinca},
  {Graziani}, \& {Volonteri}}]{Schneider23}
{Schneider}, R., {Valiante}, R., {Trinca}, A., {Graziani}, L., \& {Volonteri},
  M. 2023, arXiv e-prints, arXiv:2305.12504

\bibitem[{{Scholtz} {et~al.}(2023{\natexlab{a}}){Scholtz}, {Maiolino},
  {D'Eugenio}, {Curtis-Lake}, {Carniani}, {Charlot}, {Curti}, {Silcock},
  {Arribas}, {Baker}, {Bhatawdekar}, {Boyett}, {Bunker}, {Chevallard},
  {Circosta}, {Eisenstein}, {Hainline}, {Hausen}, {Ji}, {Ji}, {Johnson},
  {Kumari}, {Looser}, {Lyu}, {Maseda}, {Parlanti}, {Perna}, {Rieke},
  {Robertson}, {Rodr{\'\i}guez Del Pino}, {Sun}, {Tacchella}, {{\"U}bler},
  {Venturi}, {Williams}, {Willmer}, {Willott}, \& {Witstok}}]{Schotlz2023NLAGN}
{Scholtz}, J., {Maiolino}, R., {D'Eugenio}, F., {et~al.} 2023{\natexlab{a}},
  arXiv e-prints, arXiv:2311.18731

\bibitem[{{Scholtz} {et~al.}(2023{\natexlab{b}}){Scholtz}, {Witten}, {Laporte},
  {Ubler}, {Perna}, {Maiolino}, {Arribas}, {Baker}, {Bennett}, {D'Eugenio},
  {Tacchella}, {Witstok}, {Bunker}, {Carniani}, {Charlot}, {Curtis-Lake},
  {Eisenstein}, {Robertson}, {Rodriguez Del Pino}, {Simmonds}, {Smit},
  {Venturi}, {Williams}, \& {Willmer}}]{Scholtz23}
{Scholtz}, J., {Witten}, C., {Laporte}, N., {et~al.} 2023{\natexlab{b}}, arXiv
  e-prints, arXiv:2306.09142

\bibitem[{{Schulze} {et~al.}(2018){Schulze}, {Misawa}, {Zuo}, \&
  {Wu}}]{Schulze2018}
{Schulze}, A., {Misawa}, T., {Zuo}, W., \& {Wu}, X.-B. 2018, \apj, 853, 167

\bibitem[{{Scialpi} {et~al.}(2023){Scialpi}, {Mannucci}, {Marconcini},
  {Venturi}, {Pancino}, {Marconi}, {Cresci}, {Belfiore}, {Amiri}, {Bertola},
  {Carniani}, {Cicone}, {Ciurlo}, {D'Amato}, {Ginolfi}, {Lusso}, {Marasco},
  {Nardini}, {Rubinur}, {Severgnini}, {Tozzi}, {Ulivi}, {Vignali}, \&
  {Volonteri}}]{Scialpi2023}
{Scialpi}, M., {Mannucci}, F., {Marconcini}, C., {et~al.} 2023, arXiv e-prints,
  arXiv:2305.11850

\bibitem[{{Scoville} {et~al.}(2017){Scoville}, {Lee}, {Vanden Bout},
  {Diaz-Santos}, {Sanders}, {Darvish}, {Bongiorno}, {Casey}, {Murchikova},
  {Koda}, {Capak}, {Vlahakis}, {Ilbert}, {Sheth}, {Morokuma-Matsui}, {Ivison},
  {Aussel}, {Laigle}, {McCracken}, {Armus}, {Pope}, {Toft}, \&
  {Masters}}]{Scoville2017}
{Scoville}, N., {Lee}, N., {Vanden Bout}, P., {et~al.} 2017, \apj, 837, 150

\bibitem[{{Sesana} {et~al.}(2004){Sesana}, {Haardt}, {Madau}, \&
  {Volonteri}}]{Sesana2004}
{Sesana}, A., {Haardt}, F., {Madau}, P., \& {Volonteri}, M. 2004, \apj, 611,
  623

\bibitem[{{Sesana} {et~al.}(2007){Sesana}, {Volonteri}, \&
  {Haardt}}]{Sesana2007}
{Sesana}, A., {Volonteri}, M., \& {Haardt}, F. 2007, \mnras, 377, 1711

\bibitem[{{Shakura} \& {Sunyaev}(1973)}]{shakura+sunyaev1973}
{Shakura}, N.~I. \& {Sunyaev}, R.~A. 1973, \aap, 24, 337

\bibitem[{{Shi} {et~al.}(2016){Shi}, {Jiang}, {Wang}, {Zhang}, {Ji}, {Liu}, \&
  {Zhou}}]{Shi2016}
{Shi}, X.-H., {Jiang}, P., {Wang}, H.-Y., {et~al.} 2016, \apj, 829, 96

\bibitem[{{Sijacki} {et~al.}(2009){Sijacki}, {Springel}, \&
  {Haehnelt}}]{Sijacki09}
{Sijacki}, D., {Springel}, V., \& {Haehnelt}, M.~G. 2009, \mnras, 400, 100

\bibitem[{{Singh} {et~al.}(2023){Singh}, {Monaco}, \& {Tan}}]{Singh23}
{Singh}, J., {Monaco}, P., \& {Tan}, J.~C. 2023, arXiv e-prints,
  arXiv:2301.11464

\bibitem[{{Somerville} {et~al.}(2004){Somerville}, {Lee}, {Ferguson},
  {Gardner}, {Moustakas}, \& {Giavalisco}}]{Somerville2004}
{Somerville}, R.~S., {Lee}, K., {Ferguson}, H.~C., {et~al.} 2004, \apjl, 600,
  L171

\bibitem[{{Stern} \& {Laor}(2012)}]{Stern12}
{Stern}, J. \& {Laor}, A. 2012, \mnras, 426, 2703

\bibitem[{{Strait} {et~al.}(2023){Strait}, {Brammer}, {Muzzin}, {Desprez},
  {Asada}, {Abraham}, {Brada{\v{c}}}, {Iyer}, {Martis}, {Mowla}, {Noirot},
  {Sarrouh}, {Sawicki}, {Willott}, {Gould}, {Grindlay}, {Matharu}, \&
  {Rihtar{\v{s}}i{\v{c}}}}]{Strait2023}
{Strait}, V., {Brammer}, G., {Muzzin}, A., {et~al.} 2023, \apjl, 949, L23

\bibitem[{{Sturm} \& {Reines}(2024)}]{Sturm2024}
{Sturm}, M.~R. \& {Reines}, A.~E. 2024, arXiv e-prints, arXiv:2406.06675

\bibitem[{{Tacchella} {et~al.}(2023){Tacchella}, {Eisenstein}, {Hainline},
  {Johnson}, {Baker}, {Helton}, {Robertson}, {Suess}, {Chen}, {Nelson},
  {Pusk{\'a}s}, {Sun}, {Alberts}, {Egami}, {Hausen}, {Rieke}, {Rieke},
  {Shivaei}, {Williams}, {Willmer}, {Bunker}, {Cameron}, {Carniani}, {Charlot},
  {Curti}, {Curtis-Lake}, {Looser}, {Maiolino}, {Maseda}, {Rawle}, {Rix},
  {Smit}, {{\"U}bler}, {Willott}, {Witstok}, {Baum}, {Bhatawdekar}, {Boyett},
  {Danhaive}, {de Graaff}, {Endsley}, {Ji}, {Lyu}, {Sandles}, {Saxena},
  {Scholtz}, {Topping}, \& {Whitler}}]{tacchella_jades_2023}
{Tacchella}, S., {Eisenstein}, D.~J., {Hainline}, K., {et~al.} 2023, arXiv
  e-prints, arXiv:2302.07234

\bibitem[{{Tacconi} {et~al.}(2020){Tacconi}, {Genzel}, \&
  {Sternberg}}]{Tacconi2020}
{Tacconi}, L.~J., {Genzel}, R., \& {Sternberg}, A. 2020, \araa, 58, 157

\bibitem[{{Terrazas} {et~al.}(2017){Terrazas}, {Bell}, {Woo}, \&
  {Henriques}}]{Terrazas17}
{Terrazas}, B.~A., {Bell}, E.~F., {Woo}, J., \& {Henriques}, B. M.~B. 2017,
  \apj, 844, 170

\bibitem[{{Trakhtenbrot} {et~al.}(2017){Trakhtenbrot}, {Volonteri}, \&
  {Natarajan}}]{Trakhtenbrot17}
{Trakhtenbrot}, B., {Volonteri}, M., \& {Natarajan}, P. 2017, \apjl, 836, L1

\bibitem[{{Trinca} {et~al.}(2023){Trinca}, {Schneider}, {Maiolino}, {Valiante},
  {Graziani}, \& {Volonteri}}]{Trinca23seekBH}
{Trinca}, A., {Schneider}, R., {Maiolino}, R., {et~al.} 2023, \mnras, 519, 4753

\bibitem[{{Trinca} {et~al.}(2022){Trinca}, {Schneider}, {Valiante}, {Graziani},
  {Zappacosta}, \& {Shankar}}]{Trinca22}
{Trinca}, A., {Schneider}, R., {Valiante}, R., {et~al.} 2022, \mnras, 511, 616

\bibitem[{{{\"U}bler} {et~al.}(2023{\natexlab{a}}){{\"U}bler}, {Maiolino},
  {Curtis-Lake}, {P{\'e}rez-Gonz{\'a}lez}, {Curti}, {Arribas}, {Charlot},
  {Perna}, {Marshall}, {D'Eugenio}, {Scholtz}, {Bunker}, {Carniani}, {Ferruit},
  {Jakobsen}, {Rix}, {Rodr{\'\i}guez Del Pino}, {Willott}, {B{\"o}ker},
  {Cresci}, {Jones}, {Kumari}, \& {Rawle}}]{ubler+2023}
{{\"U}bler}, H., {Maiolino}, R., {Curtis-Lake}, E., {et~al.}
  2023{\natexlab{a}}, arXiv e-prints, arXiv:2302.06647

\bibitem[{{{\"U}bler} {et~al.}(2023{\natexlab{b}}){{\"U}bler}, {Maiolino},
  {P{\'e}rez-Gonz{\'a}lez}, {D'Eugenio}, {Perna}, {Curti}, {Arribas}, {Bunker},
  {Carniani}, {Charlot}, {Rodr{\'\i}guez Del Pino}, {Baker}, {B{\"o}ker},
  {Cresci}, {Dunlop}, {Grogin}, {Jones}, {Kumari}, {Lamperti}, {Laporte},
  {Marshall}, {Mazzolari}, {Parlanti}, {Rawle}, {Scholtz}, {Venturi}, \&
  {Witstok}}]{Ubler24}
{{\"U}bler}, H., {Maiolino}, R., {P{\'e}rez-Gonz{\'a}lez}, P.~G., {et~al.}
  2023{\natexlab{b}}, arXiv e-prints, arXiv:2312.03589

\bibitem[{{{\"U}bler} {et~al.}(2024){{\"U}bler}, {Maiolino},
  {P{\'e}rez-Gonz{\'a}lez}, {D'Eugenio}, {Perna}, {Curti}, {Arribas}, {Bunker},
  {Carniani}, {Charlot}, {Rodr{\'\i}guez Del Pino}, {Baker}, {B{\"o}ker},
  {Cresci}, {Dunlop}, {Grogin}, {Jones}, {Kumari}, {Lamperti}, {Laporte},
  {Marshall}, {Mazzolari}, {Parlanti}, {Rawle}, {Scholtz}, {Venturi}, \&
  {Witstok}}]{Ubler2024}
{{\"U}bler}, H., {Maiolino}, R., {P{\'e}rez-Gonz{\'a}lez}, P.~G., {et~al.}
  2024, \mnras, 531, 355

\bibitem[{{Valentini} {et~al.}(2021){Valentini}, {Gallerani}, \&
  {Ferrara}}]{valentini2021}
{Valentini}, M., {Gallerani}, S., \& {Ferrara}, A. 2021, \mnras, 507, 1

\bibitem[{{Valiante} {et~al.}(2016){Valiante}, {Schneider}, {Volonteri}, \&
  {Omukai}}]{valiant2016}
{Valiante}, R., {Schneider}, R., {Volonteri}, M., \& {Omukai}, K. 2016, \mnras,
  457, 3356

\bibitem[{{Valiante} {et~al.}(2018){Valiante}, {Schneider}, {Zappacosta},
  {Graziani}, {Pezzulli}, \& {Volonteri}}]{Valiante2018a}
{Valiante}, R., {Schneider}, R., {Zappacosta}, L., {et~al.} 2018, \mnras, 476,
  407

\bibitem[{{van der Wel} {et~al.}(2022){van der Wel}, {van Houdt}, {Bezanson},
  {Franx}, {D'Eugenio}, {Straatman}, {Bell}, {Muzzin}, {Sobral}, {Maseda}, {de
  Graaff}, \& {Holden}}]{vdWel22}
{van der Wel}, A., {van Houdt}, J., {Bezanson}, R., {et~al.} 2022, \apj, 936, 9

\bibitem[{{Vanden Berk} {et~al.}(2001){Vanden Berk}, {Richards}, {Bauer},
  {Strauss}, {Schneider}, {Heckman}, {York}, {Hall}, {Fan}, {Knapp},
  {Anderson}, {Annis}, {Bahcall}, {Bernardi}, {Briggs}, {Brinkmann}, {Brunner},
  {Burles}, {Carey}, {Castander}, {Connolly}, {Crocker}, {Csabai}, {Doi},
  {Finkbeiner}, {Friedman}, {Frieman}, {Fukugita}, {Gunn}, {Hennessy},
  {Ivezi{\'c}}, {Kent}, {Kunszt}, {Lamb}, {Leger}, {Long}, {Loveday}, {Lupton},
  {Meiksin}, {Merelli}, {Munn}, {Newberg}, {Newcomb}, {Nichol}, {Owen}, {Pier},
  {Pope}, {Rockosi}, {Schlegel}, {Siegmund}, {Smee}, {Snir}, {Stoughton},
  {Stubbs}, {SubbaRao}, {Szalay}, {Szokoly}, {Tremonti}, {Uomoto}, {Waddell},
  {Yanny}, \& {Zheng}}]{Vandenberk2001}
{Vanden Berk}, D.~E., {Richards}, G.~T., {Bauer}, A., {et~al.} 2001, \aj, 122,
  549

\bibitem[{{Vanzella} {et~al.}(2023){Vanzella}, {Loiacono}, {Bergamini},
  {Mestric}, {Castellano}, {Rosati}, {Meneghetti}, {Grillo}, {Calura},
  {Mignoli}, {Bradac}, {Adamo}, {Rihtarsic}, {Dickinson}, {Gronke}, {Zanella},
  {Annibali}, {Willott}, {Messa}, {Sani}, {Acebron}, {Bolamperti}, {Comastri},
  {Gilli}, {Caputi}, {Ricotti}, {Gruppioni}, {Ravindranath}, {Mercurio},
  {Strait}, {Martis}, {Pascale}, {Caminha}, \& {Annunziatella}}]{Vanzella23}
{Vanzella}, E., {Loiacono}, F., {Bergamini}, P., {et~al.} 2023, arXiv e-prints,
  arXiv:2305.14413

\bibitem[{{Veilleux} \& {Osterbrock}(1987)}]{Veilleux1987}
{Veilleux}, S. \& {Osterbrock}, D.~E. 1987, \apjs, 63, 295

\bibitem[{{Visbal} \& {Haiman}(2018)}]{Visbal2018}
{Visbal}, E. \& {Haiman}, Z. 2018, \apjl, 865, L9

\bibitem[{{Vito} {et~al.}(2022){Vito}, {Di Mascia}, {Gallerani}, {Zana},
  {Ferrara}, {Carniani}, \& {Gilli}}]{Vito2022}
{Vito}, F., {Di Mascia}, F., {Gallerani}, S., {et~al.} 2022, \mnras, 514, 1672

\bibitem[{{Volonteri}(2010)}]{volonteri2010}
{Volonteri}, M. 2010, \aapr, 18, 279

\bibitem[{{Volonteri} {et~al.}(2021){Volonteri}, {Habouzit}, \&
  {Colpi}}]{Volonteri21rev}
{Volonteri}, M., {Habouzit}, M., \& {Colpi}, M. 2021, Nature Reviews Physics,
  3, 732

\bibitem[{{Volonteri} {et~al.}(2023){Volonteri}, {Habouzit}, \&
  {Colpi}}]{Volonteri23}
{Volonteri}, M., {Habouzit}, M., \& {Colpi}, M. 2023, \mnras, 521, 241

\bibitem[{{Volonteri} {et~al.}(2022){Volonteri}, {Pfister}, {Beckmann},
  {Dotti}, {Dubois}, {Massonneau}, {Musoke}, \&
  {Tremmel}}]{Volonteri22_dualAGN}
{Volonteri}, M., {Pfister}, H., {Beckmann}, R., {et~al.} 2022, \mnras, 514, 640

\bibitem[{{Volonteri} {et~al.}(2020){Volonteri}, {Pfister}, {Beckmann},
  {Dubois}, {Colpi}, {Conselice}, {Dotti}, {Martin}, {Jackson}, {Kraljic},
  {Pichon}, {Trebitsch}, {Yi}, {Devriendt}, \& {Peirani}}]{Volonteri20_merging}
{Volonteri}, M., {Pfister}, H., {Beckmann}, R.~S., {et~al.} 2020, \mnras, 498,
  2219

\bibitem[{{Wang} {et~al.}(2020){Wang}, {Davies}, {Yang}, {Hennawi}, {Fan},
  {Barth}, {Jiang}, {Wu}, {Mudd}, {Ba{\~n}ados}, {Bian}, {Decarli}, {Eilers},
  {Farina}, {Venemans}, {Walter}, \& {Yue}}]{Wang20}
{Wang}, F., {Davies}, F.~B., {Yang}, J., {et~al.} 2020, \apj, 896, 23

\bibitem[{{Weller} {et~al.}(2023){Weller}, {Pacucci}, {Ni}, {Chen}, {Di
  Matteo}, {Siwek}, \& {Hernquist}}]{Weller2023}
{Weller}, E.~J., {Pacucci}, F., {Ni}, Y., {et~al.} 2023, \mnras, 520, 3955

\bibitem[{{Williams} {et~al.}(2017){Williams}, {Maiolino}, {Krongold},
  {Carniani}, {Cresci}, {Mannucci}, \& {Marconi}}]{Williams2017}
{Williams}, R.~J., {Maiolino}, R., {Krongold}, Y., {et~al.} 2017, \mnras, 467,
  3399

\bibitem[{{Willott} {et~al.}(2005){Willott}, {Percival}, {McLure}, {Crampton},
  {Hutchings}, {Jarvis}, {Sawicki}, \& {Simard}}]{Willott2015}
{Willott}, C.~J., {Percival}, W.~J., {McLure}, R.~J., {et~al.} 2005, \apj, 626,
  657

\bibitem[{{Witstok} {et~al.}(2023{\natexlab{a}}){Witstok}, {Jones}, {Maiolino},
  {Smit}, \& {Schneider}}]{Witstok23ALMA}
{Witstok}, J., {Jones}, G.~C., {Maiolino}, R., {Smit}, R., \& {Schneider}, R.
  2023{\natexlab{a}}, \mnras, 523, 3119

\bibitem[{{Witstok} {et~al.}(2023{\natexlab{b}}){Witstok}, {Shivaei}, {Smit},
  {Maiolino}, {Carniani}, {Curtis-Lake}, {Ferruit}, {Arribas}, {Bunker},
  {Cameron}, {Charlot}, {Chevallard}, {Curti}, {de Graaff}, {D'Eugenio},
  {Giardino}, {Looser}, {Rawle}, {Rodr{\'\i}guez del Pino}, {Willott},
  {Alberts}, {Baker}, {Boyett}, {Egami}, {Eisenstein}, {Endsley}, {Hainline},
  {Ji}, {Johnson}, {Kumari}, {Lyu}, {Nelson}, {Perna}, {Rieke}, {Robertson},
  {Sandles}, {Saxena}, {Scholtz}, {Sun}, {Tacchella}, {Williams}, \&
  {Willmer}}]{Witstok23UVhump}
{Witstok}, J., {Shivaei}, I., {Smit}, R., {et~al.} 2023{\natexlab{b}}, arXiv
  e-prints, arXiv:2302.05468

\bibitem[{{Woo} {et~al.}(2015){Woo}, {Yoon}, {Park}, {Park}, \&
  {Kim}}]{Woo2015}
{Woo}, J.-H., {Yoon}, Y., {Park}, S., {Park}, D., \& {Kim}, S.~C. 2015, \apj,
  801, 38

\bibitem[{{Yang} {et~al.}(2023{\natexlab{a}}){Yang}, {Caputi}, {Papovich},
  {Arrabal Haro}, {Bagley}, {Behroozi}, {Bell}, {Bisigello}, {Buat},
  {Burgarella}, {Cheng}, {Cleri}, {Dav{\'e}}, {Dickinson}, {Elbaz}, {Ferguson},
  {Finkelstein}, {Grogin}, {Hathi}, {Hirschmann}, {Holwerda},
  {Huertas-Company}, {Hutchison}, {Iani}, {Kartaltepe}, {Kirkpatrick},
  {Kocevski}, {Koekemoer}, {Kokorev}, {Larson}, {Lucas},
  {P{\'e}rez-Gonz{\'a}lez}, {Rinaldi}, {Shen}, {Trump}, {de la Vega}, {Yung},
  \& {Zavala}}]{Yang2023}
{Yang}, G., {Caputi}, K.~I., {Papovich}, C., {et~al.} 2023{\natexlab{a}},
  \apjl, 950, L5

\bibitem[{{Yang} {et~al.}(2023{\natexlab{b}}){Yang}, {Wang}, {Fan}, {Hennawi},
  {Barth}, {Ba{\~n}ados}, {Sun}, {Liu}, {Cai}, {Jiang}, {Li}, {Onoue},
  {Schindler}, {Shen}, {Wu}, {Bhowmick}, {Bieri}, {Blecha}, {Bosman},
  {Champagne}, {Colina}, {Connor}, {Costa}, {Davies}, {Decarli}, {De Rosa},
  {Drake}, {Egami}, {Eilers}, {Evans}, {Farina}, {Habouzit}, {Haiman}, {Jin},
  {Jun}, {Kakiichi}, {Khusanova}, {Kulkarni}, {Loiacono}, {Lupi},
  {Mazzucchelli}, {Pan}, {Rojas-Ruiz}, {Strauss}, {Tee}, {Trakhtenbrot},
  {Trebitsch}, {Venemans}, {Vestergaard}, {Volonteri}, {Walter}, {Xie}, {Yue},
  {Zhang}, {Zhang}, \& {Zou}}]{Yang_2023_quasar}
{Yang}, J., {Wang}, F., {Fan}, X., {et~al.} 2023{\natexlab{b}}, \apjl, 951, L5

\bibitem[{{Zhang} {et~al.}(2015){Zhang}, {Zhou}, {Wang}, {Wang}, {Shi}, {Liu},
  {Liu}, {Li}, \& {Wang}}]{Zhang2015}
{Zhang}, S., {Zhou}, H., {Wang}, T., {et~al.} 2015, \apj, 803, 58

\bibitem[{{Zhang} {et~al.}(2021){Zhang}, {Kauffmann}, {Wang}, {Chen}, {Fu}, \&
  {Wu}}]{Zhang2021}
{Zhang}, W., {Kauffmann}, G., {Wang}, J., {et~al.} 2021, \aap, 648, A25

\end{thebibliography}
